\documentclass[fleqn,usenatbib]{mnras}

\usepackage{newtxtext,newtxmath}
\usepackage[T1]{fontenc}

\DeclareRobustCommand{\VAN}[3]{#2}
\let\VANthebibliography\thebibliography
\def\thebibliography{\DeclareRobustCommand{\VAN}[3]{##3}\VANthebibliography}

\usepackage{tikz}
\usepackage{scalerel}
\usetikzlibrary{svg.path}
\definecolor{orcidlogocol}{HTML}{A6CE39}
\tikzset{orcidlogo/.pic={\fill[orcidlogocol] svg{M256,128c0,70.7-57.3,128-128,128C57.3,256,0,198.7,0,128C0,57.3,57.3,0,128,0C198.7,0,256,57.3,256,128z}; \fill[white] svg{M86.3,186.2H70.9V79.1h15.4v48.4V186.2z} svg{M108.9,79.1h41.6c39.6,0,57,28.3,57,53.6c0,27.5-21.5,53.6-56.8,53.6h-41.8V79.1z M124.3,172.4h24.5c34.9,0,42.9-26.5,42.9-39.7c0-21.5-13.7-39.7-43.7-39.7h-23.7V172.4z} svg{M88.7,56.8c0,5.5-4.5,10.1-10.1,10.1c-5.6,0-10.1-4.6-10.1-10.1c0-5.6,4.5-10.1,10.1-10.1C84.2,46.7,88.7,51.3,88.7,56.8z};}}
\newcommand\orcidicon[1]{\href{https://orcid.org/#1}{\mbox{\scalerel*{
\begin{tikzpicture}[yscale=-1,transform shape]\pic{orcidlogo};
\end{tikzpicture}}{|}}}}

\usepackage{algorithm}
\usepackage{algorithmic}
\usepackage{array}
\usepackage{graphicx}	% Including figure files

\usepackage{amsmath}	% Advanced maths commands
\usepackage{mathtools}
\usepackage{multicol}
\usepackage{caption}
\usepackage{subcaption}
\usepackage{lipsum}
\usepackage[referable]{threeparttablex}

\newcommand{\mach}{\mathcal{M}}
\newcommand{\macha}{\mathcal{M}_\mathrm{A}}
\newcommand{\tturb}{t_{\mathrm{turb}}}
\newcommand{\Ekin}{E_{\mathrm{kin}}}
\newcommand{\Emag}{E_{\mathrm{mag}}}
\newcommand{\Pkin}{P_{\mathrm{kin}}}
\newcommand{\Pmag}{P_{\mathrm{mag}}}
\newcommand{\Akin}{A_{\mathrm{kin}}}
\newcommand{\Amag}{A_{\mathrm{mag}}}
\newcommand{\pkin}{p_{\mathrm{kin}}}
\newcommand{\pmag}{p_{\mathrm{mag}}}
\newcommand{\knu}{{k_\nu}}
\newcommand{\knut}{\Tilde{k}_\nu}
\newcommand{\keta}{{k_\eta}}
\newcommand{\ketat}{\Tilde{k}_\eta}
\newcommand{\kbn}{k_{\mathrm{bn}}}
\newcommand{\pnu}{p_\nu}
\newcommand{\peta}{p_\eta}
\newcommand{\pbn}{p_{\mathrm{bn}}}
\newcommand{\rek}{{\mathrm{Re}}}
\newcommand{\rekt}{{\widetilde{\rek}}}
\newcommand{\rekpre}{{\rek}_{\mathrm{pred}}}
\newcommand{\rektilpre}{{\rekt}_{\mathrm{pred}}}
\newcommand{\rem}{{\mathrm{Rm}}}
\newcommand{\pem}{{\mathrm{Pm}}}
\newcommand{\kturb}{k_\mathrm{turb}}
\newcommand{\kbox}{k_\mathrm{box}}
\newcommand{\uturb}{u_\mathrm{turb}}
\newcommand{\va}{v_\mathrm{A}}
\newcommand{\cs}{c_\mathrm{s}}
\newcommand{\lturb}{\ell_{\mathrm{turb}}}
\newcommand{\cnu}{c_\nu}
\newcommand{\ceta}{c_\eta}
\newcommand{\clin}{c_{\mathrm{lin}}}
\newcommand{\nrek}{N_\rek}
\newcommand{\nrem}{N_\rem}
\newcommand{\prek}{p_\rek}
\newcommand{\prekth}{p_{\rek_{\mathrm{theo}}}}
\newcommand{\prem}{p_\rem}

\newcommand{\ppem}{p_\pem}
\newcommand{\pmnot}{\pem_{0}}
\newcommand{\nunum}{\nu_{N}}
\newcommand{\etanum}{\eta_{N}}

\newcommand{\rmp}{Reviews of Modern Physics}
\newcommand{\jfm}{Journal of Fluid Mechanics}

\title[Numerical viscosity and resistivity]{Numerical viscosity and resistivity in MHD turbulence simulations}

\author[L.~Malvadi Shivakumar \& C.~Federrath]{
Lakshmi Malvadi Shivakumar,$^{\orcidicon{0009-0003-7789-4516}\,1,2}$\thanks{E-mail: \url{lakshmimalva@iisc.ac.in}}
Christoph Federrath$^{\orcidicon{0000-0002-0706-2306}\,1,3}$\thanks{E-mail: \url{christoph.federrath@anu.edu.au}}
\\
$^{1}$Research School of Astronomy \& Astrophysics, Australian National University, Canberra, ACT 2611, Australia\\
$^{2}$Indian Institute of Science, Bangalore
560012, India\\
$^{3}$Australian Research Council Centre of Excellence in All Sky Astrophysics (ASTRO3D), Canberra, ACT 2611, Australia\\
}

\date{Accepted XXX. Received YYY; in original form ZZZ}

\pubyear{2023}

\begin{document}
\label{firstpage}
\pagerange{\pageref{firstpage}--\pageref{lastpage}}
\maketitle

% Abstract of the paper
\begin{abstract}
For magnetohydrodynamical (MHD) turbulence simulations to accurately capture the underlying physics, we must understand numerical dissipation. Here we quantify numerical viscosity and resistivity in the subsonic and supersonic turbulence regimes, with Mach numbers $\mach = 0.1$ and $10$, respectively. We find that the hydrodynamic ($\rek$) and magnetic Reynolds numbers ($\rem$) on the turbulence driving scale $\lturb$ in a cubic domain of side length $L$ with a total of $N^3$ resolution elements are well-described by $\rek=[2(N/\nrek)(\lturb/L)]^{\prek}$ and $\rem=[2(N/\nrem)(\lturb/L)]^{\prem}$. We provide two sets of fit values of $(\nrek, \prek, \nrem, \prem)$: one with $\prek$ and $\prem$ fixed at their theoretical values, and the other one allowing all 4~parameters to vary. The sets for $\mach=0.1$ are $(1.57_{-0.12}^{+0.10},4/3,1.55_{-0.14}^{+0.45},4/3)$ and $(0.83_{-0.08}^{+0.09},1.20_{-0.02}^{+0.02},4.19_{-4.05}^{+2.95},1.60_{-0.33}^{+0.18})$, respectively. For $\mach=10$, they are $(3.55_{-0.56}^{+0.78},3/2,1.03_{-0.11}^{+0.12},3/2)$ and $(10.46_{-0.85}^{+0.96},1.90_{-0.04}^{+0.04},0.44_{-0.23}^{+0.61},1.32_{-0.09}^{+0.17})$. The resulting magnetic Prandtl numbers ($\pem=\rem/\rek$) are consistent with constant values of $1.0_{-0.2}^{+0.3}$ for $\mach = 0.1$, and $6.2_{-4.8}^{+5.6}$ for $\mach = 10$. These results apply when the magnetic energy ($\Emag$) is $\lesssim10\%$ of the turbulent kinetic energy ($\Ekin$). When $\Emag/\Ekin\sim0.1-1$, $\rem$ is reduced by a factor $\sim3$ (implying an increase in $\nrem$ by a factor $\sim2$) for $\mach=0.1$, while $\rem$ for $\mach=10$ and $\rek$ (for any $\mach$) remain largely unaffected. We compare our \mbox{$\rek$--$N$} relation with 14~other simulations from the literature, which use a large range of different numerical methods (with and without Riemann solvers, different reconstruction schemes and orders, and smoothed particle hydrodynamics), and find that they all agree with the \mbox{$\rek$--$N$} relations above to within a factor of $3$. We further compare these results to target $\rek$ and $\rem$ values in simulations using explicit dissipation from the literature. These literature comparisons and our relations allow users to assess what value of $\rek$ and $\rem$ can be reached at a given $N$, ensuring that physical dissipation dominates over numerical dissipation.
\end{abstract}

\begin{keywords}
methods:numerical -- viscosity -- resistivity -- turbulence -- MHD -- dynamo -- magnetic fields  
\end{keywords}

%%%%%%%%%%%%%%%%% BODY OF PAPER %%%%%%%%%%%%%%%%%%

\section{Introduction}

Turbulence is a critical physical process, with relevance in weather prediction, climate models, the oceans \citep{Gargett1989,SohailEtAl2019}, in engines, industrial burners, and turbines \citep{GiustiMastorakos2019,LeggettZhaoSandberg2023}, as well as in the aerospace \citep{ParkLintonThornber2022} and automotive industries \citep{IgaliEtAl2019}. However, turbulence also plays a crucial role in astrophysical systems, especially in the interstellar medium of galaxies \citep{2001RvMP...73.1031F,ElmegreenScalo2004,MacLowKlessen2004,2007ARA&A..45..565M,2012A&ARv..20...55H}, with particular relevance for star formation \citep{PadoanEtAl2014,2018PhT....71f..38F}.

Turbulence is intrinsically three-dimensional (3D) and complex, with non-linear interactions being a key to generating turbulent flows, which are impossible to tackle via analytic calculations. Thus, we rely on numerical solutions of the turbulent systems by solving the governing hydrodynamical equations on a set of particles or grid cells \citep[e.g.,][]{PriceFederrath2010}, via discretisation of the continuous fluid equations. Discretisation of the hydrodynamical equations in time (time-step) and space (grid-size) introduces second-order difference terms that have viscous-like effects, particularly in regions where strong gradients (shearing motions or shocks) are present. These numerical effects are often referred to as numerical dissipation, because their mode of operation is similar to that of physical dissipation. The physical (explicit) terms may be included in numerical simulations \citep{laney1998computational,2007nmai.conf.....B,2011A&A...528A.106S,2011JCoPh.230.3331W,2019IJNMF..89...16D}, however, the numerical (implicit) contributions to dissipation are always present as well. 

In addition to purely hydrodynamical turbulence, magnetic fields are an essential component in astrophysical systems over a wide range of scales spanning from planets \citep{2003E&PSL.208....1S,2011AnRFM..43..583J,2021A&ARv..29....1K} to stars \citep{2015MNRAS.450.4035F,2021MNRAS.503.2014S}, galaxies \citep{1988ASSL..133.....R,2013pss5.book..641B,2021ApJ...907....2S}, accretion discs around black holes \citep{2010MNRAS.408..752P,2014Natur.510..126Z}, and even the early Universe \citep{2010AN....331..110S,2010ApJ...721L.134S}. These magnetic fields are subject to interaction with the fluid motions, including reconnection events \citep{LazarianEtAl2020}. Furthermore, magnetic fields can be amplified in turbulent systems by a mechanism called the 'turbulent dynamo', by converting turbulent kinetic energy to magnetic energy \citep{2002RvMP...74..775W,2005PhR...417....1B,2008RPPh...71d6901K,2016JPlPh..82f5301F,2019JPlPh..85d2001R,2021PhRvL.126i1103A,2020MNRAS.499.2076S,2022MNRAS.514..957S,HewFederrath2023}. The turbulent dynamo is usually simulated in turbulent systems using the governing magneto-hydrodynamical (MHD) equations \citep[although there is more recent work using weakly-collisional plasma conditions][]{StOngeKunz2018,AchikanathEtAl2024}. In the early stages of the turbulent dynamo, called the kinematic phase, the turbulent fluid motion induces stretching, twisting, folding, and merging of the magnetic field lines, leading to the amplification of the magnetic field \citep{2005PhR...417....1B,2004PhRvL..92e4502S,2016JPlPh..82f5301F}. As the magnetic field grows, it begins to influence the fluid motion by exerting magnetic forces (Lorentz force) on the fluid. These magnetic forces modify the turbulent motions, ultimately leading to the saturation of the dynamo. During the saturation phase, the magnetic field can reaches a level where the energy gained from the turbulent motions is balanced by the energy dissipated through various processes, such as dissipation and diffusion. Therefore, the back-reaction of the magnetic fields through the Lorentz force is significant enough to suppress further amplification of the magnetic field.

In turbulent fluids, energy transfers from large to small scales \citep{1991RSPSA.434....9K,1995tlan.book.....F,2008JFM...605..355G}. As the energy cascades to smaller scales, it eventually reaches a spatial scale where it is dissipated due to the effects of viscosity and resistivity \citep{1991RSPSA.434...15K,ertesvaag2000eddy}. The Fourier wave numbers associated with these dissipation scales are called viscous dissipation wave number $\knu$ and resistive dissipation wave number $\keta$, respectively. The corresponding hydrodynamic Reynolds number ($\rek$) is defined as
\begin{equation}
    \rek = \frac{u_\mathrm{turb}\,\lturb}{\nu},
    \label{eq:Re}
\end{equation}
where $u_\mathrm{turb}$ is the fluid turbulent velocity dispersion at the turbulence driving scale $\lturb = 2\pi/\kturb$, with $\kturb$ being the turbulence driving wave number, and $\nu$ is the kinematic viscosity of the fluid. Eq.~(\ref{eq:Re}) characterises the relative contribution of inertial forces to viscous forces in a fluid flow. 

Similarly, the magnetic Reynolds number ($\rem$) is defined as
\begin{equation}
    \rem = \frac{u_\mathrm{turb}\,\lturb}{\eta},
    \label{eq:Rm}
\end{equation}
where $\eta$ is the magnetic resistivity. Eq.~(\ref{eq:Rm}) represents the ratio between the induction forces and magnetic dissipation.

The magnetic Prandtl number ($\pem$), which controls the scale separation between $\knu$ and $\keta$, is given by
\begin{equation}
    \pem = \frac{\rem}{\rek}.
    \label{eq:pm_rel}
\end{equation}

Together, $\rek$, $\rem$, and $\pem$, represent the main plasma parameters, crucially controlling the dynamics, structure, and evolution of magnetised turbulent flows, and thus, it is critical to know them.

The degree of numerical viscosity~($\nunum$) or resistivity~($\etanum$) must be sufficiently lower than the chosen explicit viscosity~($\nu$) or resistivity~($\eta$), i.e., $\nu>\nunum$ and $\eta>\etanum$, for a simulation to avoid excessive smearing of features or over-damping of the flow. Otherwise, the physical state of the system significantly deviates from the expectations set by explicit viscosity or resistivity \citep{2007JCoPh.226.1902T}. This implies that the associated numerical Reynolds numbers must be greater than the explicit Reynolds numbers. 

The problem faced in numerical simulations is that the lower the grid resolution, the greater will be the effects of numerical dissipation. We know that the spatial scales $l_{\nu} = 2\pi/\knu$ and $l_{\eta} = 2\pi/\keta$, at which kinetic and magnetic dissipation occurs, are directly related to the corresponding dissipation terms (viscosity $\nu$ and resistivity $\eta$). Therefore, the lower the target viscosity, the greater the grid resolution required to resolve the spatial scale to capture the dissipation range of the turbulent fluid \citep{2004PhRvE..70a6308H,BalsaraEtAl2004,2005ApJ...625L.115S,2011ApJ...731...62F}. One way to estimate the effects of numerical viscosity and resistivity of a simulation is to perform convergence tests. By running several simulations with increasing $N$, the resulting numerical $\nunum$ and $\etanum$ are reduced until the target explicit $\nunum<\nu$ and $\etanum<\eta$. However, resolution studies are time-consuming and computationally costly. 

Thus, even though the target (explicit) Reynolds number is set for a simulation, we have to ensure that the grid resolution of the simulation is sufficient to capture the dissipation range set by the explicit dissipation terms. If not, the effects of the numerical dissipation will dominate and the Reynolds number of the simulation will technically be the numerical Reynolds number and not the target Reynolds number. Therefore, it is critical to have an estimate of the numerical viscosity or resistivity corresponding to a particular grid resolution.

Therefore, in this study, we determine the Reynolds numbers and characteristic dissipation wave numbers associated with numerical viscosity and resistivity for a given linear grid resolution $N$, by studying their variations with $N$, and establishing empirical relations for ideal MHD simulations. This provides us with estimates of the numerical viscosity and resistivity for a given $N$, thereby verifying whether the target Reynolds numbers as per expectation can be achieved. Similar studies have been performed for other physical applications than MHD turbulence, showing that dissipation depends on the number of resolution elements employed \citep{ZhouEtAl2014,RembiaszEtAl2017,NixonPringlePringle2024}, therefore providing motivation for the present study on MHD turbulence in the subsonic and supersonic regimes of turbulence \citep[extending work by][]{2023ApJ...942L..34G}.

In Section~\ref{sec:methods}, we introduce our simulation methods, suite of simulation models, and describe the formulations for fitting kinetic and magnetic spectra. Section~\ref{sec:results} presents the main simulation results, including analyses of the time evolution, the spatial structure of the gas, and the spectra from which we extract the viscous and resistive dissipation scales. In Section~\ref{sec:rermpm}, we convert the measured dissipation scales to their respective Reynolds numbers and provide models for their dependence on the numerical grid resolution. While most of the study focuses on MHD turbulence in the regime of weak magnetic fields (kinematic phase of the turbulent dynamo), we also study the dissipation properties of a subset of our simulations in the strong-field regime (saturation phase of the dynamo) in Sec.~\ref{ssec:sat_results} and Appendix~\ref{ap:sat_reg_time_evol}. We further extend our main results, which focus on the subsonic (Mach~0.1) and supersonic (Mach~10) regimes of turbulence, to also study the Mach number dependence of the dissipation properties with Mach numbers ranging from $\sim10^{-2}$ to $10^2$ in Appendix~\ref{ap:ronmach}. In Section~\ref{sec:link} we compare our results to other works in the literature, including simulations that explicitly aim to control the Reynolds numbers, and in \S\ref{ssec:num_scheme} we quantify the dependence of our relations on the numerical scheme/method, by comparing to several different codes/schemes/methods, including smoothed particle hydrodynamics (SPH). We summarise our results in Section~\ref{sec:conclusions}.

\section{Methods} \label{sec:methods}

\subsection{Simulations}
\label{ssec:simul}
\subsubsection{Basic Equations}
We consider the ideal, compressible magnetohydrodynamic equations (MHD) for isothermal plasma, given by
\begin{align}
    \frac{\partial\rho}{\partial t} + \nabla\cdot(\rho\textbf{u}) &= 0
    \label{eq:dens_rate}, \\
    \rho\left(\frac{\partial}{\partial t}+\textbf{u}\cdot\nabla\right)\textbf{u} &= \frac{1}{4\pi}(\textbf{B}\cdot\nabla)\textbf{B} - \nabla P_{\mathrm{tot}} + \nabla\cdot(2\nu\rho\textbf{S}) + \rho\textbf{F}
    \label{eq:mom_eq}, \\
    \frac{\partial\textbf{B}}{\partial t} &= \nabla\times(\textbf{u}\times\textbf{B})+\eta\nabla^2\textbf{B}
    \label{eq:B_rate}, \\
    \nabla\cdot\textbf{B} &= 0
    \label{eq:B_div}, 
\end{align}
where $\nu$ is the kinematic viscosity, $\eta$ is the magnetic resistivity, $\rho$ is the fluid density, $\textbf{u}$ is the fluid velocity, $\textbf{B}$ is the magnetic field, $P_{\mathrm{tot}} = P_{\mathrm{th}}+|\textbf{B}|^2/(8\pi)$ is the total pressure, with the thermal pressure $P_{\mathrm{th}} = \cs^2\rho$ ($\cs$ is the constant speed of sound) and the magnetic pressure ${|\textbf{B}|^2}/{8\pi}$, and $\textbf{F}$ is the turbulent acceleration field (discussed in \S\ref{sssec:turb_driv}). The strain rate tensor $S_{ij} = (1/2)(\partial_i u_j+\partial_j u_i)-(1/3){\delta_{ij}\nabla\cdot\textbf{u}}$ in the momentum equation, Eq.~(\ref{eq:mom_eq}), incorporates the viscous dissipation rate. The energy equation is not included because the fluid is isothermal.

In the ideal-MHD case, $\nu$ and $\eta$ are set to zero in the equations above, i.e., the respective dissipation terms are not included. However, as discussed in the introduction, numerical dissipation is always present, because the system of equations is solved in a discretised way, i.e., here on a numerical grid with $N^3$ grid cells. These cells have a finite spacing, which gives rise to numerical viscosity and resistivity. However, the numerical dissipation in grid-based codes has a similar effect as to what is mathematically described by the physical dissipation terms in Eqs.~(\ref{eq:mom_eq}) and (\ref{eq:B_rate}), i.e., $\nabla\cdot(2\nu\rho\textbf{S})$ and $\eta\nabla^2\textbf{B}$, however, with effective values of $\nu$ and $\eta$ that depend on $N$. Therefore, throughout our study, $\nu$ and $\eta$ refer to the numerical viscosity and resistivity, respectively, unless otherwise stated.

We solve Eqs.~(\ref{eq:dens_rate})--(\ref{eq:B_div}) with a modified version of the {\fontfamily{qcr}\selectfont FLASH} code on a uniformly discretised, triply-periodic, 3D grid with a box size length of $L$, for 6~numerical resolutions (total number of grid cells $N^3$), with $N=144$, 288, 576, 1152 and 2304, by utilising the HLL5R, 5-wave, approximate Riemann solver \citep{bouchut2007multiwave,bouchut2010multiwave,2011JCoPh.230.3331W}. This scheme is 2nd-order accurate in space and time, and uses an approximate Riemann solver, which has been compared to the more widely known Roe and HLLE Riemann solvers \citep{2011JCoPh.230.3331W}.

\subsubsection{Turbulence Driving}
\label{sssec:turb_driv}
To drive turbulence, we use the Ornstein-Uhlenbeck process \citep{EswaranPope1988,SchmidtHillebrandtNiemeyer2006,2010A&A...512A..81F} implemented in the publicly available \texttt{TurbGen} \citep{FederrathEtAl2022ascl} code, which generates the turbulent acceleration field $\textbf{F}$ in Eq.~(\ref{eq:mom_eq}).

The turbulence driving amplitude is a paraboloid in Fourier space (with wave number $k$), which peaks at $k = \kturb = 2$, where we measure $k$ in units of $\kbox = 2\pi/L$, and the amplitude is set to 0 for $k\leq1$ and $k\geq3$. The driving amplitudes are adjusted such that the desired sonic Mach number ($\mach=\uturb/\cs$) is 0.1 for the subsonic regime and $\mach=10$ for the supersonic regime, respectively.

Here we use purely solenoidal (divergence-free) driving of the turbulence, as it is traditionally used for subsonic (incompressible) studies of turbulence, and we aim to compare to such studies later in \S\ref{sec:link}. For consistency we use the same driving (solenoidal) for the supersonic set of simulations. We note that we do not expect that our results depend on the turbulence driving mode, and leave a possible investigation of this aspect to a future study.

\subsubsection{Initial Conditions and Simulation Parameters}
\label{sssec:init_cond}
We describe the following physical quantities in dimensionless units, with $\cs=1$ and $\rho_0=1$, the latter being the initial uniform density. Thus, $\textbf{u}$ is in units of $\cs$, $\rho$ is in units of $\rho_0$, $\textbf{B}$ is in units of $\cs\rho_0^{1/2}$, and the dissipation wave numbers $\knu$ and $\keta$ are in units of $\kbox$. In fact, throughout the study, all wave numbers are reported in units of $\kbox$ unless stated otherwise.

We initialise a turbulent magnetic field with zero net flux, i.e., without a mean field. The fluctuating magnetic field is generated with \texttt{TurbGen} \citep{FederrathEtAl2022ascl}, in analogy to the turbulence driving field, i.e., with a parabolic Fourier amplitude spectrum over $1\leq k\leq3$ (as described in \S\ref{sssec:turb_driv}). We set the root-mean-squared magnetic field value to $B=3.545\times10^{-11}$ and $B=3.545\times10^{-9}$ for the $\mach=0.1$ and $\mach=10$ simulations, respectively. These values give an Alfv\'en Mach number (ratio of turbulent velocity to Alfv\'en speed) of $\macha=\uturb/\va=10^{10}$ for all simulations, based on $\uturb=0.1$ and $10$ for the subsonic and supersonic simulations sets, respectively. The respective values of plasma beta are $\beta = P_{\mathrm{th}}/P_{\mathrm{mag}}=6.5\times10^{22}$ for $\mach=0.1$ (subsonic regime) and $6.5\times10^{18}$ for $\mach=10$ (supersonic regime). The resulting peak initial magnetic energy is $E_{\mathrm{mag},0} = 5\times10^{-23}$ in the subsonic simulation set and $E_{\mathrm{mag},0} = 5\times10^{-19}$ in the supersonic simulation set, which means that the magnetic field is initially extremely weak, such that we can observe turbulent dynamo amplification. We note that as long as the field is weak, the initial field strength and structure do not affect the properties of the magnetic field generated by the dynamo \citep{2020MNRAS.499.2076S}.

Apart from the distinction of the subsonic and supersonic regimes of turbulence, which we parameterise with $\mach=0.1$ and $10$, respectively, our primary focus is on the effects of the numerical resolution $N$. Thus, we run simulations with $N=144$, 288, 576, 1152 and 2304, for the subsonic and supersonic simulation sets. Table~\ref{tab:fit_params} lists these simulations in the kinematic (exponential growth) phase of the dynamo and the corresponding main analysis results, which we discuss in Sections~\ref{sec:results} and~\ref{sec:rermpm}. Similarly, Tab.~\ref{tab:sat_regime} in Sec.~\ref{ssec:sat_results} lists the simulation results of two of the simulations that were followed into the saturation phase of the dynamo, which are discussed in \S\ref{ssec:sat_results}, showing that the dissipation scales do not change significantly upon entering the saturation regime.

\begingroup
\setlength{\tabcolsep}{1.7pt} % 
\renewcommand{\arraystretch}{1.3}
\begin{table*}
	\caption{Simulation parameters and main results of the kinematic phase of the dynamo.}
	\label{tab:fit_params}
\begin{threeparttable}
\begin{tabular}{r|c|ccccc|ccc|ccccc}
\hline
\multicolumn{2}{c}{} & \multicolumn{5}{c}{\textbf{From $P_{\mathrm{kin}}$}} & \multicolumn{3}{c}{\textbf{From $P_{\mathrm{mag}}$}} & \multicolumn{5}{c}{\textbf{Derived}}\\
\hline
\multicolumn{1}{c|}{$N$} & \multicolumn{1}{c|}{$\Gamma$} & \multicolumn{1}{c}{$\pkin$} & \multicolumn{1}{c}{$\pbn$} & \multicolumn{1}{c}{$\pnu$} & \multicolumn{1}{c}{$\kbn$} & \multicolumn{1}{c|}{$\knut$} & \multicolumn{1}{c}{$\pmag$} & \multicolumn{1}{c}{$\peta$} & \multicolumn{1}{c|}{$\ketat$} & \multicolumn{1}{c}{$\knu$} & \multicolumn{1}{c}{$\keta$} & \multicolumn{1}{c}{$\rek$} & \multicolumn{1}{c}{$\pem$} & \multicolumn{1}{c}{$\rem$}\\
\multicolumn{1}{c|}{(1)} & \multicolumn{1}{c|}{(2)} & \multicolumn{1}{c}{(3)} & \multicolumn{1}{c}{(4)} & \multicolumn{1}{c}{(5)} & \multicolumn{1}{c}{(6)} & \multicolumn{1}{c|}{(7)} & \multicolumn{1}{c}{(8)} & \multicolumn{1}{c}{(9)} & \multicolumn{1}{c|}{(10)} & \multicolumn{1}{c}{(11)} & \multicolumn{1}{c}{(12)} & \multicolumn{1}{c}{(13)} & \multicolumn{1}{c}{(14)} & \multicolumn{1}{c}{(15)}\\
\hline
\multicolumn{15}{c}{$\mach = 0.1$}  \\
\hline
2304 & $5.2\!\pm\!0.4$  & -1.7 & $0.32\!\pm\!0.01$ & 1.0 & $39.9\!\pm\!0.4$ & $65.0\!\pm\!0.2$  & $2.71\!\pm\!0.01$ & $0.83\!\pm\!0.01$ & $34.4\!\pm\!0.6$  & $65.0\!\pm\!0.2$  & $69.8\!\pm\!2.8$   & $1.4_{-0.2}^{+0.2}\!\times\!10^{4}$ & $1.5_{-0.5}^{+1.0}$                  & $2.1_{-0.8}^{+2.0}\!\times\!10^{4}$  \\
1152 & $3.3\!\pm\!0.1$  & -1.7 & $0.38\!\pm\!0.02$ & 1.0 & $21.6\!\pm\!0.3$ & $34.1\!\pm\!0.1$  & $2.59\!\pm\!0.01$  & $0.88\!\pm\!0.01$ & $22.6\!\pm\!0.5$ & $34.1\!\pm\!0.1$  & $34.7\!\pm\!1.6$   & $6.0_{-0.8}^{+1.0}\!\times\!10^{3}$ & $1.3_{-0.5}^{+1.0}$                  & $8.0_{-3.0}^{+6.0}\!\times\!10^{3}$  \\
576  & $2.1\!\pm\!0.2$  & -1.7 & $0.42\!\pm\!0.02$ & 1.0 & $11.5\!\pm\!0.2$ & $18.1\!\pm\!0.1$ & $2.56\!\pm\!0.02$   & $0.89\!\pm\!0.01$ & $12.8\!\pm\!0.4$ & $18.1\!\pm\!0.1$ & $17.5\!\pm\!1.2$   & $2.6_{-0.3}^{+0.4}\!\times\!10^{3}$ & $1.2_{-0.4}^{+1.0}$                  & $3.1_{-1.0}^{+2.0}\!\times\!10^{3}$  \\
288  & $1.3\!\pm\!0.1$  & -1.7 & $0.39\!\pm\!0.02$  & 1.0 & $5.80\!\pm\!0.10$  & $9.71\!\pm\!0.05$ & $2.52\!\pm\!0.04$  & $0.90\!\pm\!0.02$   & $7.13\!\pm\!0.46$ & $9.71\!\pm\!0.05$ & $8.97\!\pm\!1.00$   & $1.1_{-0.1}^{+0.2}\!\times\!10^{3}$ & $1.1_{-0.4}^{+0.9}$                  & $1.2_{-0.5}^{+1.0}\!\times\!10^{3}$  \\
144  & $0.85\!\pm\!0.01$ & -1.7 & $0.30\!\pm\!0.03$  & 1.0 & $2.35\!\pm\!0.11$ & $5.34\!\pm\!0.04$ & $2.82\!\pm\!0.10$    & $0.79\!\pm\!0.03$   & $2.39\!\pm\!0.40$  & $5.34\!\pm\!0.04$ & $3.04\!\pm\!0.78$  & $5.1_{-0.7}^{+0.8}\!\times\!10^{2}$ & $0.41_{-0.2}^{+0.4}$ & $2.1_{-1.0}^{+2.0}\!\times\!10^{2}$  \\
\hline
\multicolumn{15}{c}{$\mach = 10$}  \\
\hline
2304 & $0.82\!\pm\!0.02$ & -2.0 & $0.00\!\pm\!0.16$  & 0.7 & $30.2\!\pm\!0.6$ & $34.6\!\pm\!0.5$  & $1.52\!\pm\!0.01$  & $0.81\!\pm\!0.01$ & $56.7\!\pm\!1.9$  & $158.0\!\pm\!3.1$  & $144.0\!\pm\!13.0$ & $3.0_{-0.6}^{+1.0}\!\times\!10^{4}$ & $2.7_{-2.0}^{+7.0}$                  & 	$8.3_{-5.0}^{+20.0}\!\times\! 10^{4}$ \\
1152 & $0.71\!\pm\!0.01$ & -2.0 & $0.00\!\pm\!0.01$  & 0.7 & $17.6\!\pm\!0.4$ & $18.4\!\pm\!0.3$  & $1.47\!\pm\!0.02$  & $0.84\!\pm\!0.01$  & $36.5\!\pm\!1.8$  & $64.0\!\pm\!1.7$   & $71.8\!\pm\!9.0$   & $7.6_{-2.0}^{+3.0}\!\times\!10^{3}$ & $4.2_{-2.0}^{+10.0}$                  & 	$3.2_{-2.0}^{+9.0}\!\times\! 10^{4}$  \\
576  & $0.57\!\pm\!0.02$ & -2.0 & $0.00\!\pm\!0.02$  & 0.7 & $10.5\!\pm\!0.3$ & $9.89\!\pm\!0.25$  & $1.43\!\pm\!0.03$  & $0.86\!\pm\!0.02$  & $22.9\!\pm\!1.7$  & $26.4\!\pm\!1.0$  & $37.8\!\pm\!6.6$   & $2.0_{-0.4}^{+0.7}\!\times\!10^{3}$ & $6.7_{-4.0}^{+20.0}$                  & 	$1.4_{-0.7}^{+4.0}\!\times\! 10^{4}$  \\
288  & $0.53\!\pm\!0.01$ & -2.0 & $0.00\!\pm\!0.03$  & 0.7 & $6.42\!\pm\!0.25$ & $5.37\!\pm\!0.18$  & $1.43\!\pm\!0.03$  & $0.83\!\pm\!0.02$  & $12.4\!\pm\!1.2$  & $11.0\!\pm\!0.5$  & $20.5\!\pm\!4.3$   & $5.5_{-1.0}^{+2.0}\!\times\!10^{2}$ & $11_{-7.0}^{+30}$                  & 	$6.1_{-3.0}^{+20.0}\!\times\! 10^{3}$ \\
144  & $0.43\!\pm\!0.01$ & -2.0 & $0.00\!\pm\!0.03$  & 0.7 & $4.01\!\pm\!0.23$ & $2.95\!\pm\!0.14$  & $1.52\!\pm\!0.05$  & $0.75\!\pm\!0.03$  & $5.23\!\pm\!0.70$  & $4.68\!\pm\!0.33$  & $8.98\!\pm\!2.30$   & $1.5_{-0.4}^{+0.5}\!\times\! 10^{2}$   & 	$12_{-8.0}^{+40}$   & 	$1.8_{-1.0}^{+5.0}\!\times\! 10^{3}$ \\
\hline
\end{tabular}
\begin{tablenotes}[flushleft]
\note{All measured and derived parameters were obtained by time averaging over the kinematic (exponential growth) phase of the dynamo, from $t \geq 4\tturb$ to when $\Emag/\Ekin \leq 10^{-3}$, for both the subsonic ($\mach = 0.1$) and supersonic ($\mach = 10$) regimes (see \S\ref{sssec:time_avg_win}).
Columns: \textbf{(1)} Linear grid resolution $N$ for our series of simulations. \textbf{(2)} The exponent of the time-rate of change $\Gamma$ of $\Emag/\Ekin$, i.e., the dynamo growth rate (see Eq.~\ref{eq:time_rate}). The following columns are the measured parameters (fixed parameter values are shown without errors) from fitting Eq.~(\ref{eq:kinspectra}) to $\Pkin$ from our numerical simulations. \textbf{(3)} Power-law exponent of the scaling range of $\Pkin$. \textbf{(4)} Exponent of the bottleneck effect of $\Pkin$. \textbf{(5)} Exponent of the dissipation term of $\Pkin$. \textbf{(6)} Scaling wave number of the bottleneck effect of $\Pkin$. \textbf{(7)} Viscous dissipation wave number if $\pnu = 1$. The following columns are the measured parameters from fitting Eq.~(\ref{eq:magspectra}) to $\Pmag$. \textbf{(8)} Power-law exponent of the scaling range of $\Pmag$. \textbf{(9)} Exponent of the dissipation term of $\Pmag$. \textbf{(10)} Resistive dissipation wave number if $\peta = 1$. The following columns are the derived quantities from the previous columns. \textbf{(11)} Viscous dissipation wave number derived from column~7 (see Eq.~\ref{eq:knu_actual}). \textbf{(12)} Resistive dissipation wave number derived from column~10 (see Eq.~\ref{eq:keta_actual}). \textbf{(13)} Hydrodynamic Reynolds number derived from column~11 (see Eq.~\ref{eq:knugeneral}). \textbf{(14)} Magnetic Prandtl number derived from columns~11 and 12 (see Eq.~\ref{eq:ketageneral}). \textbf{(15)} Magnetic Reynolds number derived from columns~13 and~14 (see Eq.~\ref{eq:pm_rel}). The wave numbers are reported in units of $\kbox$ throughout the study unless mentioned otherwise (see \S\ref{sssec:init_cond}). The error bars reported are two-sigma variations of the corresponding parameter.}
\end{tablenotes}
\end{threeparttable}
\end{table*}
\endgroup

\subsection{Spectral Fitting} \label{ssec:spec_fit}
In order to determine the dissipation scales of the turbulence, we directly fit a functional form to the kinetic and magnetic energy spectra, where the dissipation scales are fit parameters. All fits in this work are done using \texttt{LMFIT} \citep{https://doi.org/10.5281/zenodo.598352}.
We consider the model for the kinetic spectrum as a function of wave number $k$, defined in \citet{2022MNRAS.513.2457K},
\begin{equation}
    \Pkin(k) = \Akin k^{\pkin}\exp{\left(-\frac{k}{\knu}\right)}.
    \label{eq:krielkin}
\end{equation}
Similarly, the functional form of the magnetic spectrum in \citet{2022MNRAS.513.2457K} is defined as
\begin{equation}
    \Pmag(k) = \Amag k^{\pmag}K_0\left(\frac{k}{\keta}\right),
    \label{eq:krielmag}
\end{equation}
In these equations, $\Akin$ and $\Amag$ are amplitude coefficients, $\pkin$ and $\pmag$ are slopes of the power-law parts of the spectra, and $\knu$ and $\keta$ are the characteristic wave numbers where the dissipation terms dominate in $\Pkin$ and $\Pmag$, respectively. The function $K_0(x)$ is the modified Bessel function of the second kind and order~0.

Here we extend the model by \citet{2022MNRAS.513.2457K} for $\Pkin$ to include the bottleneck effect \citep{1994PhFl....6.1411F, 2004astro.ph..7616S, 2007JPhA...40.4401V}. With an assumption of local energy transfer \citep{1994PhFl....6.1411F}, the bottleneck effect signifies the suppression of non-linear interactions by dissipative modes, which decreases the efficiency of the energy cascade around that scale, resulting in a pile-up of kinetic energy in this range. Therefore, we define the modified functional form of the kinetic spectrum, including the bottleneck effect, as
\begin{equation}
    \Pkin(k) = \Akin\left[\left(\frac{k}{\kbn}\right)^{\pkin}+\left(\frac{k}{\kbn}\right)^{\pbn}\right]\exp{\left[-\left(\frac{k}{\knut}\right)^{\pnu}\right]}.
    \label{eq:kinspectra}
\end{equation}
In this model, the wave number is scaled by the additional fit parameters $\kbn$ and $\pbn$, in order to account for the different extents and strengths of the observed bottleneck effect across different linear grid resolutions, studied below. Finally, we generalise the exponential dissipation term with the exponent $\pnu$, in order to account for a slower or faster exponential decay compared to Eq.~(\ref{eq:krielkin}), close to $\knu$, which we observe in our numerical simulation results below. 

Likewise, we modify the functional form of the magnetic spectra by adding the exponent $\peta$ in the Bessel function to account for a potentially slower or faster decay around $\keta$ compared to Eq.~(\ref{eq:krielmag}), and obtain
\begin{equation}
    \Pmag(k) = \Amag k^{\pmag}K_0\left[\left(\frac{k}{\ketat}\right)^{\peta}\right].
    \label{eq:magspectra}
\end{equation}

The generalisations of the dissipation functions in the spectral models, i.e., the additions of $\pnu$ and $\peta$ as exponents in the dissipation terms, thereby account for different levels of sharpness or smoothness of the transitions from the inertial range into the dissipation range. In order to obtain the characteristic wave numbers equivalent to the results in \citet{2022MNRAS.513.2457K}, the effect of the generalisation has to be reversed. Therefore, the value of the viscous dissipation wave number for the kinetic energy spectrum is computed as
\begin{equation}
    \knu = \knut^{1/\pnu},
    \label{eq:knu_actual}
\end{equation}
such that $\knu$ represents a value comparable to the one measured in \citet{2022MNRAS.513.2457K}, where $\pnu=1$. Similarly, the resulting value of the resistive dissipation wave number for the magnetic energy spectrum is given by
\begin{equation}
    \keta = \ketat^{1/\peta}.
    \label{eq:keta_actual}
\end{equation}

In order to fit these models to the simulation data, we must first compute time-averaged spectra from the simulations. For the kinematic phase of the turbulent dynamo, i.e., when the turbulence is fully developed and the magnetic energy grows exponentially, this is achieved by time-averaging the power spectra from $t/\tturb\ge4$, to safely start when the turbulence is fully developed \citep{2023MNRAS.524.3201B} in both the subsonic and supersonic regime of turbulence. Moreover, the end of the time-averaging window is defined such that the ratio of magnetic energy ($\Emag = {|\textbf{B}|^2}/{8\pi}$) to kinetic energy ($\Ekin = \rho_0 \uturb^2/2$) is $\Emag/\Ekin \leq 10^{-3}$, in order to exclude the transition from the kinematic into the linear and saturated regimes of the dynamo.

Additionally, due to the growth of magnetic energy in the kinematic phase of the dynamo, in order to be able to time-average the magnetic energy, $\Pmag$ is first normalised by its total magnetic energy, which is the integral of $\Pmag(k)$ over all $k$, at each time frame. Thus, we effectively time-average the shape of $\Pmag(k)$ compensated by the amplitude increase over time, allowing for a robust measurement of $\keta$.

All spectra ($\Pkin$ and $\Pmag$) are fitted from $k\ge3$ to only consider the range of fully-developed turbulence, excluding the turbulence driving scales. The upper limit of the fit in wave-number space is half the maximum $k$ for every $N$, i.e., $k_{\mathrm{max}}/2=N/4$. This is to exclude spurious numerical effects on scales smaller than 2 grid cell lengths.

For the kinetic spectra, $\pkin$ is chosen to be $\pkin=-1.7$ as per Kolmogorov's theory \citep{Kolmogorov1941c,Kolmogorov1962,SheLeveque1994,FederrathEtAl2021} for our subsonic set of simulations ($\mach = 0.1$), and $\pkin=-2.0$ as per Burgers turbulence \citep{Burgers1948,1995PhRvE..52.3656B,MacLowKlessen2004,KritsukEtAl2007,2013MNRAS.436.1245F,FederrathEtAl2021} for our set of supersonic simulations ($\mach = 10$). While self-consistently fitting these exponents is possible for the highest-resolution simulation used here ($N=2304$), it is practically impossible to do so for resolutions $N\lesssim1000$ \citep{KitsionasEtAl2009,2010A&A...512A..81F,PriceFederrath2010,2011ApJ...737...13K,2013MNRAS.436.1245F,FederrathEtAl2021}. Thus, we fix these scaling exponents to allow for a robust determination of the dissipation wavenumbers.

The viscous dissipation exponent $\pnu$ is fixed at 1.0 and 0.7 for the $\mach=0.1$ and~10 simulation sets, respectively. We fix these values to prohibit variations in this parameter for different numerical resolutions $N$, therefore preventing systematic dependencies on $\pnu$, while still providing excellent fits in both the subsonic and supersonic regimes of turbulence (see Appendix~\ref{ap:comp}).

The dissipation wave numbers $\knu$ and $\keta$, along with their errors, are extracted from weighted fits of $\Pkin$ and $\Pmag$, respectively, taking into account the one-sigma time variations at each $k$. We emphasise that, unless mentioned otherwise, all extracted wave-numbers, i.e., $\kbn$, $\knut$, $\ketat$, $\knu$ and $\keta$, are in units of $\kbox = 2\pi/L$.

% === Results ===
\section{Results} \label{sec:results}

In this section, we aim to measure the effects of the magnetic energy amplification mechanism and the dissipation effects on the time-averaged kinetic and magnetic power spectra. We are particularly interested in their dependence on the linear grid resolution $N$. We always distinguish between the subsonic and supersonic regimes of turbulence. We first concentrate on the kinematic phase of the dynamo. We begin by developing a qualitative understanding of the implications of the grid resolution by studying the morphology of the gas density, and the kinetic and magnetic energies. We then determine the dissipation wavenumbers from fits to the kinetic ($\Pkin$) and magnetic ($\Pmag$) power spectra across different $N$. Later, we compare these results from the kinematic phase to the saturation phase of the dynamo.

\subsection{Time evolution}
\label{ssec:time_evol}

%%%%%%%%%%%%%%%%%%%%%%%%%%%%%%%%%%%%%%%%%%%%%%%%%%
\begin{figure*}
    \centering
    \includegraphics[width=1.0\linewidth]{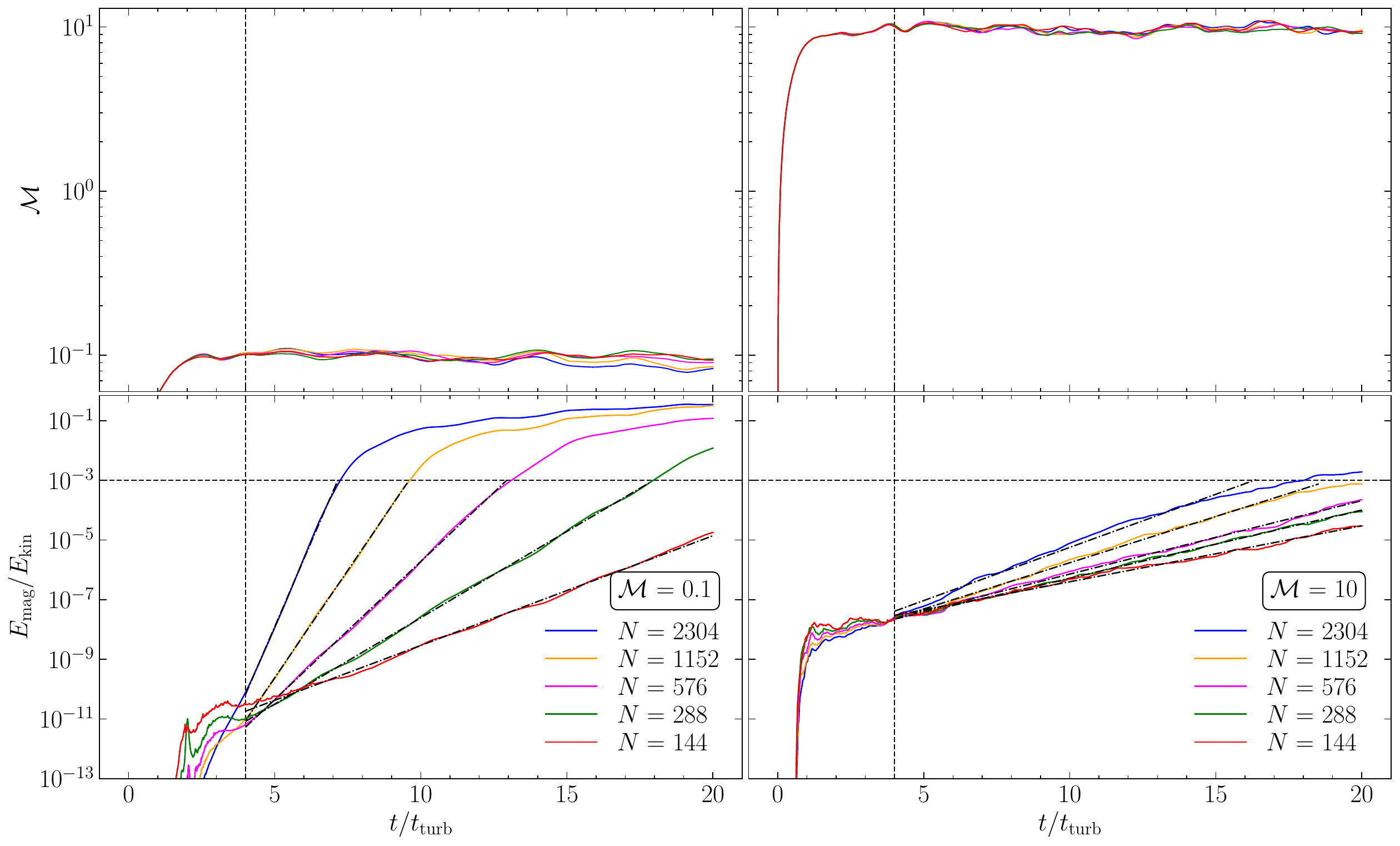}
    \caption{Time evolution of the sonic Mach number ($\mach$; top panels), and the ratio of magnetic to kinetic energy ($\Emag/\Ekin$; bottom panels) in the kinematic phase of the dynamo. The left panels show our set of subsonic simulation models ($\mach = 0.1$) and the right panels show our set of supersonic simulation models ($\mach = 10$). After the initial transient phase, $t \lesssim 4\tturb$, (marked by the vertical lines), the turbulence in all our simulations is fully developed and the magnetic field has entered the exponential (kinematic) growth phase of the turbulent dynamo. $\Emag/\Ekin = 10^{-3}$ (marked by the horizontal lines in the bottom panels) safely separates the kinematic phase from the linear (and saturated) phase of the dynamo, for all simulations in our series of $N$. Therefore, we only consider times $t \geq 4\tturb$ and when $\Emag/\Ekin \leq 10^{-3}$ for time-averaging the power spectra (c.f., \S\ref{ssec:spec_fit}). The growth rate of $\Emag/\Ekin$ (marked by black dashed lines overlaid in the bottom panels) is obtained from fitting Eq.~(\ref{eq:time_rate}) in the kinematic phase of the dynamo, and is listed in column~2 of Tab.~\ref{tab:fit_params}. }
    \label{fig:Time_evol}
\end{figure*}
%%%%%%%%%%%%%%%%%%%%%%%%%%%%%%%%%%%%%%%%%%%%%%%%%%

We start by explaining our choice of the time-averaging window of the power spectra in the kinematic phase of the dynamo as described in \S\ref{ssec:spec_fit}, by analysing the time frames corresponding to the different phases of our simulations represented in Fig.~\ref{fig:Time_evol}. Fig.~\ref{fig:Time_evol} shows the time evolution of the sonic Mach number ($\mach$) in the top two panels, and the ratio of magnetic to kinetic energy ($\Emag/\Ekin$) in the bottom panels, for a series of linear grid resolutions ($N$) obtained from our numerical simulations in the subsonic ($\mach=0.1$; left panels) and supersonic ($\mach=10$; right panels) regimes. 

\subsubsection{Defining the time-averaging window for spectral analysis in the kinematic phase}
\label{sssec:time_avg_win}
The region of fully-developed turbulence in the time evolution plots of the sonic Mach numbers is the region after the transition from the initial transient rise to the state where the time-variations in $\mach$ for all our simulations lie within 10--20\% of our target $\mach$, in both the subsonic and supersonic regimes. For the time evolution of $\Emag/\Ekin$, the region prior to fully-developed turbulence, referred to as the initial transient phase, is signified by the irregularities in $\Emag/\Ekin$ before the transition to the kinematic phase of the dynamo. We observe from the time evolution of $\Emag/\Ekin$ that the kinematic phase starts at $t \approx 4\tturb$ in both the subsonic and supersonic regimes (indicated by vertical dashed lines in Fig.~\ref{fig:Time_evol}). 

We emphasise that most of our simulations do not reach the saturation phase of the dynamo. The transition from the kinematic to the linear phase can only be observed in the subsonic regime for the top three resolutions in our series of simulations. For all other models, in particular the supersonic runs, it is evident that the magnetic field continues to grow beyond $t/\tturb = 20$ (the time at which we choose to stop the simulations due to computational costs, and the fact that this study focuses on the dissipation scales in the kinematic phase of the dynamo). Hence, the region below the horizontal line at $\Emag/\Ekin = 10^{-3}$ and right of the vertical line at $t/\tturb = 4$ marks the kinematic phase of the dynamo, safely away from the initial transient phase and the saturation phase, as shown in the bottom panels. Thus, this defines our time-averaging window for the spectral analysis below.

\subsubsection{Basic time evolution}
We start by comparing the magnetic amplification in the kinematic phase of the dynamo in the subsonic and supersonic regimes and for different numerical grid resolutions $N$. In order to do so, we fit an exponential model to $\Emag/\Ekin$ in the time-averaging window (see \S\ref{sssec:time_avg_win}) to the data shown in the bottom panels on Fig.~\ref{fig:Time_evol}, to quantify its growth rate, 
\begin{equation}
    \Emag/\Ekin = (\Emag/\Ekin)_0 e^{\Gamma (t-4\tturb)},
    \label{eq:time_rate}
\end{equation}
where $(\Emag/\Ekin)_0$ is the value of $\Emag/\Ekin$ at $t = 4\tturb$. The growth rate, $\Gamma$, is measured in units of $\tturb^{-1}$. The extracted $\Gamma$ from the fits is listed in column~2 of Tab.~\ref{tab:fit_params}. We see that $\Gamma$ increases with $N$. We expect this because higher grid resolutions better capture magnetic field amplification by vorticity, which is dominant in smaller-scale turbulent motions of the fluid elements \citep{FederrathEtAl2011,2020MNRAS.499.2076S,2021PhRvL.126i1103A}. In addition, \citet{2012PhRvE..85b6303S,BovinoEtAl2013} have shown that $\Gamma$ positively scales with $\rek$ and $\rem$, respectively. Taken together, this may imply that the Reynolds numbers depend on $N$, serving as additional motivation to determine the exact dependence. Comparing the growth rates between the subsonic and supersonic regimes for a particular grid resolution, we find that the magnetic field amplification is higher for $\mach = 0.1$ than for $\mach = 10$ as in previous studies \citep{FederrathEtAl2011,2020MNRAS.499.2076S,2021PhRvL.126i1103A}.

For the resolutions 2304, 1152 and 576 in subsonic simulations, we also observe a drop in the sonic Mach numbers due to the back-reaction on the fluid from the amplified magnetic field $B$ in the region where $\Emag/\Ekin \sim 10^{-1}$. This back-reaction on the velocity field is negligible in the kinematic phase of the dynamo, because the corresponding magnetic field is very weak.

\subsection{Kinetic and Magnetic Energy Structure}
\label{ssec:ener_struc}

%%%%%%%%%%%%%%%%%%%%%%%%%%%%%%%%%%%%%%%%%%%%%%%%%%
\begin{figure*} 
\centering
\includegraphics[width=1.0\linewidth]{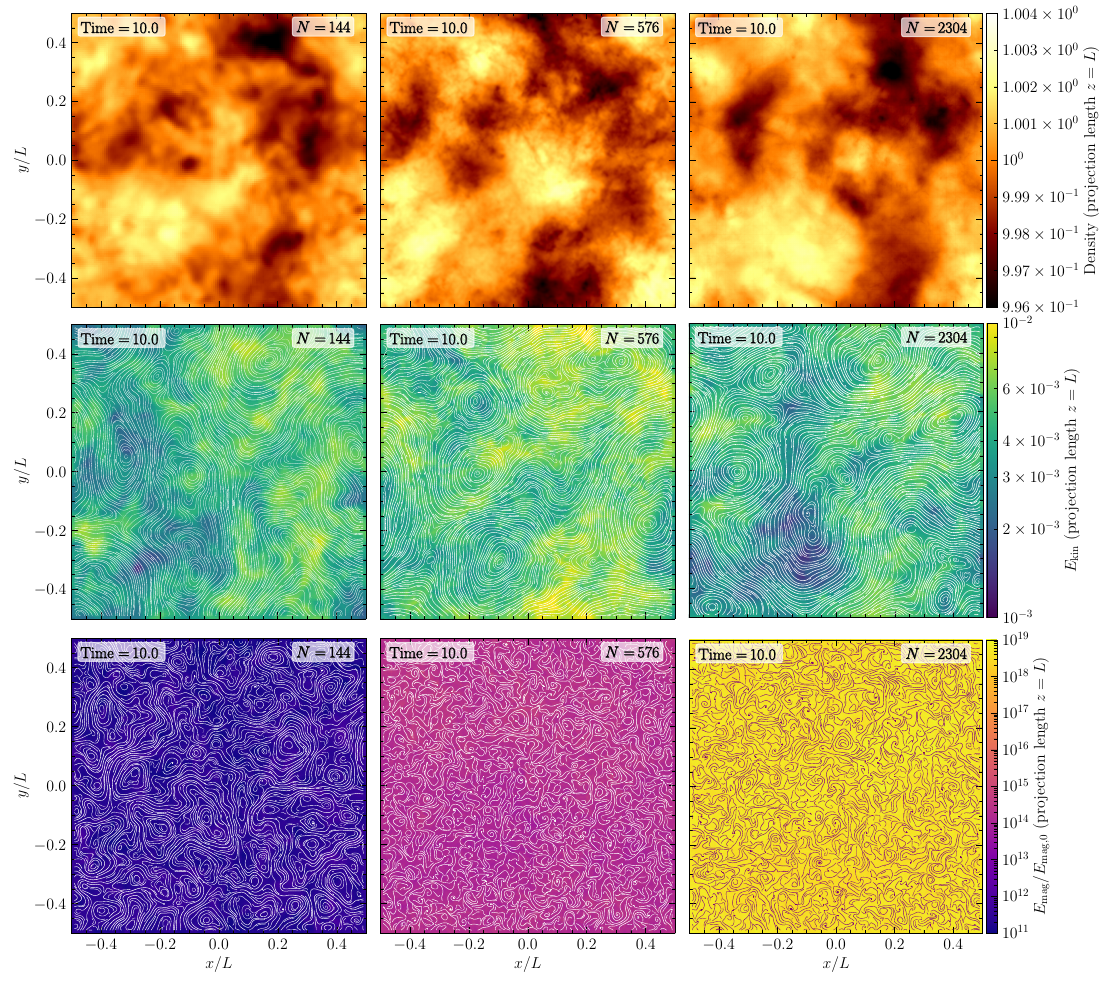}
\caption{Two-dimensional projections (spatial average along the entire $z$-axis), in the subsonic regime ($\mach=0.1$) of fluid density (top panels), kinetic energy ($\Ekin$; middle panels), and magnetic energy ($\Emag/E_{\mathrm{mag},0}$; bottom panels) in the kinematic phase of the dynamo at $t = 10\tturb$. Left to right panels correspond to 3 of our 6~grid resolutions ($N$), with the leftmost, middle, and rightmost panels showing $N = 144$, 576, and 2304, respectively. The lines in $\Ekin$ and $\Emag/E_{\mathrm{mag},0}$ panels represent streamlines of the velocity and magnetic field, respectively. \textbf{(i) Top panels:} The density variations are of the order of $1\%$ as expected for $\mach\sim0.1$. \textbf{(ii) Middle panels:} Large-scale eddies dominate the kinetic energy, and hence, $\Ekin$ does not show a significant resolution dependence. \textbf{(iii) Bottom panels:} Vorticity, predominant in the smallest-scale eddies, is principally responsible for the exponential magnetic field amplification in the subsonic regime. We see that increasing $N$ leads to resolving smaller-scale magnetic field loops.}
\label{fig:slices0p1}
\end{figure*} 
%%%%%%%%%%%%%%%%%%%%%%%%%%%%%%%%%%
\begin{figure*}
\centering
\includegraphics[width=1.0\linewidth]{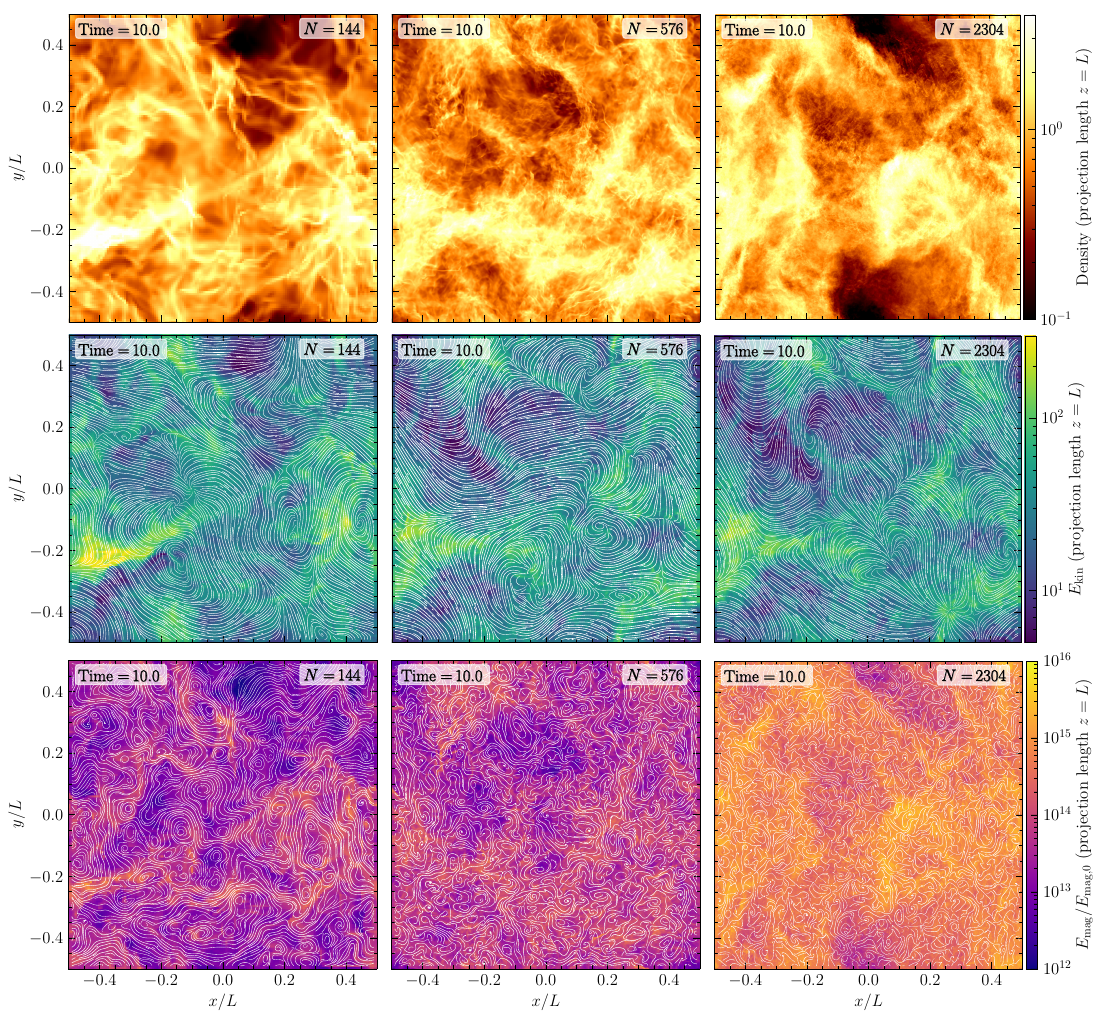}
\caption{Same as Fig.~\ref{fig:slices0p1}, but for the supersonic ($\mach=10$) simulations. \textbf{(i) Top panels:} In contrast to the subsonic simulations, here the density variations occupy several orders of magnitude as expected for supersonic turbulence. \textbf{(ii) Middle panels:} Similar to the subsonic regime, $\Ekin$ does not show a significant resolution dependence due to the dominant contribution of large-scale eddies and shocks. \textbf{(iii) Bottom panels:} The process of shock production due to the compression of fluid elements, which primarily happens on considerably larger scales (length of shocks) than vorticity, is prominently occurring in the supersonic regime. The shocks modify the structure of the magnetic field and vorticity production \citep[primarily reducing it compared to the subsonic regime][]{FederrathEtAl2011}. Due to the presence of shocks, we expect $\Pmag$ to be higher for small $k$ in $\mach=10$ compared to $\mach=0.1$. 
For the same reason, regions of high magnitude of magnetic and kinetic fields correspond to regions of relatively higher density. The greater resolution of magnetic field scales with $N$ and the associated exponential amplification implies that the turbulent dynamo still operates via stretch-twist-fold-merge and vorticity, even in the presence of shocks.}
\label{fig:slices10} 
\end{figure*} 

%%%%%%%%%%%%%%%%%%%%%%%%%%%%%%%%%%

Before going into the details of the spectral fitting analysis to determine the dissipation scales $\knu$ and $\keta$, we first look at some of the spatial features of our simulation sets, in order to get a qualitative sense of the effect of the numerical resolution $N$ in the kinematic phase. Figs.~\ref{fig:slices0p1} and~\ref{fig:slices10} show two-dimensional projections from our simulations, of the fluid density (top panels), kinetic energy ($\Ekin$; middle panels), and magnetic energy ($\Emag/E_{\mathrm{mag},0}$; bottom panels), for the $\mach=0.1$ and $\mach=10$ simulation sets, respectively. The values on the colour bar are the spatial average through the entire line of sight along the $z$-axis of the simulation domain and at $t = 10\tturb$, which lies in the kinematic phase of the dynamo. The panels from left to right correspond to 3 of our 6~grid resolutions, here for $N = 144$, $576$, and $2304$.

In the subsonic regime, the density variations are only of the order of $\sim1\%$. In contrast to this, we see very large (order-of-magnitude) density variations in the supersonic simulation sets, shown in the top panels of Fig.~\ref{fig:slices10}, as expected from the shock-jump condition $d\rho\sim\mach^2$. It is also evident how smaller-scale variations in the density are better resolved at higher numerical resolution. 

The kinetic energy does not display a strong resolution dependence, because the kinetic energy is dominated by contributions of large-scale eddies, which implies that $\Ekin$ has converged on relatively larger spatial scales, which is indeed what we see in the middle panels of Figs.~\ref{fig:slices0p1} and~\ref{fig:slices10}. 

In contrast to the structure of $\Ekin$, the scale of the magnetic field lines (bottom panels) decreases with increasing resolution. This is because magnetic amplification due to vorticity (deformations of the magnetic field lines by local stretching, twisting and folding of fluid elements) is driven by the turbulent eddies on relatively smaller spatial scales. Therefore, with an increase in grid resolution, smaller-scale vortices are captured to a greater extent, causing an increase in $\Emag$. For the $\mach=10$ case, regions of higher magnetic and kinetic fields are correlated with high-density regions. This suggests the occurrence of magnetic energy amplification due to the process of shock creation (rapid compression of fluid elements), which primarily occurs on relatively larger scales compared to vorticity, in the supersonic regime. However, as we shall show later, the dominant mechanism for magnetic energy amplification in both subsonic and supersonic regimes is due to vorticity (see Appendix~\ref{ap:decomp_spec}). This suggests that the small-scale eddies live inside the large-scale shocks created.

\subsection{Kinetic and Magnetic Power Spectra}
\label{ssec:power_spectra}

%%%%%%%%%%%%%%%%%%%%%%%%%%%%%%%%%%
\begin{figure*}
    \centering
    \includegraphics[width=1.0\linewidth]{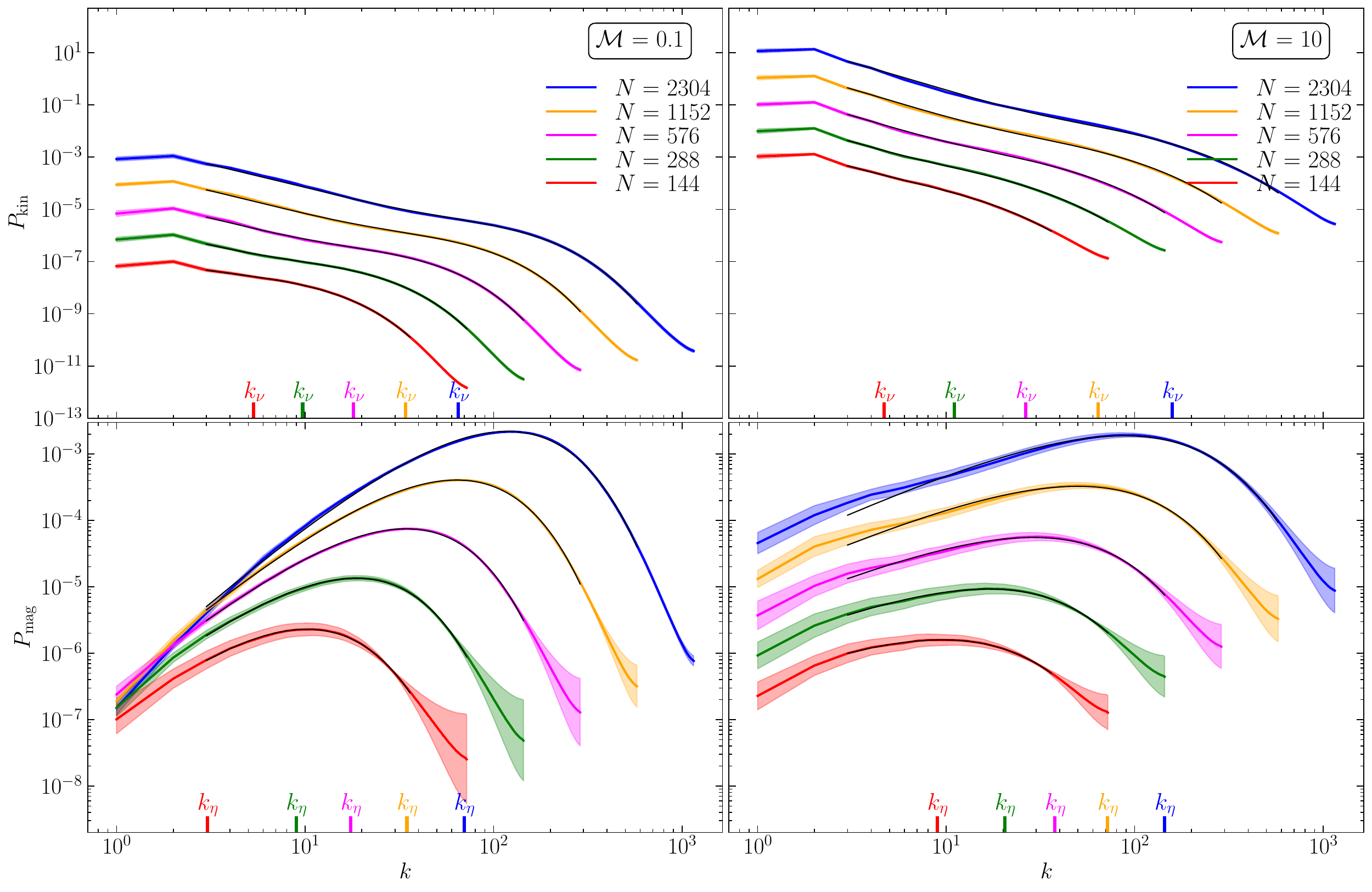}
    \caption{Time-averaged kinetic power spectra ($\Pkin$; top panels) and magnetic power spectra ($\Pmag$; bottom panels) in the kinematic phase of the dynamo for the series of linear grid resolutions $N$ (see legend). Wave numbers ($k$) are in units of $\kbox = 2\pi/L$. The left panels are for $\mach = 0.1$ and the right panels are for $\mach = 10$. $\Pkin$ and $\Pmag$ for every $N$ except $N = 2304$ is multiplied by a factor of 0.1 relative to the next-higher resolution. We overplot our fitted spectrum models (Eqs.~\ref{eq:kinspectra} and~\ref{eq:magspectra}) as thin black lines. Tab.~\ref{tab:fit_params} lists the fit parameters with columns~11 and 12 for the viscous dissipation wave number $\knu$ and the resistive dissipation wave number $\keta$, respectively, shown as coloured tick markers on the $k$-axis corresponding to the colour code of the power spectra.}
    \label{fig:spectra}
\end{figure*}
%%%%%%%%%%%%%%%%%%%%%%

Fig.~\ref{fig:spectra} shows the time-averaged kinetic power spectra ($\Pkin$; top panels), and magnetic power spectra ($\Pmag$; bottom panels), plotted against wave number $k$, along with their error bars (time variations from the time-averaged power for a particular $k$), for the series of $N$, obtained from our simulations. The left panels belong to the subsonic regime and the right panels belong to the supersonic regime. Eqs.~(\ref{eq:kinspectra}) and (\ref{eq:magspectra}) are fitted to $\Pkin$ and $\Pmag$ (taking into account the one-sigma time variations at each $k$), respectively, in the kinematic phase of the dynamo as described in \S\ref{ssec:spec_fit}. Except for the highest resolution ($N = 2304$), each power spectrum and its corresponding fit are multiplied by 0.1 relative to its nearest higher resolution case for ease of visualisation, i.e., to allow for a visual separation of the curves. The extracted dissipation wave numbers from $\Pkin$ and $\Pmag$ ($\knu$ and $\keta$, respectively) are displayed as tick markers on the wave-number axis, with colour code according to that of the power spectra.

We see that the spectrum models given by Eqs.~(\ref{eq:kinspectra}) and~(\ref{eq:magspectra}) fit $\Pkin$ and $\Pmag$ well within their error bars. The fit parameters obtained from spectral fitting are listed in Tab.~\ref{tab:fit_params}. The parameter errors listed in the table are the two-sigma errors. We emphasise that we have only fitted data within the range $3\leq k \leq k_{\mathrm{max}}/2$ for each resolution, to safely exclude the low-$k$ turbulence driving range, and the slight upturn observed at high $k$, which is caused by numerical effects (note that limiting the fit to $k \leq k_{\mathrm{max}}/2$ means that we exclude data on the scale of a couple of grid cells).

We now comment on the fit model parameters.

\subsubsection{Kinetic power spectra ($\Pkin$)}

\paragraph{Power-law scaling exponent $\pkin$.}\label{para:pkin_kine}
$\pkin$ is the power-law exponent of the self-similar scaling range of the turbulence. We note that measuring $\pkin$ requires very high resolution $N$. For $N=2304$, we find exponents of $\pkin\sim-1.7$ and $\pkin\sim-2.0$ for the subsonic and supersonic regimes, respectively. However, we cannot determine $\pkin$ self-consistently for the low-resolution models. Thus, we fix $\pkin$ to the values determined for the highest-resolution sets, and consistent with the theoretical expectations for $\pkin\sim-1.7$ in the subsonic regime \citep{1987PhFl...30.2030K,1992PhFlA...4.1086S,SheLeveque1994,smith1991dissipation,FederrathEtAl2021}, and $\pkin\sim-2.0$ in the supersonic regime \citep{KritsukEtAl2007, 2010A&A...512A..81F,2013MNRAS.436.1245F,FederrathEtAl2021}. However, we have repeated the analyses by varying $\pkin$ between $-1.6$ and $-1.8$ (i.e., $-1.7\pm0.1$) for $\mach=0.1$, and $-1.9$ and $-2.1$ (i.e., $-2.0\pm0.1$) for $\mach=10$, which are reasonable ranges in the respective regimes of turbulence. As we are interested in the dissipation wave numbers, we calculated the change in $\knu$ and found that it is always $\lesssim1.5\%$ in both the subsonic and supersonic regimes. As this variation is less than that induced by the $N$-dependence (column~11 of Tab.~\ref{tab:fit_params}), we proceed by fixing $\pkin$ for the main analyses.

\paragraph{Bottleneck exponent ($\pbn$).} 
We find that the bottleneck effect is more pronounced in the subsonic regime than in the supersonic regime, which is reflected in the higher values of $\pbn$ for $\mach = 0.1$ compared to $\mach = 10$. Indeed, we find that $\pbn\sim0$ in the supersonic regime. In the subsonic regime, we find \mbox{$\pbn\sim0.27$--$0.42$}, without a systematic dependence on $N$.

\paragraph{Sharpness of viscous dissipation transition ($\pnu$).}\label{para:pnu_kine}
Concerning the sharpness of the transition from the power-law scaling range into the dissipation range (modelled with $\pnu$), we find that it is somewhat softer (extending over a somewhat larger range of scales) for $\mach = 10$ compared to $\mach = 0.1$. Therefore, we choose $\pnu=0.7$ for $\mach=10$, compared to $\pnu=1.0$ for $\mach=0.1$, which provides very good fits of the transition, for all $N$. However, as for $\pkin$, here too, we have repeated the analyses by varying $\pnu$ from $0.9$ to $1.1$ (i.e., $1\pm0.1$) for $\mach=0.1$, and from $0.6$ to $0.8$ (i.e., $0.7\pm0.1$) for $\mach=10$. This change induces a difference in $\knu$ of $\lesssim10\%$ for most $N$. The maximum change in $\knu$ is $41\%$ for $\pnu=0.8$ ($\mach=10$) and $N=144$, i.e., our lowest-resolution case. In conclusion, as changes induced by varying $\pnu$ are significantly less than those induced by $N$, we proceed by fixing $\pnu$ close to unity. 

\paragraph{Bottleneck and viscous dissipation scales ($\kbn$ and $\knu$).}

We find that $\kbn$ and $\knu$ (determined from $\knut$ via Eq.~\ref{eq:knu_actual}) increase systematically with $N$, in both the subsonic and supersonic regimes. From Tab.~\ref{tab:fit_params}, we see that as $N$ increases by a factor of 2, $\kbn$ and $\knu$ approximately double for $\mach = 0.1$, and they increase by a factor of $\sim 2.5$ for $\mach = 10$. Therefore, we expect $\knu$ to vary approximately linearly with $N$, at least in the subsonic regime. However, we will quantify this in detail below (see \S\ref{ssec:kvsN}). We also note that $\kbn\lesssim20$ for $N\lesssim1000$, which means that the bottleneck (and the transition into dissipation) starts on relatively large scales for low $N$, reflecting the challenges in measuring $\pkin$ for low $N$. In other words, there is practically no scaling range over which $\pkin$ can be reliably measured if $N\lesssim1000$ \citep{KritsukEtAl2007,KitsionasEtAl2009,2010A&A...512A..81F,PriceFederrath2010,2013MNRAS.436.1245F,FederrathEtAl2021}. 

\subsubsection{Magnetic power spectra ($\Pmag$)}
We now turn our attention to the magnetic power spectra in the bottom panels of Fig.~\ref{fig:spectra}, in order to understand the properties of $\keta$. We find that $\Pmag$ is higher for small wave numbers in the supersonic regime compared to the subsonic regime, for all grid resolutions. This signifies the role of shocks compressing the magnetic field, which have relatively large length scales (overall size of the shocks, as opposed to their width). However, we note that the magnetic field is amplified also at larger $k$ in the supersonic regime, which implies the existence of magnetic field amplification due to the turbulent dynamo and the role of vorticity, taking place on smaller spatial scales. Moreover, shocks are very thin and can compress the field to small scales inside shocks. A more quantitative discussion of these effects is needed in future studies comparing subsonic and supersonic turbulence \citep{KrielEtAl2025}, while here we focus on measuring the dissipation scales.

\paragraph{Power-law scaling exponent $\pmag$.}
We find \mbox{$\pmag\sim2.7$--3.0} for $\mach=0.1$ and \mbox{$\pmag\sim1.5$--2.1} for $\mach=10$, with slight variations with $N$. Based on \citet{KazantsevEtAl1985}, one theoretically expects $\pmag=3/2$ for the turbulent dynamo. However, due to limited resolution, we may not be able to find this theoretical value. It is also possible that the theoretical model for $\pmag$ does not fully describe the physical system, and may not apply equally in the subsonic and supersonic regimes. Due to these caveats, we allowed this parameter to vary and fitted it (instead of fixing it as for $\pkin$). While we caution numerical-resolution effects on $\pmag$, we find significant differences in $\pmag$ between the subsonic and supersonic regimes of turbulence, which may be a consequence of the presence of shocks, leading to more magnetic power on larger scales as compared to the subsonic regime. A dedicated study with extremely high resolution is needed to determine the physically converged value of $\pmag$ in the subsonic and supersonic regimes of the turbulent dynamo.

\paragraph{Sharpness of resisitive dissipation transition $\peta$.}
We find relatively robust values of $\peta$, with $\peta\sim0.6$--$0.9$ for both the subsonic and supersonic regimes, with only a weak dependence on $N$. Thus, since $\peta\lesssim1$, the transition is somewhat smoother than for $\pmag=1$, but only very mildly so.

\paragraph{Resistive dissipation scale ($\keta$).}
First, the resistive dissipation wave number ($\keta$) is determined from $\ketat$ via Eq.~(\ref{eq:keta_actual}) as discussed in Sec.~\ref{ssec:spec_fit}.
We find that the onset of the dissipation range of $\Pmag$ for a particular $N$ takes place at a larger $k$ in the supersonic regime compared to the subsonic regime, with $\keta$ about twice as large for $\mach = 10$ compared to $\mach = 0.1$. For a particular sonic Mach number, $\keta$ approximately doubles as $N$ increases by a factor of 2. Therefore, we expect $\keta$ to vary linearly with $N$, as we will quantify in detail below.

\subsection{Saturation phase}
\label{ssec:sat_results}
Here we extend the analyses described in \S\ref{sec:methods} to the saturation phase of the turbulent dynamo. Due to the computational cost of running these simulations into the saturation phase, we only focus on the $N=576$ runs in both the subsonic ($\mach=0.1$) and supersonic ($\mach=10$) regimes of turbulence, allowing us to compare the dissipation-scale results for this resolution in the kinematic and saturation phases of the dynamo.

\begin{figure*}
    \centering   
    \includegraphics[width=1.0\linewidth]{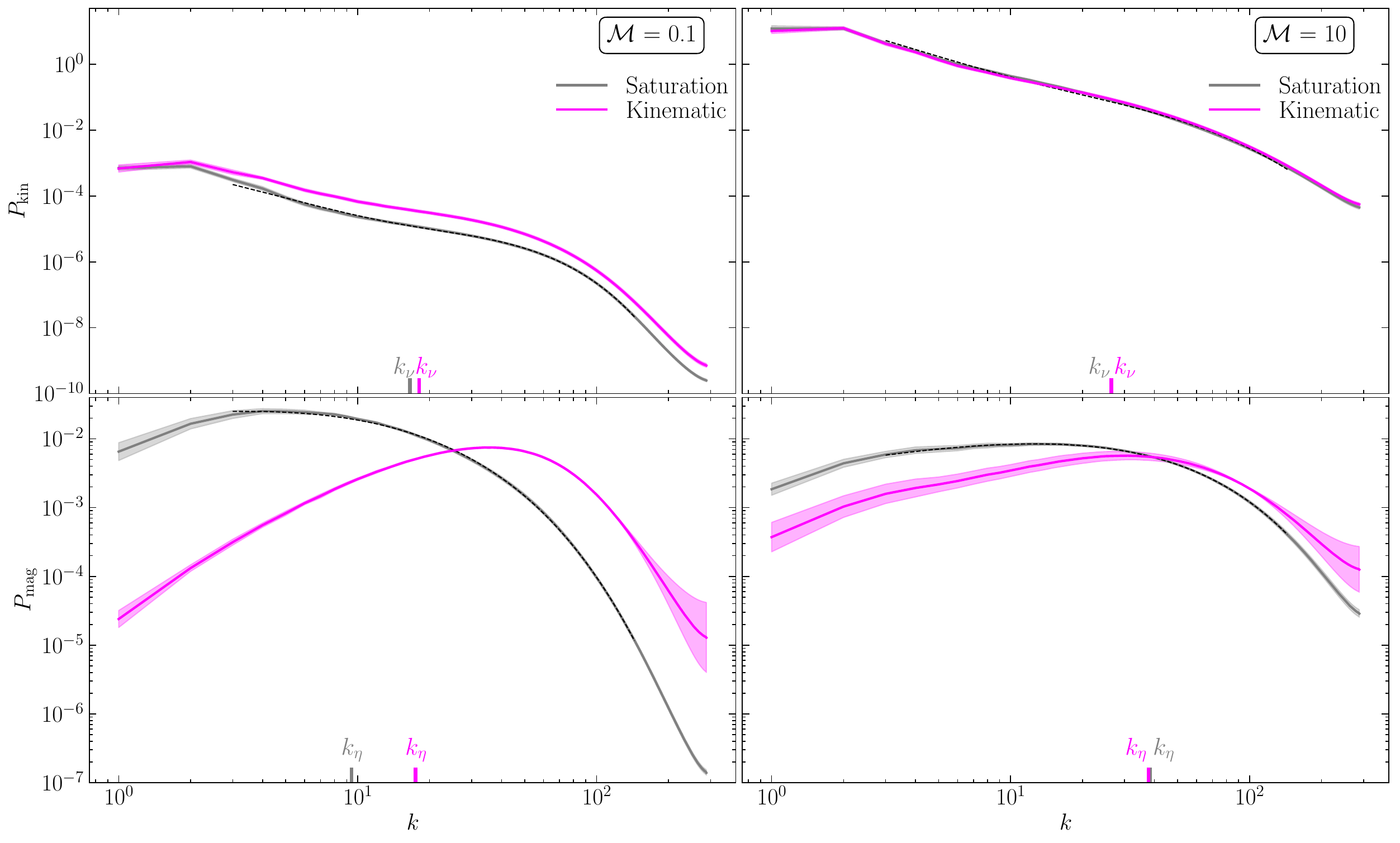}
    \caption{Same as Fig.~\ref{fig:spectra}, but for grid resolution $N=576$ only, comparing the spectra in the saturated phase (grey) with the previously-discussed spectra in the kinematic phase (magenta). We over-plot the fitted spectrum models (Eqs.~\ref{eq:kinspectra} and~\ref{eq:magspectra}) in the saturation phase as thin black dashed lines. The viscous and resistive dissipation wave numbers, $\knu$ and $\keta$, obtained from the saturation and kinematic phases are shown as tick markers on the $k$-axis in their respective colour. The full set of fit values is listed in Tab.~\ref{tab:sat_regime}. We see that the dissipation scales are very similar between the kinematic and saturation phase, with the exception of $\Pmag$ in the subsonic regime, showing a strong inversion of the magnetic spectrum as the magnetic peak moves to larger scales, as expected. Despite the significant difference in the spectral shape of $\Pmag$ in the saturation phase, $\keta$ is only reduced by a factor of $\sim2$ compared to the kinematic phase.
    }
    \label{fig:saturation_regime}
\end{figure*}

The comparison of the kinematic and saturated energy spectra is presented in Fig.~\ref{fig:saturation_regime}. We fit the same models, Eqs.~(\ref{eq:kinspectra}) and~(\ref{eq:magspectra}), used in the kinematic phase, to the averaged kinetic and magnetic spectra, respectively. The time-window for averaging in the saturation phase, $37\leq t/\tturb \leq47$, is determined in Appendix~\ref{ap:sat_reg_time_evol}. The fits are overlaid (black dashed lines) on the saturated-regime spectra (grey) in Fig.~\ref{fig:saturation_regime}. Table~\ref{tab:sat_regime} lists the corresponding parameters (columns~3--10) and the characteristic wave numbers, $\knu$ and $\keta$ (columns~11 and~12, respectively). We now compare the saturation-phase results with those obtained in the kinematic phase (Tab.~\ref{tab:fit_params}).

Focusing first on the kinetic energy spectra (top panels of Fig.~\ref{fig:saturation_regime}), we see that they are practically identical for $\mach=10$ in the kinematic and saturation regimes. For $\mach=0.1$, the back-reaction of the magnetic field via the Lorentz force, due to the high saturation level in that case, causes the kinetic spectra to experience a slight reduction in amplitude, which is also reflected in the sonic Mach number evolution (Fig.~\ref{fig:time_evol_sat}) upon entering saturation. Despite this back-reaction, the spectral shapes are similar (as are the fit parameters--see Tab.~\ref{tab:sat_regime}), and both the $\mach=0.1$ and $10$ simulations, respectively, exhibit very similar viscous dissipation wave numbers between their kinematic and saturation regimes, with $\knu$ values identical to within $10\%$. 

The magnetic spectra (bottom panels of Fig.~\ref{fig:saturation_regime}) show a substantially stronger change upon saturation. The peak of the magnetic spectra is shifted to larger scales (smaller $k$), as expected in both the subsonic and supersonic regimes \citep{2005PhR...417....1B,2014ApJ...797L..19F}. Thus, the power-law exponent of the magnetic spectra, $\pmag\approx1$ (see Tab.~\ref{tab:sat_regime}) in the saturation phase, smaller than that in the kinematic phase. However, the other fit parameters are largely similar between the kinematic and saturation phases. Comparing the two phases, the main derived parameter, $\keta$, is almost identical in the supersonic regime, and despite the major differences in the spectral shape of $\Pmag$ in the subsonic regime, $\keta$ is only a factor of $\sim2$ smaller in the saturation regime, compared to the kinematic regime, while the magnetic peaks are about an order of magnitude different. While this factor of $\sim2$ difference may reflect a physical effect in that the back-reaction of the field may lead to dissipation on somewhat larger scales, we cannot exclude the possibility that the fit in the saturation regime slightly underestimates $\keta$, because of the reduced dynamic range in the power-law part of $\Pmag$ upon saturation.

In conclusion, the transition from the kinematic to the saturation phase of the dynamo affects the magnetic spectra more than the kinetic spectra. However, despite the major differences in the subsonic magnetic spectra upon saturation, the derived $\keta$ is only a factor of $\sim2$ smaller compared to the kinematic regime. Thus, our main results for the resolution dependence of numerical dissipation derived in the kinematic phase still hold, with a slight modification of the normalisation constant in the relation between $\keta$ and $N$. These relations are derived in the following section.

\begingroup
\setlength{\tabcolsep}{1.0pt} % 
\renewcommand{\arraystretch}{1.3}
\begin{table*}
	\caption{Simulation parameters and main results of the saturation phase for $N = 576$.}
	\label{tab:sat_regime}
\begin{threeparttable}
\begin{tabular}{l|c|ccccc|ccc|ccccc}
\hline
\multicolumn{1}{c}{} & \multicolumn{1}{c}{} & \multicolumn{5}{c}{\textbf{From $P_{\mathrm{kin}}$}} & \multicolumn{3}{c}{\textbf{From $P_{\mathrm{mag}}$}} & \multicolumn{5}{c}{\textbf{Derived}}\\
\hline
\multicolumn{1}{c|}{$\mach$} & \multicolumn{1}{c|}{$(\Emag/\Ekin)_\mathrm{sat}$} & \multicolumn{1}{c}{$\pkin$} & \multicolumn{1}{c}{$\pbn$} & \multicolumn{1}{c}{$\pnu$} & \multicolumn{1}{c}{$\kbn$} & \multicolumn{1}{c|}{$\knut$} & \multicolumn{1}{c}{$\pmag$} & \multicolumn{1}{c}{$\peta$} & \multicolumn{1}{c|}{$\ketat$} & \multicolumn{1}{c}{$\knu$} & \multicolumn{1}{c}{$\keta$} & \multicolumn{1}{c}{$\rek$} & \multicolumn{1}{c}{$\pem$} & \multicolumn{1}{c}{$\rem$}\\
\multicolumn{1}{c|}{(1)} & \multicolumn{1}{c|}{(2)} & \multicolumn{1}{c}{(3)} & \multicolumn{1}{c}{(4)} & \multicolumn{1}{c}{(5)} & \multicolumn{1}{c}{(6)} & \multicolumn{1}{c|}{(7)} & \multicolumn{1}{c}{(8)} & \multicolumn{1}{c}{(9)} & \multicolumn{1}{c|}{(10)} & \multicolumn{1}{c}{(11)} & \multicolumn{1}{c}{(12)} & \multicolumn{1}{c}{(13)} & \multicolumn{1}{c}{(14)} & \multicolumn{1}{c}{(15)}\\
\hline
0.1 & $(4.63\pm0.01)\!\times\!10^{-1}$ & -1.7 & $0.88\!\pm\!0.03$ & 1.0 & $14.8\!\pm\!0.2$ & $16.5\!\pm\!0.1$ & $0.91\!\pm\!0.03$  & $0.68\!\pm\!0.01$ & $4.46\!\pm\!0.24$ & $16.5\!\pm\!0.1$ & $8.99\!\pm\!0.93$ & $2.3_{-0.3}^{+0.4}\!\times\!10^{3}$ & $0.42_{-0.2}^{+0.4}$ & 	$9.5_{-3.0}^{+8.0}\!\times\! 10^{2}$ \\
10 & $(8.65\!\pm\!0.02)\!\times\!10^{-3}$ & -2.0 & $0.00\!\pm\!0.00$ & 0.7 & $12.7\!\pm\!0.4$ & $9.91\!\pm\!0.38$ & $0.94\!\pm\!0.01$ & $0.78\!\pm\!0.01$ & $17.2\!\pm\!0.4$ & $26.5\!\pm\!1.4$ & $38.5\!\pm\!2.0$ & $2.0_{-0.5}^{+0.7}\!\times\! 10^{3}$ & $7.0_{-4.0}^{+20.0}$ & 	$1.4_{-0.8}^{+4.0}\!\times\! 10^{4}$ \\
\hline
\end{tabular}
\begin{tablenotes}[flushleft]
\note{All measured and derived parameters were obtained by time averaging over the saturation phase of the dynamo, $37 \leq t/\tturb \leq 47$, for both the subsonic ($\mach = 0.1$) and supersonic ($\mach = 10$) regimes (see Appendix~\ref{ap:sat_reg_time_evol}). The methodologies employed here are the same as in \S\ref{sec:methods}, and the parameters listed hold the same definition as in Tab.~\ref{tab:fit_params}, except for the saturation level, $(\Emag/\Ekin)_{\mathrm{sat}}$ (column~2--see Appendix~\ref{ap:sat_reg_time_evol} for its measurement), which replaces the dynamo growth rate in col.~2 of Tab.~\ref{tab:fit_params}.}

\end{tablenotes}
\end{threeparttable}
\end{table*}
\endgroup

\section{Effective Reynolds Numbers ($\rek$, $\rem$) in Ideal MHD} \label{sec:rermpm}

Our main goal here is to determine the dissipation properties of grid-based ideal MHD turbulence simulations as a function of the grid resolution $N$. We do so in two layers. We first determine the variations of the viscous dissipation wave number ($\knu$) and the resistive dissipation wave number ($\keta$) with $N$, which are the scales at which kinetic and magnetic energy begin to dissipate, respectively. Using the relations established in \citet{2022MNRAS.513.2457K}, we calculate the hydrodynamic Reynolds number ($\rek$) and the magnetic Reynolds number ($\rem$) as a function of $N$.

\subsection{Dependence of dissipation wave numbers, $\knu$ and $\keta$, on $N$}
\label{ssec:kvsN}

%%%%%%%%%%%%%%%%%%%%%%%%%%%%%%%%%%
\begin{figure*}
    \centering
    \includegraphics[width=1.0\linewidth]{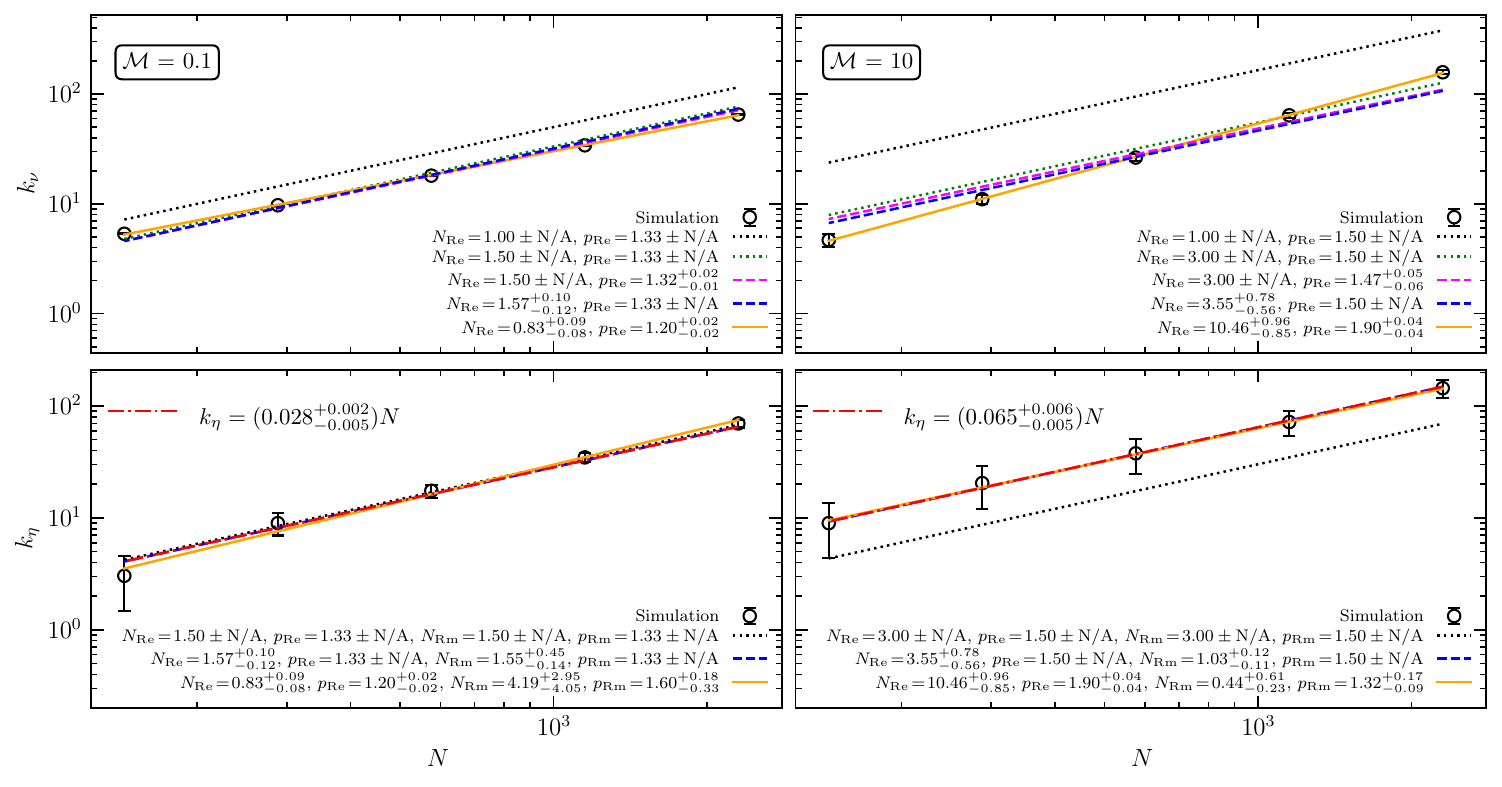}
    \caption{Viscous dissipation wave number $\knu$ obtained from $\Pkin$ (top panels; column~11 in Tab.~\ref{tab:fit_params}), and the resistive dissipation wave number $\keta$ obtained from $\Pmag$ (bottom panels; column~12 in Tab.~\ref{tab:fit_params}), in units of $\kbox$, against linear grid resolution $N$, along with their 2-sigma variations displayed as error bars. The left and right panels belong to the subsonic and supersonic regimes, respectively. We find that $\knu$ for $\mach = 0.1$ and $\keta$ for both subsonic and supersonic regimes vary linearly with $N$, well within the error bars, and $\knu$ for $\mach = 10$ varies slightly super-linearly with $N$. We fit Eq.~(\ref{eq:ketalin}) to $\keta$ to determine the proportionality constant in the linear relation between $\keta$ and $N$ (see legend in the top left corner of the bottom panels). Eqs.~(\ref{eq:knuvsN}) and (\ref{eq:ketavsN}) are fitted (taking the 2-sigma variations into account) to $\knu$ and $\keta$, respectively, to extract the corresponding parameters (shown in the legends). Various different combinations of free and fixed fit parameters (where fixed parameters are indicated with N/A on the respective parameter error quoted in the legend; the lower and upper error bars correspond to the 16th and 84th percentiles, respectively) are presented with different colours and line styles (see legends; discussed in the text). }
    \label{fig:kvsN}
\end{figure*}
%%%%%%%%%%%%%%%%%%%%%%%%%%%%%%%%%%

%%%%%%%%%%%%%%%%%%%%%%%%%%%%%%%%%%
\begin{figure*} 
    \centering
    \includegraphics[width=1.0\linewidth]{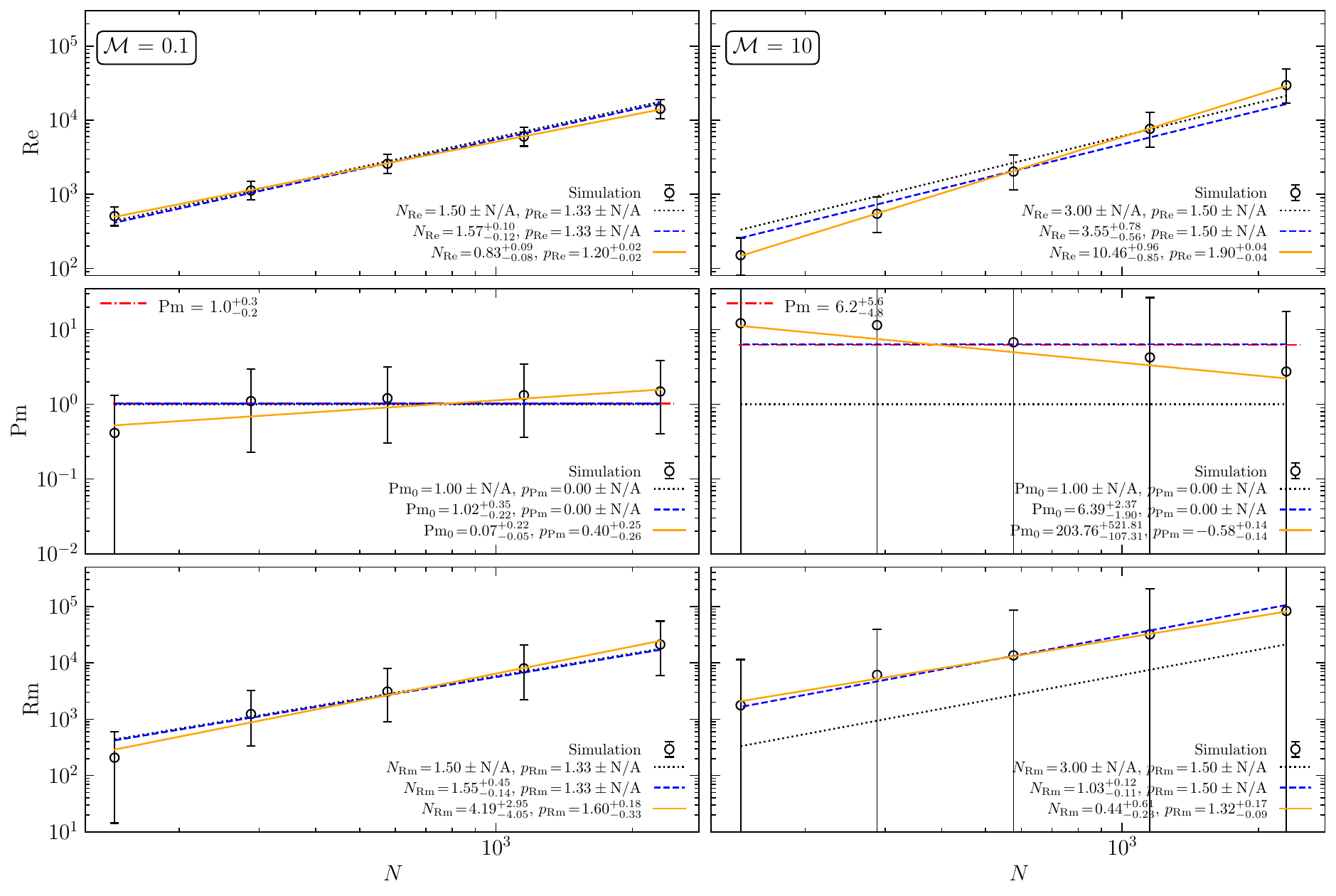}
    \caption{Hydrodynamic Reynolds number $\rek$ calculated from $\knu$ (top panels; column~13 of Tab.~\ref{tab:fit_params}), magnetic Prandtl number $\pem$ calculated from $\knu$ and $\keta$ (middle panels; column~14 of Tab.~\ref{tab:fit_params}), and magnetic Reynolds number $\rem$ calculated from $\rek$ and $\pem$ (bottom panels; column~15 of Tab.~\ref{tab:fit_params}), against $N$. Eqs.~(\ref{eq:fitRevsN}), (\ref{eq:fitRmvsN}), and (\ref{eq:fitPmvsN}) are plotted with fixed parameters from Fig.~\ref{fig:kvsN}. A completely free fit would suggest that $\pem\sim N^{0.40}$ and $\sim N^{-0.58}$ in the subsonic and the supersonic regime, respectively. However, assuming $\pem = $ constant resulting in $\pem=1.0_{-0.2}^{+0.3}$ for $\mach = 0.1$, and $\pem=6.2_{-4.8}^{+5.6}$ for $\mach = 10$ (red dash-dotted lines) also provides reasonable fits, considering the uncertainties in $\pem$.}
    \label{fig:reyvsN}
\end{figure*}
%%%%%%%%%%%%%%%%%%%%%%%%%%%%%%%%%%

\subsubsection{Formulation and results} \label{sssec:reykrels}
We start by developing models for $\knu$ and $\keta$ as a function of $N$, which are all intended to apply for $\rek>100$, based on the works by \citet{2022MNRAS.513.2457K,KrielEtAl2025}. As discussed previously in \S\ref{ssec:power_spectra}, we expect $\knu$ and $\keta$ to vary linearly with $N$. We consider the following relation between $\knu$ and $\rek$ from \citet{2022MNRAS.513.2457K},
\begin{equation}
\knu = \cnu\kturb\rek^{1/\prekth},
\label{eq:knugeneral}
\end{equation}
with the coefficient $\cnu = 0.025^{+0.005}_{-0.006}$ in the subsonic case as determined in \citet{2022MNRAS.513.2457K}. Using fig.~6 from \citet{KrielEtAl2025}, we derive $\cnu = 0.083^{+0.027}_{-0.031}$ for the supersonic regime (see Appendix~\ref{ap:cons_calc}). The theoretical scaling exponents $\prekth=4/3$ in the subsonic regime ($\mach = 0.1$) \citep{1991RSPSA.434....9K,1995tlan.book.....F,2022MNRAS.513.2457K} and $\prekth=3/2$ in the supersonic regime ($\mach = 10$)\footnote{The exponent $\prekth=3/2$, since $\rek \propto u(\ell)\ell \propto \ell^{3/2} \propto N^{3/2}$ (see Eq.~\ref{eq:Re}), because $u(\ell) \propto \ell^{1/2}$ in the supersonic regime \citep{2013MNRAS.436.1245F,FederrathEtAl2021}.}.
This implies a relation between $\rek$ and $N$, postulated as
\begin{equation}
\rek = \left(\frac{N}{\nrek}\right)^{\prek},
\label{eq:fitRevsN}
\end{equation}
where $\nrek$ is the grid resolution at which $\rek=1$, i.e., occurring at the Kolmogorov micro-scale, and $\prek$ is the power-law exponent in that relation. Theoretically, we expect $\prek=\prekth$, but we will measure $\prek$ directly from the simulations. We note that $\prek=\prekth$ is required for $\knu$ to depend linearly on $N$. We emphasise that Eq.~\ref{eq:fitRevsN} is for $\kturb = 2$. A generalised transferrable relation in terms of $\kturb$ is
\begin{equation}
\rek = \left[\left(\frac{N}{\nrek}\right)\left(\frac{2\kbox}{\kturb}\right)\right]^{\prek} = \left[\left(\frac{N}{\nrek}\right)\left(\frac{2\lturb}{L}\right)\right]^{\prek},
\label{eq:RevsN_kt}
\end{equation}
where the latter is in terms of $\lturb = 2\pi/\kturb$.
In our study, we use Eq.~(\ref{eq:fitRevsN}) for further calculations. All resultant formulations can be obtained for any other $\kturb$ by using Eq.~(\ref{eq:RevsN_kt}). 

The relation for $\knu$ on $N$ is the resulting model obtained from plugging Eq.~(\ref{eq:fitRevsN}) into  Eq.~(\ref{eq:knugeneral}), which results in
\begin{equation}
    \knu = \cnu\kturb\left(\frac{N}{\nrek}\right)^{\prek/\prekth}.
    \label{eq:knuvsN}
\end{equation}

In order to determine the values of $\prek$ and $\nrek$, we fit Eq.~(\ref{eq:knuvsN}) to $\knu$ obtained from the simulations' power spectra as a function of $N$, with 2-sigma variations taking into account during fitting (see column~11 in Tab.~\ref{tab:fit_params}). The resulting plot is shown in the top panels of Fig.~\ref{fig:kvsN}, with the left and right panels for the subsonic and supersonic regimes, respectively. The fit parameter values obtained are presented in the figure legends, with an error of N/A indicating that the corresponding parameter was maintained fixed for the associated fit. The lower and upper error bars correspond to the 16th and 84th percentiles, respectively. We discuss the behaviour of the fit parameters in \S\ref{sssec:effecK_dic}.

A clear and robust result from these simulations is that we find that $\keta$ varies linearly with $N$ in both the subsonic and supersonic regimes of turbulence. We note that the lowest-resolution case ($N=144$) is the only simulations that show some deviations from this, which is expected, given that those models are extremely low in resolution.

Therefore, we analyse the dependence of the resistive dissipation wave number on grid resolution by fitting the linear relation,
\begin{equation}
    \keta = \clin N,
    \label{eq:ketalin}
\end{equation}
to $\keta$ obtained from the simulations (with 2-sigma variations taken into account; see column~12 in Tab.~\ref{tab:fit_params}), and $\clin$ is displayed in the legend in the bottom panels of Fig.~\ref{fig:kvsN}.

In order to link this linear relation to the commonly established relation for $\keta$ in the literature, we now consider the following relation \citep{2022MNRAS.513.2457K}:
\begin{equation}
\keta = \ceta\knu\pem^{1/2}.
\label{eq:ketageneral}
\end{equation}
The coefficient $\ceta = 0.88^{+0.21}_{-0.23}$ in the subsonic case as measured in \citet{2022MNRAS.513.2457K}. Similar to $\cnu$, from fig.~6 in \citet{KrielEtAl2025}, we derive $\ceta = 0.55^{+0.25}_{-0.27}$ for the supersonic regime (see Appendix~\ref{ap:cons_calc}).

We propose the following model for $\rem$ as a function of $N$ for $\kturb = 2$, in analogy to Eq.~(\ref{eq:fitRevsN}), i.e., $\rem$ is defined identically to $\rek$, with the only difference being the use of resistivity instead of viscosity in the denominator (c.f., Eqs.~\ref{eq:Re} and~\ref{eq:Rm}),
\begin{equation}
\rem = \left(\frac{N}{\nrem}\right)^{\prem}.
\label{eq:fitRmvsN}
\end{equation}
Including the dependence on $\kturb$ or $\lturb$, the equation reads
\begin{equation}
\rem = \left[\left(\frac{N}{\nrem}\right)\left(\frac{2\kbox}{\kturb}\right)\right]^{\prem} = \left[\left(\frac{N}{\nrem}\right)\left(\frac{2\lturb}{L}\right)\right]^{\prem}.
\label{eq:RmvsN_kt}
\end{equation}
As previously mentioned, we use Eq.~(\ref{eq:fitRmvsN}) with $\kturb=2$ for further calculations, and Eq.~(\ref{eq:RmvsN_kt}) can be used to scale these results to any other $\kturb$. 

The resulting formulation for the relation between $\keta$ and $N$ obtained by substituting Eqs.~(\ref{eq:fitRmvsN}), (\ref{eq:knugeneral}) and (\ref{eq:fitRevsN}) into Eq.~(\ref{eq:ketageneral}), is
\begin{equation}
    \keta = \cnu\ceta\kturb\frac{N^{\prek/\prekth-\prek/2+\prem/2}}{\nrek^{\prek/\prekth-\prek/2}\,\nrem^{\prem/2}}.
    \label{eq:ketavsN}
\end{equation}
We fit Eq.~(\ref{eq:ketavsN}) to $\keta$, and extract $\prem$ and $\nrem$. The resulting plot is shown in the bottom panels of Fig.~\ref{fig:kvsN}. We discuss the characteristics of $\keta$ as a function of $N$ in \S\ref{sssec:effecK_dic}. 

\subsubsection{Discussion}
\label{sssec:effecK_dic}
In this section, we aim to understand the effects of linear grid resolution $N$ on the dissipation scales. Hence, we study the properties of the dependence of the dissipation wave numbers $\knu$ and $\keta$ on $N$. Before that, we reason out the reliability of the simulation results based on their numerical resolution. Columns~13 and~14 of Tab.~\ref{tab:fit_params} list $\rek$ and $\pem$ calculated from $\knu$ and $\keta$ obtained from the power spectra (columns~11 and 12) using Eqs.~(\ref{eq:knugeneral}) and (\ref{eq:ketageneral}), respectively. The $\rem$ (column~15) is calculated from the resulting $\rek$ and $\pem$ using Eq.~(\ref{eq:pm_rel}). For magnetic amplification to dominate over magnetic dissipation in the kinematic phase of the dynamo, $\rem$ has to exceed a critical value of $\mathrm{Rm}_\mathrm{crit} \simeq 100$, depending on the sonic Mach number and $\pem$ of the fluid \citep{2004PhRvL..92e4502S,2004ApJ...612..276S,2004PhRvE..70c6408H,2005PhR...417....1B,2012PhRvE..85b6303S,2014ApJ...797L..19F,2021PhRvL.126i1103A}. We find that $\rem \gtrsim 100$ for $N\gtrsim 100$. We also know that, generally, turbulence fully develops for \mbox{$\rek\gtrsim 100$--$1000$} \citep{1995tlan.book.....F,2014PNAS..11110961S}, although it depends on the geometry of the system. We find that $\rek < 1000$ for $N = 144$, for both $\mach = 0.1$ and 10, while $\rek \gtrsim 1000$ for \mbox{$N\gtrsim 200$--$300$}. This implies that fluid motions for the lowest resolution are borderline turbulent, and the intensity of turbulence, characterised by $\rek$, increases with $N$. Therefore, the lowest-resolution runs ($N=144$) are expected to show deviations from the scaling relations that we establish in this work. Indeed, we see in Fig.~\ref{fig:kvsN} that this resolution shows deviations from the power-law scaling relations with $N$, particularly for $\keta$ (bottom panels).

The dotted lines in Fig.~\ref{fig:kvsN} represent the basic theoretical relations for $\knu$ and $\keta$, i.e., the formulated models plotted with fixed parameters given by theory, as displayed. The dashed lines are fits of the dissipation wave numbers with at least one (but not all) parameter left free to vary, and the solid lines are fitted with all the parameters left free to vary. 

\paragraph{Viscous dissipation scale ($\knu$):}

We start by discussing the dependence of the viscous dissipation wave number $\knu$ on the grid resolution $N$. We note that (see Appendix~\ref{ap:p_re_kin})
\begin{equation}
    \prek = \frac{1-\pkin}{2}.
    \label{eq:pre_kin}
\end{equation}
Therefore, using theoretical values for $\pkin$, with $\pkin = -5/3$ from Kolmogorov turbulence (subsonic regime), and $\pkin = -2$ for Burgers turbulence (supersonic regime), the theoretical predictions are $\prek=4/3$ and 3/2 for $\mach = 0.1$ and 10, respectively.

From the fits in Fig.~\ref{fig:kvsN}, we find that in the subsonic regime, Eq.~(\ref{eq:knuvsN}) with $\nrek = 1.50$ and $\prek = 4/3$ is a reasonable model for $\knu$ as a function of $N$, with the slope of the data being slightly shallower. A fit with $\nrek = 1.50$ gives $\prek = 1.32_{-0.01}^{+0.02} \approx \prekth$ (magenta-dashed line). We find $\nrek=1.57_{-0.12}^{+0.10}$ for a fit with $\prek = \prekth$ (blue-dashed line), quantifying the data being shallower than our predictions. Therefore, even though the best fit (yellow-solid line) provides $\nrek=0.83_{-0.08}^{+0.09}$ and $\prek=1.20_{-0.02}^{+0.02}$, we consider the blue-dashed line to be our preferred fit for further analysis, as it provides a reasonable model (with only 1~fit parameter), especially for $N\geq144$ (as discussed above), and follows the theoretical prediction of $\prek = \prekth = 4/3$ reasonably well. 

Considering the supersonic regime, we observe that $\knu$ as a function of $N$ is steeper than either of our theoretical predictions, i.e., the black-dotted line ($\nrek = 1.0$; $\prek = 3/2$) and the green-dotted line ($\nrek = 3.0$; $\prek = 3/2$). Therefore, we find $\prek>\prekth$ in the supersonic regime. This is quantified by the best-fit line with $\prek = 1.90_{-0.04}^{+0.04}$ being $\sim28\%$ larger than $\prekth=3/2$, and $\nrek=10.46_{-0.85}^{+0.96}$. This result suggests that $\knu$ in the supersonic regime varies super-linearly with $N$, with $\rem = 1$ occurring at a larger number of cells ($\nrek\sim4$) for $\mach=10$ than for $\mach=0.1$ ($\nrek\sim1.5$).

In conclusion, while the results of the dependence of $\knu$ on $N$ in the subsonic regime agree with the predicted value of $\prek$, $\prek$ is found to be $\sim28\%$ larger than the theoretical value (3/2) for $\mach = 10$. We discuss this further in Appendix~\ref{ap:decomp_spec}.

We can categorise $\nrek$ and $\prek$ into two reasonable sets. The first one is for $\prek = \prekth$, and the second one leaves $\prek$ free to fit (yellow lines). The reason for such a characterisation is because the green, magenta, and blue lines all provide reasonably similar and good fits for $\prek \sim \prekth$, while the yellow lines formally provide the best fits when leaving also $\prek$ to vary. Therefore, the sets for the subsonic regime are $(\nrek, \prek) = (1.57_{-0.12}^{+0.10},4/3)$ and $(0.83_{-0.08}^{+0.09}, 1.20_{-0.02}^{+0.02})$, and for the supersonic regime they are $(3.55_{-0.56}^{+0.78},3/2)$ and $(10.46_{-0.85}^{+0.96},1.90_{-0.04}^{+0.04})$. 

\paragraph{Resistive dissipation scale ($\keta$).}

As discussed above, we find that $\keta$ varies linearly with $N$, irrespective of $\mach$ (see red dot-dashed lines in the bottom panels of Fig.~\ref{fig:kvsN}). From Eq.~(\ref{eq:ketavsN}), we proceed to obtain Eqs.~(\ref{eq:p_rel_sub}) and (\ref{eq:p_rel_sup}) between $\prem$ and $\prek$ for $\mach=0.1$ and 10, respectively, which obeys the linear relation between $\keta$ and $N$ (see Appendix~\ref{ap:p_re_rm}). This implies that for $\prek = \prekth$, we must have $\prem = \prek$ (irrespective of $\mach$).

We note that in the bottom panels of Fig.~\ref{fig:kvsN}, the errors in $\nrek$ and $\prek$ do not imply that these parameters were allowed to vary in that particular fit, but are the errors carried over from the corresponding parameter values in the top panels. For the plots of $\keta$ vs.~$N$, the dotted lines represent the theoretical predictions with $\prek = \prem = \prekth$ and $\nrek = \nrem$ (values taken from the green-dotted lines in $\knu$ vs.~$N$). The blue-dashed and yellow-solid lines correspond to the two characterised sets of $(\nrek, \prek)$ mentioned above, where either $\prem = \prekth$ (blue-dashed line) or $\prem$ was left free to fit (yellow-solid line).

Considering the linear dependence of $\keta$ on $N$, we note that the coefficients of the fit are $0.028_{-0.005}^{+0.002}$ and $0.065_{-0.005}^{+0.006}$ in the subsonic and supersonic regimes, respectively. We also note in this context that $\keta$ has a weaker dependence on $\rek$ than on $\rem$, i.e., $\keta\propto\knu\pem^{1/2}\propto\rek^{1/\prek-1/2}\rem^{1/2}$ (Eqs.~\ref{eq:ketageneral} and \ref{eq:knugeneral}). Therefore, $\keta$ is not as sensitive to $\prek$ compared to $\knu$, and thus, we obtain good linear relations for $\keta$ with $N$ even for the supersonic case, where the data in the top right panel of Fig.~\ref{fig:kvsN} suggests $\prek\sim1.9$. This is further reflected in the $\prem$ values obtained from the yellow fits, which abide by Eq.~(\ref{eq:p_rel_sup}) within the 1-sigma uncertainties. Eq.~(\ref{eq:p_rel_sup}) predicts $\prem = 1.40$ for $\prek = 1.20$ in the subsonic regime, and $\prem = 1.37$ for $\prek = 1.90$ in the supersonic regime, consistent with the $\prem$ values obtained from the yellow fits within error bars.

\paragraph{Summary} \label{para:summ_krels}
We now summarise all fit parameters, distinguishing two sets (one with fixed $\prek$ and $\prem$ at their theoretical values, and the other where they are free to vary), in the order $(\nrek, \prek, \nrem, \prem)$:
\begin{itemize}
    \item Subsonic regime: $(1.57_{-0.12}^{+0.10}, 4/3, 1.55_{-0.14}^{+0.45}, 4/3)$ and $(0.83_{-0.08}^{+0.09}, 1.20_{-0.02}^{+0.02}, 4.19_{-4.05}^{+2.95}, 1.60_{-0.33}^{+0.18})$.
    \item Supersonic regime: $(3.55_{-0.56}^{+0.78},3/2,1.03_{-0.11}^{+0.12},3/2)$ and $(10.46_{-0.85}^{+0.96},1.90_{-0.04}^{+0.04},0.44_{-0.23}^{+0.61},1.32_{-0.09}^{+0.17})$.
\end{itemize}
In the subsonic regime, as the model abiding by the theoretical prediction (blue-dashed line) fits the data within 2-sigma variations, we use this set: $(1.57_{-0.12}^{+0.10}, 4/3, 1.55_{-0.14}^{+0.45}, 4/3)$ as our primary model for further investigations. In contrast, our primary set for the supersonic regime is $(10.46_{-0.85}^{+0.96},1.90_{-0.04}^{+0.04},0.44_{-0.23}^{+0.61},1.32_{-0.09}^{+0.17})$ as it provides the best fit to the data, but with $\prek\sim1.9$, steeper than the theoretical expectation. Appendix~\ref{ap:decomp_spec} discusses reasons for this steeper-than-theoretical dependence of $\knu$ on $N$. While we here primarily distinguish the subsonic and supersonic regime, represented by $\mach=0.1$ and 10, respectively, we provide a study of the $\mach$-number dependence in Appendix~\ref{ap:ronmach}. Finally, we find that these relations depend only mildly on whether the magnetic field is saturated or not; see \S\ref{ssec:sat_results}.

%%%%%%%%%%%%%%%%%%%%%%%%%%%%%%%%%%
\subsection{Dependence of $\rek$, $\rem$ and $\pem$ on $N$}
\label{ssec:RvsN}
Here we determine the dependence of the hydrodynamic Reynolds number $\rek$ and the magnetic Reynolds number $\rem$ on the grid resolution $N$ in ideal-MHD simulations, which quantify the numerical viscous and resistive dissipation, respectively. 

$\rek$ as a function of $\knu$, obtained by rearranging Eq.~(\ref{eq:knugeneral}), is given by
\begin{equation}
    \rek = \left(\frac{\knu}{\cnu\kturb}\right)^{\prekth}.
    \label{eq:re_knu}
\end{equation}

Similarly, the magnetic Prandtl number $\pem$ as a function of $\keta$ and $\knu$, obtained by rearranging Eq.~(\ref{eq:ketageneral}), is given by
\begin{equation}
    \pem = \left(\frac{\keta}{\ceta\knu}\right)^2.
    \label{eq:pm_keta}
\end{equation}

From Eqs.~(\ref{eq:re_knu}), (\ref{eq:pm_keta}), and~(\ref{eq:pm_rel}), we compute the values of $\rek$, $\pem$ and the corresponding $\rem$, along with their errors, using Monte-Carlo error propagation. The results are listed in columns~13,~14 and~15 of Tab.~\ref{tab:fit_params}, respectively (see~\S\ref{ssec:kvsN}). 

The resulting values of $\rek$, $\rem$, and $\pem$ are shown in Fig.~\ref{fig:reyvsN}, following the same basic figure organisation as in Fig.~\ref{fig:kvsN}. The black-dotted, blue-dashed, and yellow-solid lines in Fig.~\ref{fig:reyvsN} hold the same meaning as those in the $\keta$-plot (see bottom panel of Fig.~\ref{fig:kvsN}). The values and errors of the parameters shown in the panel legends are obtained using Monte-Carlo error propagation from the fits done in Fig.~\ref{fig:kvsN}.

\subsubsection{Hydrodynamic Reynolds number ($\rek$)} 

We start by considering $\rek$ as a function of $N$. This is shown in the top panels of Fig.~\ref{fig:reyvsN}, with the left and right panels belonging to $\mach=0.1$ and 10, respectively. We observe that the parameters obtained for the dependence of $\knu$ on $N$ fit that of $\rek$, within the error bars. Therefore, in the subsonic case, the exponent of Eq.~(\ref{eq:fitRevsN}) ($\prek$) follows the prediction of Eq.~(\ref{eq:pre_kin}), whereas in the supersonic case, $\prek$ is $28\%$ higher than the expectation from Burgers turbulence.

\subsubsection{Magnetic Prandtl number ($\pem$)} 

To study any potential (albeit weak) variation of $\pem$ with $N$, we define
\begin{equation}
\pem = \pmnot\,N^{\ppem}, \label{eq:fitPmvsN}
\end{equation}
where $\pmnot = \pem$ when $N=1$. From Eqs.~(\ref{eq:fitRevsN}) and (\ref{eq:fitRmvsN}), we obtain
\begin{align}
    \ppem &= \prem - \prek 
    \label{eq:ppm}, \\
    \pmnot &= {\nrek^{\prek}}/{\nrem^{\prem}}.
\end{align}
With no further fitting done, we calculate $\ppem$ and $\pmnot$, and their corresponding errors. We consider $\nrek$, $\prek$, $\nrem$ and $\prem$ from the bottom panels of Fig.~\ref{fig:reyvsN} for this process, to produce the blue and yellow lines of $\pem$, respectively, shown in the middle panels of Fig.~\ref{fig:reyvsN}. We note that from Eq.~(\ref{eq:ppm}), if $\prek = \prem$, it automatically follows that $\pem$ is independent of $N$. This is depicted by the black dotted lines for both $\mach=0.1$ and $10$, where we also picked $\nrek = \nrem$ ($1.50$ for $\mach=0.1$ and $3.0$ for $\mach=10$), resulting in $\pem=1$. Considering the yellow-solid line, we note that $\pem$ scales as $\sim N^{0.40_{-0.26}^{+0.25}}$ and $\sim N^{-0.58_{-0.14}^{+0.14}}$ in the subsonic and supersonic regime, respectively. Having explored the weak dependence of $\pem$ on $N$, we note that overall, $\pem$ is largely consistent with a constant value in the subsonic and supersonic regimes, within the 2-sigma variations. Therefore, we also fit a constant (i.e., $\pem = $ constant) and obtain $1.0_{-0.2}^{+0.3}$ for $\mach = 0.1$, and $6.2_{-4.8}^{+5.6}$ for $\mach = 10$. These are shown as red dash-dotted lines in Fig.~\ref{fig:reyvsN}.

\subsubsection{Magnetic Reynolds number ($\rem$)} 

The bottom panels of Fig.~\ref{fig:reyvsN} show $\rem$ based on Eq.~(\ref{eq:fitRmvsN}) with parameters $\nrem$ and $\prem$ taken from $\keta$ vs.~$N$. For both subsonic and supersonic regimes, we note that both the blue-dashed ($\prem = \prekth$) and the yellow-solid lines ($\prem = 1.60_{-0.33}^{+0.18}$ for $\mach=0.1$; $\prem = 1.32_{-0.09}^{+0.17}$ for $\mach=10$) are good fits.

% === SECTION 5 ===
\section{Comparison to simulation data in the literature}
\label{sec:link}
\newcounter{savecntr}

Up to this point, we determined (c.f., Fig.~\ref{fig:reyvsN}) the dependence of the Reynolds numbers ($\rek$ and $\rem$), and the magnetic Prandtl number ($\pem$), on the numerical resolution ($N$) in the simulations developed in this work, resulting in the relations given by Eqs.~(\ref{eq:RevsN_kt}) and~(\ref{eq:RmvsN_kt}) with fit parameters provided in Sec.~\ref{para:summ_krels}. In this section, we aim to better understand the significance of the present findings and how they relate to other works. We start by comparing our relations with simulations using explicit viscosity and resistivity, setting target $\rek$ and $\rem$ for a given $N$. We also compare our relations with those from \citet{2023ApJ...942L..34G} in the subsonic regime, finding good agreement. We note that different numerical schemes and solvers for hydrodynamics use different approximations and methods for computing solutions to the HD/MHD equations. Therefore, it is important to determine the dependence of the relations (and associated $\nrek$, $\prek$, $\nrem$, and $\prem$) on the numerical scheme/solver/method. Therefore, we quantify the variations in the $\rek$-$N$ relations across 13~different state-of-the-art schemes/codes including grid-based methods and smoothed-particle-hydrodynamics (SPH) from turbulence simulations in the literature. Before doing that, we compare our results to simulations using explicit viscosity and resistivity.

\subsection{Comparison with simulations using explicit dissipation}
%%%%%%%%%%%%%%%%%%%%%%%%%%%%%%%%%%
\begin{figure*}
    \centering
    \includegraphics[width=1.0\linewidth]{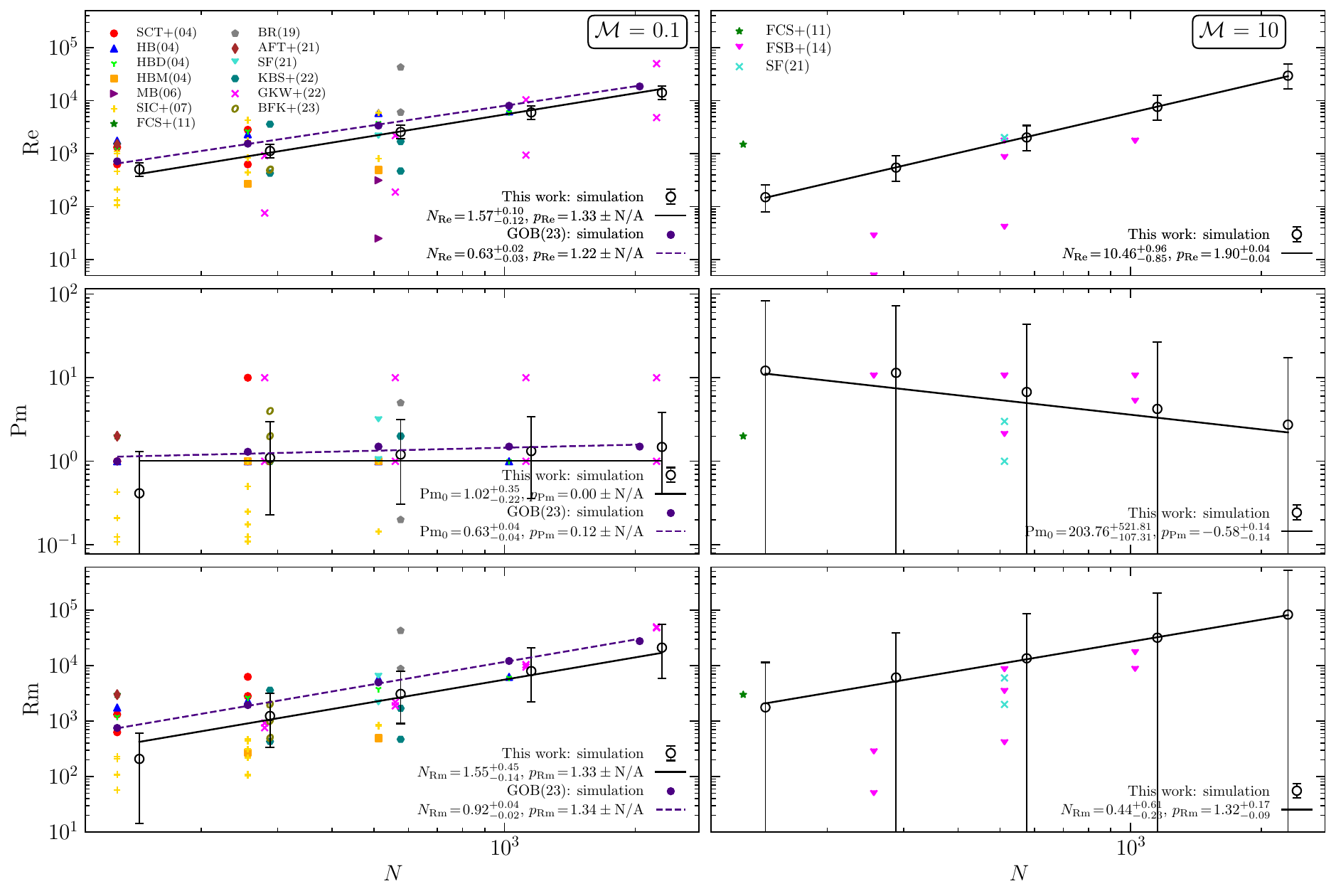}
    \caption{Hydrodynamic Reynolds number ($\rek$, top panels), magnetic Prandtl number ($\pem$, middle panels), and magnetic Reynolds number ($\rem$, bottom panels) as a function of $N$ in published simulation studies with explicit dissipation (see legends in the top-left corner of the top panels), plotted alongside the numerical $\rek$, $\pem$, and $\rem$ models determined in the present work (corresponding parameters shown in the bottom-right corner legends of each panel), for the subsonic regime (left panels) and supersonic regime (right panels), respectively. Data points from the literature in the order displayed in the legend are -- for the subsonic regime: SCT+(04): \citep{2004ApJ...612..276S}, HB(04): \citep{2004PhRvE..70c6408H}, HBD(04): \citep{2004PhRvE..70a6308H}, HBM(04): \citep{2004MNRAS.353..947H}, MB(06): \citep{2006MNRAS.370..415M}, SIC+(07): \citep{2007NJPh....9..300S}, FCS+(11): \citep{FederrathEtAl2011}, BR(19): \citep{2019ApJ...879...57B}, AFT+(21): \citep{2021PhRvL.126i1103A}, SF(11): \citep{2021PhRvF...6j3701S}, KBS+(22): \citep{2022MNRAS.513.2457K}, GKW+(22): \citep{2022PhRvX..12d1027G}, and BFK+(23): \citep{2023MNRAS.524.3201B}; and for the supersonic regime: FCS+(11): \citep{FederrathEtAl2011}, FSB+(14): \citep{2014ApJ...797L..19F}, and SF(11): \citep{2021PhRvF...6j3701S}. For the subsonic regime we also show $\rek$, $\pem$, and $\rem$ obtained from ILES in \citet{2023ApJ...942L..34G} (GOB(23)) as indigo-filled circles, and their respective fit relations are shown as indigo dashed lines, for comparison with the respective ILES results obtained in the present study (black open circles and black solid lines) -- see legends in the bottom-right corners of the left-hand panels.}
    \label{fig:link}
\end{figure*}
%%%%%%%%%%%%%%%%%%%%%%%%%%%%%%%%%%

In order to understand the importance of reliable estimates of the numerical Reynolds numbers, we show $\rek$, $\pem$, and $\rem$ values that were set as target values (by explicitly setting $\nu$ and $\eta$ in Eqs.~\ref{eq:mom_eq} and~\ref{eq:B_rate}) in simulations of various studies in the literature. Fig.~\ref{fig:link} shows these simulation data with abbreviations of the literature sources in the top-left legend of the top panels for $\mach = 0.1$ and $10$. These literature sources, in the order displayed (increasing order of publication year) for the subsonic regime (left-hand panels) are SCT+(04): \citep{2004ApJ...612..276S}, HB(04): \citep{2004PhRvE..70c6408H}, HBD(04): \citep{2004PhRvE..70a6308H}, HBM(04): \citep{2004MNRAS.353..947H}, MB(06): \citep{2006MNRAS.370..415M}, SIC+(07): \citep{2007NJPh....9..300S}, FCS+(11): \citep{FederrathEtAl2011}, BR(19): \citep{2019ApJ...879...57B}, AFT+(21): \citep{2021PhRvL.126i1103A}, SF(21): \citep{2021PhRvF...6j3701S}, KBS+(22): \citep{2022MNRAS.513.2457K}, GKW+(22): \citep{2022PhRvX..12d1027G}, and BFK+(23): \citep{2023MNRAS.524.3201B}. Similarly, for the supersonic regime, the works included are FCS+(11): \citep{FederrathEtAl2011}, FSB+(14): \citep{2014ApJ...797L..19F}, and SF(21): \citep{2021PhRvF...6j3701S}.

We see that most of the target $\rek$ and $\rem$ explicitly set in these simulations are close to their maximum achievable values, i.e., the numerical $\rek$ and $\rem$ for a particular $N$ based on our relations. However, some published simulations set target $\rek$ and $\rem$ values below the numerical ones, while other simulations have targets exceeding the numerical values of $\rek$ and $\rem$ for a given $N$. The former have dissipation and $\rek$ and $\rem$ well resolved, while the latter may effectively have lower $\rek$ and $\rem$ than their target values set in the simulations. However, the details of this will depend on the numerical method used. Thus, we next compare our relations with those from other simulation works in the literature to understand by how much the numerical $\rek$--$N$ relations depend on the numerical method.

\subsection{Comparing numerical dissipation formulations}

In order to better understand the significance and robustness of our relations, we compare to recent simulations in \citet{2023ApJ...942L..34G}. \citet{2023ApJ...942L..34G} find relations between the numerical viscosity $\nu$ and resistivity $\eta$, and the grid spacing $\Delta_x$, i.e., $\nu \propto \Delta_x^{1.22}$ and $\eta \propto \Delta_x^{1.34}$, obtained from implicit large eddy simulations (ILES), similar to ours, but only in the subsonic regime, similar to our $\mach=0.1$ simulations (see their figure~5). With $\uturb$ and $\lturb$ fixed in an MHD simulation, we can rewrite these relations, using Eqs.~(\ref{eq:Re}) and~(\ref{eq:Rm}), as
\begin{align}
    & \nu \propto \Delta_x^{\prek} \propto N^{-\prek} \propto \rek^{-1}, \label{eq:nu_Re_greta}\\
    & \eta \propto \Delta_x^{\prem} \propto N^{-\prem} \propto \rem^{-1},
    \label{eq:eta_Rm_greta}
\end{align}
as $N = L/{\Delta_x}$, where $L$ is the side length of the box (computational domain).

Thus, \citet{2023ApJ...942L..34G} found $\prek=1.22$ and $\prem=1.34$, which is in agreement with $\prek\in[1.18,1.34]$ and $\prem\in[1.27,1.78]$ obtained from our simulations (see Fig.~\ref{fig:reyvsN}) in the subsonic regime. In order to find the parameters $\nrek$ and $\nrem$ for \citet{2023ApJ...942L..34G}'s data, we fit $\rek$ and $\rem$ to \citet{2023ApJ...942L..34G}'s simulations, as a function of $N$, using Eqs.~(\ref{eq:fitRevsN}) and (\ref{eq:fitRmvsN}), respectively. We plot the resulting model as the indigo-dashed line in Fig.~\ref{fig:link}, alongside \citet{2023ApJ...942L..34G}'s data, and our data and the best-fit model (c.f., Fig.~\ref{fig:reyvsN}; fit parameters are displayed in the legends in the bottom-right corner of Fig.~\ref{fig:link}). We see that the $\nrek\in[0.75,1.67]$ and $\nrem\in[0.14,7.14]$ obtained from our simulations are larger than $\nrek = 0.63_{-0.03}^{+0.02}$ and $\nrem = 0.92_{-0.02}^{+0.04}$, obtained from the fits to \citet{2023ApJ...942L..34G}'s data, by about a factor of $\sim2$. This difference in $\nrek$ and $\nrem$ between our simulations and \citeauthor{2023ApJ...942L..34G}'s simulations is a result of the MHD solver and code used in \citet{2023ApJ...942L..34G} compared to the numerical methods used here, a comparison that will be expanded in \S\ref{ssec:num_scheme} below. While \citet{2023ApJ...942L..34G} use a predictor-corrector, van Leer-type integrator, with the HLLD Riemann solver, and constrained transport, we use a modified version of the FLASH code utilising a 5-wave, approximate Riemann solver with divergence cleaning (see \S\ref{ssec:simul}). This suggests that the numerical MHD scheme used in \citet{2023ApJ...942L..34G} is somewhat less dissipative than the numerical scheme used here. However, focusing on $\mach = 0.1$, there is about an order of magnitude difference in $\rek$ across a factor of $16$ in $N$, as opposed to a factor of $\sim2$ difference in $\rek$ between our own and \citet{2023ApJ...942L..34G}'s numerical scheme and code for any particular $N$. Similarly, there is a $\sim2$ orders-of-magnitude difference in $\rem$ across $N$, as opposed to only a factor of $\sim2$ difference between the two numerical schemes and codes. This indicates that the numerical viscosity and resistivity depend primarily on the numerical resolution $N$ rather than the numerical scheme employed. However, here we only compared 2~different schemes, and we will now expand this comparison to another 13~different simulations that use various different numerical schemes.

\subsection{Dependence on the numerical scheme/method} \label{ssec:num_scheme}

\begingroup
\setlength{\tabcolsep}{6.0pt} % 
\renewcommand{\arraystretch}{1.2}
\begin{table*}
	\caption{Simulation parameters and main results of the code comparison study shown in Fig.~\ref{fig:compare_lit}.}
	\label{tab:compare_tab}
\begin{threeparttable}
\begin{tabular}{l|cc|cccccc|cc}
\hline
\multicolumn{3}{c}{From literature} & \multicolumn{6}{c}{\textbf{From $P_{\mathrm{kin}}$}} & \multicolumn{2}{c}{\textbf{Derived}}\\
\hline
\multicolumn{1}{c|}{Literature} & \multicolumn{1}{c}{$N$} & \multicolumn{1}{c|}{$\kturb$} & \multicolumn{1}{c}{$\pkin$} & \multicolumn{1}{c}{$\pbn$} & \multicolumn{1}{c}{$\pnu$} & \multicolumn{1}{c}{$\kbn$} & \multicolumn{1}{c}{$\knu$} & \multicolumn{1}{c|}{$\rek$} & \multicolumn{1}{c}{$\rek/\rekpre$} & \multicolumn{1}{c}{$\rek/\rektilpre$}\\
\multicolumn{1}{c|}{(1)} & \multicolumn{1}{c}{(2)} & \multicolumn{1}{c|}{(3)} & \multicolumn{1}{c}{(4)} & \multicolumn{1}{c}{(5)} & \multicolumn{1}{c}{(6)} & \multicolumn{1}{c}{(7)} & \multicolumn{1}{c}{(8)} & \multicolumn{1}{c|}{(9)} & \multicolumn{1}{c}{(10)} & \multicolumn{1}{c}{(11)} \\
\hline
\;\,(1)\quad KNC11-ENZO    & 512  & 1.5 & -2.00  & 0.30  & 0.73  & $6.03\pm0.06$ & $12.9\pm0.4$  & $1.1_{-2.0}^{+0.5}\!\times\! 10^{3}$ &	$0.5_{-0.3}^{+0.4}$ &	$0.49_{-0.3}^{+0.4}$ \\
\;\,(2)\quad KNC11-FLASH   & 512  & 1.5 & -2.10  & 0.20  & 0.66  & $5.39\pm0.05$ & $12.2\pm0.4$ & $1.0_{-2.0}^{+0.5}\!\times\! 10^{3}$ &	$0.46_{-0.3}^{+0.4}$ &	$0.45_{-0.3}^{+0.3}$ \\
\;\,(3)\quad KNC11-KT-MHD  & 512  & 1.5 & -1.95  & 0.38  & 0.75  & $6.31\pm0.09$  & $12.9\pm0.5$ & $1.1_{-2.0}^{+0.5}\!\times\! 10^{3}$ &	$0.5_{-0.3}^{+0.4}$ &	$0.49_{-0.3}^{+0.4}$ \\
\;\,(4)\quad KNC11-LL-MHD  & 512  & 1.5 & -2.00  & -0.15 & 0.94  & $7.51\pm0.15$  & $23.7\pm1.1$  & $2.8_{-5.0}^{+1.0}\!\times\! 10^{3}$ &	$1.3_{-0.8}^{+1.0}$ &	$1.2_{-0.8}^{+0.9}$ \\
\;\,(5)\quad KNC11-PLUTO   & 512  & 1.5 & -1.95  & -0.15 & 0.90  & $7.23\pm0.22$  & $26.8\pm1.8$  & $3.4_{-6.0}^{+2.0}\!\times\! 10^{3}$ &	$1.5_{-1.0}^{+1.0}$ &	$1.5_{-0.9}^{+1.0}$  \\
\;\,(6)\quad KNC11-PPML    & 512  & 1.5 & -1.89  & 0.20  & 0.70  & $8.00\pm0.25$   & $22.7\pm2.3$  & $2.6_{-4.0}^{+1.0}\!\times\! 10^{3}$ &	$1.2_{-0.7}^{+0.9}$ &	$1.1_{-0.7}^{+0.9}$  \\
\;\,(7)\quad KNC11-RAMSES  & 512  & 1.5 & -1.97  & 0.01  & 0.94  & $8.27\pm0.12$  & $22.6\pm0.7$ & $2.6_{-4.0}^{+1.0}\!\times\! 10^{3}$ &	$1.2_{-0.7}^{+0.9}$ &	$1.1_{-0.7}^{+0.9}$ \\
\;\,(8)\quad KNC11-STAGGER & 512  & 1.5 & -2.30  & -0.24 & 0.55  & $4.79\pm0.10$ & $21.8\pm2.3$  & $2.5_{-4.0}^{+1.0}\!\times\! 10^{3}$ &	$1.1_{-0.7}^{+0.9}$ &	$1.1_{-0.7}^{+0.8}$  \\
\;\,(9)\quad KNC11-ZEUS    & 512  & 1.5 & -2.00  & 0.41  & 0.70  & $6.28\pm0.12$  & $11.4\pm0.6$ & $9.4_{-20.0}^{+4.0}\!\times\! 10^{2}$ &	$0.42_{-0.3}^{+0.3}$ &	$0.41_{-0.3}^{+0.3}$ \\
(10)\quad PF10-PHANTOM   & 512  & 2.0 & -1.96  & -0.25 & 0.70  & $7.16\pm0.47$  & $13.1\pm1.7$  & $7.0_{-3.0}^{+2.0}\times 10^{2}$  & $0.41_{-0.1}^{+0.1}$ & $0.43_{-0.1}^{+0.1}$ \\
(11)\quad PF10-FLASH     & 512  & 2.0 & -1.90  & -0.55 & 0.55  & $6.81\pm1.13$   & $13.2\pm4.8$  & $7.1_{-5.0}^{+4.0}\times 10^{2}$  & $0.43_{-0.2}^{+0.2}$ & $0.45_{-0.2}^{+0.3}$ \\
(12)\quad BRS23-RUN-G    & 512  & 1.5 & -1.76  & 0.35  & 0.69  & $7.11\pm0.25$  & $12.0\pm1.0$  & $2.3_{-2.0}^{+0.9}\!\times\! 10^{3}$ &	$0.74_{-0.4}^{+0.5}$ &	$0.76_{-0.4}^{+0.5}$ \\
(13)\quad BRS23-RUN-M'   & 2048 & 1.5 & -1.70  & -0.01 & 0.94  & $40.0\pm25.6$   & $78.1\pm52.3$  & $3.0_{-5.0}^{+2.0}\!\times\! 10^{4}$ &	$1.8_{-1.0}^{+1.0}$ &	$2.2_{-1.0}^{+2.0}$  \\
\hline
\end{tabular}
\begin{tablenotes}[flushleft]
\note{Column~1: 13~simulations from the literature -- the first 9 are from \citet{2011ApJ...737...13K} (KNC+11), the next 2~are from \citet{PriceFederrath2010} (PF10), and the last 2~are from \citet{2023MNRAS.518.6367B} (BRS23). Column~2: number of resolution elements per spatial dimension (linear grid resolution; $N$). Column~3: turbulent driving wave number ($\kturb$). Columns~4-9 hold the same meaning as the corresponding parameters in Tab.~\ref{tab:fit_params}. The wave numbers are reported in units of $\kbox$. Columns~10 and~11: ratios of $\rek$ obtained from spectral fitting (column~9) to $\rek$ derived from our work using parameters in Fig.~\ref{fig:reyvsN}, for the two sets of fit parameters distinguished in Sec.~\ref{para:summ_krels}, i.e., $\rek/\rekpre$ (column~10) uses the theoretical value of $\prek$, and $\rek/\rektilpre$ (column~11) uses the fully unconstrained fit. The ratios in the last two columns measure the factor difference between the simulation $\rek$ and that predicted by our relations, i.e., a unity ratio would correspond to perfect agreement.}
\end{tablenotes}
\end{threeparttable}
\end{table*}
\endgroup

%%%%%%%%%%%%%%%%%%%%%%%%%%%%%%%%%%%%%%%%%%%%%%%%%%
\begin{figure}
    \centering
    \includegraphics[width=1.0\linewidth]{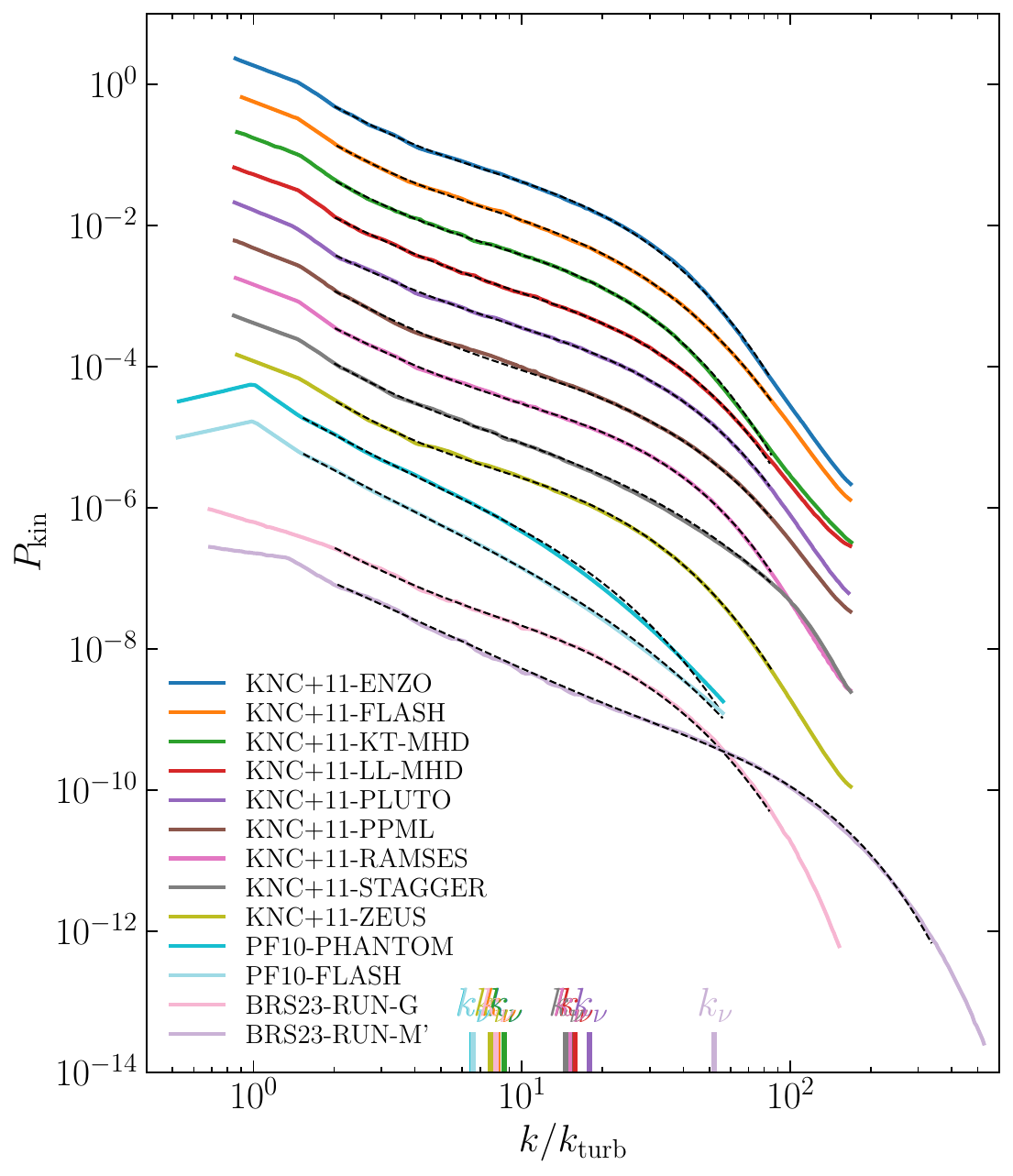}
    \caption{Velocity power spectra ($\Pkin$) obtained in simulations using a wide range of different numerical methods/codes/solvers from the literature (column~1 in Tab.~\ref{tab:compare_tab}; see legend) along with fits using Eq.~(\ref{eq:kinspectra}) (black-dashed lines). The spectra along with their fits are shifted arbitrarily along the $\Pkin$ axis to separate the curves for clarity. Tab.~\ref{tab:compare_tab} lists the fit parameters, with the viscous dissipation wave number ($\knu$) in column~8, shown as tick markers coloured as in the legend, on the $k$-axis (here displayed in units of $\kturb$ to allow for a uniform presentation, as $\kturb$ varies in different simulations; see col.~3 in Tab.~\ref{tab:compare_tab}). 
    }
    \label{fig:compare_lit}
\end{figure}
%%%%%%%%%%%%%%%%%%%%%%%%%%%%%%%%%%%%%%%%%%%%%%%%%%

So far we only discussed the variations in the exact value of numerical dissipation due to the specifics of the numerical solver employed between \citet{2023ApJ...942L..34G}'s and our numerical scheme. Here, we quantify the dependence of the relations between $\rek$ and $N$ across a wide range of state-of-the-art numerical schemes. For this purpose, we extract the velocity spectra, $\Pkin$, from \citet{2011ApJ...737...13K} and \citet{PriceFederrath2010} in the supersonic regime, and from \citet{2023MNRAS.518.6367B} in the subsonic regime using \href{https://automeris.io/}{WebPlotDigitizer}. We repeat our analyses (see \S\ref{ssec:spec_fit} and~\ref{sssec:reykrels}), fitting the velocity spectra to measure $\knu$ to ultimately obtain $\rek$, depending on the specific $\kturb$ and $N$ used in the respective simulation (see Eq.~\ref{eq:RevsN_kt}). We compare a total of 13~simulations, where 9~are ILES from \citet{2011ApJ...737...13K}, 2~are ILES from \citet{PriceFederrath2010}, and another 2~are simulations with explicit viscosity and resistivity from \citet{2023MNRAS.518.6367B}, covering various different numerical methods including schemes with and without Riemann solvers, different reconstruction orders, and smoothed particle hydrodynamics (SPH). A complete list of the 13~simulations and derived results are listed in Tab.~\ref{tab:compare_tab}.

We refer to the literature sources as KNC+11, PF10, and BRS23, which correspond to \citet{2011ApJ...737...13K}, \citet{PriceFederrath2010}, and \citet{2023MNRAS.518.6367B}, respectively. The numerical schemes include grid-based codes/methods (ENZO, FLASH, KT-MHD, LL-MHD, PLUTO, PPML, RAMSES, STAGGER, ZEUS, and PENCIL) and a particle-based (SPH) code/method (PHANTOM). We note that FLASH is a code that supports multiple solvers. \citet{PriceFederrath2010}, \citet{2011ApJ...737...13K}, and our work (see \S\ref{sec:methods}) all use different solvers within FLASH, and this is reflected in the slight differences of the results with those different FLASH solvers below.

Fig.~\ref{fig:compare_lit} shows a comparison of the spectral fits across the 13~different simulations. We find that our $\Pkin$ model Eq.~(\ref{eq:kinspectra}) provides reasonably good fits to the spectra across all 13~simulations. Using the fitted $\knu$, we calculate $\rek$ from Eq.~(\ref{eq:knugeneral}). We then compute the Reynolds number predicted from our relations, with $\nrek$ and $\prek$ taken from the two sets of fit parameters derived in \S\ref{sec:rermpm}. The predicted Reynolds number calculated from the first set ($\prek=\prekth$) and the second set (where $\prek$ was free to vary) are referred to as $\rekpre$ and $\rektilpre$, respectively, in Tab.~\ref{tab:compare_tab}. Additionally, both these values are calibrated to the driving scale $\kturb$ of the corresponding literature source. Therefore, the last two columns in Tab.~\ref{tab:compare_tab} list $\rek/\rekpre$ and $\rek/\rektilpre$, to quantify the variations of our relation between $\rek$ and $N$ across different numerical schemes.

Before discussing the main results of this comparison, we note that spectra belonging to KNC+11 are from decaying turbulence simulations. Hence, even though $\mach=9$ at $t=0$, the spectra we extract are at a time when the Mach number has dropped to $\mach\sim2-3$. We do this, because transient features due to the initial conditions shared between all the ILES studied in KNC+11 have decayed by that time, and we can objectively compare the different codes/schemes. As noted in Appendix~\ref{ap:ronmach}, considering the supersonic regime in the range $2\lesssim\mach\lesssim10$, $\knu$ decreases slightly as $\mach$ decreases, accounting for a $\sim10\%$ reduction in $\knu$ (or $\sim15\%$ reduction in $\rek$) relative to our main relations established at $\mach=10$ (see Fig.~\ref{fig:spectra_mach}). Moreover, in KNC+11 and BRS23, the driving amplitudes are box functions in $k/\kbox=[1,2]$, as opposed to driving amplitudes with a more localised maximum at $k=\kturb$, e.g., by using a parabolic driving spectrum as in our work (see \S\ref{sssec:turb_driv}) or PF10. Therefore, it is not possible to determine the exact value of $\kturb$ in KNC+11 and BRS23, because the same power is injected over the range $k/\kbox=[1,2]$. To incorporate this uncertainty in $\kturb$ into the analysis, we chose an arithmetic mean of $\kturb = 1.5\pm0.5$, which contributes $40-50\%$ uncertainty in the resulting $\rek$ values.

Considering these uncertainties, we concentrate on the mean values of $\rek/\rekpre$ and $\rek/\rektilpre$ as a quantification of the variations in $\rek$ across the wide range of numerical schemes represented by the 13~simulations in Tab.~\ref{tab:compare_tab}. Averaging over this set of simulations, we find $\rek/\rekpre=0.89\pm0.46$ and $\rek/\rektilpre=0.90\pm0.52$, i.e., on average, the different schemes agree with our main work to within a factor of 2--3. The largest deviations from unity are given by PF10-PHANTOM with $\rek/\rekpre=0.41$ and KNC+11-ZEUS with $\rek/\rektilpre=0.41$ on the lower end, and BRS23-RUN-M' with $\rek/\rekpre=1.8$ and $\rek/\rektilpre=2.2$ on the upper end. Thus, the largest deviation from our relations (see \S\ref{para:summ_krels} and Fig.~\ref{fig:reyvsN}) is less than a factor of $3$ across all simulations compared here, which is a representative sample of state-of-the-art schemes used for modelling turbulence, with the notable exception of spectral methods\footnote{While spectral schemes formally resolve all modes down to the Nyquist frequency, they too need large $N$ to model high-$\rek$ turbulence, so we expect spectral methods to follow similar basic principles as the numerical schemes compared here, possibly with $\nrek$ and $\nrem$ values closer to unity. We encourage future work to determine the effective Reynolds numbers of spectral methods, following a similar systematic approach.}.

\subsection{Summary} \label{ssec:num_summ}  

As a general rule, in order to keep the numerical dissipation effects minimal, the explicit dissipation terms have to be chosen such that they exceed the numerical viscosity and resistivity, e.g., the target explicit Reynolds numbers for a particular $N$ must be set to a value smaller than the corresponding numerical value of $\rek$ and $\rem$ determined by an equivalent ILES. The ILES relations obtained here and in \citeauthor{2023ApJ...942L..34G}'s work can serve as a good benchmark, as we have seen in the previous subsection that our relations agree across several different numerical schemes to within a factor of 3. Using these relations, one can ensure that the target $\rek$ and $\rem$ are actually resolved for a given $N$. On the contrary, if $N$ is too small for a target $\rek$ and $\rem$, then the actual $\rek$ and $\rem$ will be limited by numerical dissipation as quantified here. While there is clearly a dependence on the specifics of the numerical method at a factor $<3$ level, the principal dependence of the numerical $\rek$, $\rem$, and $\pem$ is on the number of resolution elements $N$, as demonstrated by the systematic relations with $N$ obtained here, and associated order-of-magnitude variations in the Reynolds numbers with $N$, as opposed to factor $<3$ variations across a wide range of numerical methods (see \S\ref{ssec:num_scheme}). This demonstrates that numerical dissipation is primarily $N$-dependent. In other words, while a better numerical scheme can achieve factors of a few higher Reynolds numbers at the same $N$ compared to a more dissipative numerical scheme, one ultimately has to increase $N$ to reach high-$\rek$ flows. While this is qualitatively known in the turbulence simulation community, the present study provides a systematic quantification of the dependence of the Reynolds numbers on $N$ in various regimes of MHD turbulence (subsonic vs.~supersonic, and weak-field vs.~strong-field regimes) to achieve target explicit Reynolds numbers for their specific turbulence application or to understand the effective Reynolds number when no explicit dissipation is used.

%We have established that the average variation of the exact value of $\rek$ and $\rem$ across state-of-the-art grid-based and SPH numerical schemes from our models is a factor of $1.74$ (due to the dependence of $\nrek$ and $\nrem$ on the details of the numerical solvers and MHD schemes used in a particular code; see \S\ref{ssec:num_scheme}), as compared to the orders-of-magnitude variation with the grid resolution. Therefore, numerical dissipation primarily depends on the resolution $N$ (see \S\ref{ssec:summ} and discussion above). The present study provides relations for the numerical $\rek$ and $\rem$ as a function of $N$, which provides users with an estimate of the grid resolution required to achieve target explicit Reynolds numbers.

\section{Summary and Conclusions}
\label{sec:conclusions}

We determined the effective hydrodynamic Reynolds number ($\rek$), the magnetic Prandlt number ($\pem$), and the magnetic Reynolds number ($\rem$) as a function of linear grid resolution ($N$), from MHD simulations with purely numerical viscosity and resistivity, i.e., in Implicit Large Eddy Simulations (ILES). To do so, we studied the kinetic and magnetic power spectra of the turbulent dynamo in detail. Throughout the study, we distinguish the subsonic ($\mach = 0.1$) and the supersonic ($\mach = 10$) regime of turbulence. We summarise our results as follows:
\begin{itemize}
    \item Through the time evolution of the magnetic-to-kinetic energy ratio ($\Emag/\Ekin$) in the kinematic phase of the dynamo (see Fig.~\ref{fig:Time_evol}), we show that the rate of growth of the magnetic field increases with the linear grid resolution $N$ (c.f., column~2 of Tab.~\ref{tab:fit_params}). The qualitative differences in the magnetic field morphology between the subsonic and supersonic regimes in Figs.~\ref{fig:slices0p1} and~\ref{fig:slices10} demonstrate that smaller-scale magnetic field structures are increasingly resolved with improved grid resolution. This implies that the amplification of the magnetic field increases with increasing $N$.

    \item In order to quantify these qualitative findings, we perform spectral (Fourier) analysis on the time-averaged kinetic ($\Pkin$) and magnetic ($\Pmag$) power spectra (see Fig.~\ref{fig:spectra}) obtained from our simulations in the kinematic regime of the dynamo. To determine the viscous ($\knu$) and resistive ($\keta$) dissipation wave numbers, we fit model spectra to the simulations (see Eqs.~\ref{eq:kinspectra} and~\ref{eq:magspectra}). The extracted $\knu$ and $\keta$ from $\Pkin$ and $\Pmag$ (see columns~11 and~12 of Tab.~\ref{tab:fit_params}) quantify our previous findings, i.e., the dissipation scales and the associated Reynolds numbers depend on the grid resolution $N$.

    \item We repeat the above analyses on $\Pkin$ and $\Pmag$ in the saturation regime of the dynamo (see Fig.~\ref{fig:saturation_regime}) and extract $\knu$ and $\keta$ (see columns~11 and~12 of Tab.~\ref{tab:sat_regime}). Consistent with earlier works, we find that the kinetic-to-magnetic energy conversion is more efficient in the subsonic regime compared to the supersonic regime (see column~2 of Tab.~\ref{tab:sat_regime}), such that the magnetic field induces a back-reaction onto the kinetic flow. The derived $\knu$ is similar to within $10\%$ between the kinematic and saturated regime in all cases. The magnetic spectra experience a significant change upon saturation, with the peak of the magnetic spectrum shifting to larger scales (smaller wave numbers). However, despite the major change in the spectral shape, especially in the subsonic regime, $\keta$ is only reduced by a factor of $\sim2$ in the saturation phase compared to the kinematic phase. For the supersonic regime, the reduction in $\keta$ is only $2\%$. Therefore, the $N$-dependence of the numerical dissipation quantified in detail for the kinematic phase still holds for the saturation phase, with differences by a factor of $\sim2$ in the coefficient of the derived $\keta$ relation for the subsonic regime.

    \item In order to quantify the $N$-dependence of the dissipation wave numbers and the associated Reynolds numbers, we fit $\knu$ and $\keta$ in the kinematic regime of the dynamo as functions of $N$ (see Fig.~\ref{fig:kvsN}) with the formulations proposed from the theory of turbulence (Kolmogorov and Burgers turbulence for the subsonic and supersonic regimes, respectively; see Eqs.~\ref{eq:knuvsN} and~\ref{eq:ketavsN}), and translate that information to Fig.~\ref{fig:reyvsN} to show $\rek$, $\pem$, and $\rem$ as functions of $N$. 

    \item While the results of $\knu$ and $\rek$ as functions of $N$ are in agreement with the predictions of Kolmogorov turbulence ($\prek = 4/3$) in the subsonic regime, the measured $\prek = 1.90_{-0.04}^{+0.04}$ for the supersonic regime exceeds the expectation of Burgers turbulence ($\prek = 3/2$) by $\sim28\%$, which is further explored by a decomposition into longitudinal and transverse velocity modes. We further find that $\keta$ is a linear function of $N$ with the coefficients being $0.028_{-0.005}^{+0.002}$ in the subsonic regime and $0.065_{-0.005}^{+0.006}$ in the supersonic regime.

    \item Overall, we find the relations $\rek=[2(N/\nrek)(\lturb/L)]^{\prek}$ and $\rem=[2(N/\nrem)(\lturb/L)]^{\prem}$, with the ratio of driving scale ($\lturb$) to simulation box scale ($L$), with two sets of $(\nrek, \prek, \nrem, \prem)$ as follows:
    \begin{itemize}
        \item Subsonic regime: $(1.57_{-0.12}^{+0.10}, 4/3, 1.55_{-0.14}^{+0.45}, 4/3)$ and $(0.83_{-0.08}^{+0.09}, 1.20_{-0.02}^{+0.02}, 4.19_{-4.05}^{+2.95}, 1.60_{-0.33}^{+0.18})$.
        \item Supersonic regime: $(3.55_{-0.56}^{+0.78}, 3/2, 1.03_{-0.11}^{+0.12}, 3/2)$ and $(10.46_{-0.85}^{+0.96},1.90_{-0.04}^{+0.04},0.44_{-0.23}^{+0.61},1.32_{-0.09}^{+0.17})$.
    \end{itemize}

    \item Formally, we find a dependence of $\pem$ on $N$. However, our data is also consistent with constant values of $\pem=1.0_{-0.2}^{+0.3}$ for $\mach = 0.1$, and $6.2_{-4.8}^{+5.6}$ for $\mach = 10$.

    \item In order to put these results into a broader context, we compare them to figure~5 in \citet{2023ApJ...942L..34G}, restricted to the subsonic regime (\citealt{2023ApJ...942L..34G} only investigated subsonic turbulence). We find that while $\nrek$ and $\nrem$ obtained from our simulations are larger than the ones obtained from our fits to \citeauthor{2023ApJ...942L..34G}'s data by a factor of $\sim2$, which is due to the differences in the MHD solver and code used in both works, the dependence of $\rek$ and $\rem$ on $N$, i.e., $\prek$ and $\prem$, agrees well in both data sets. 

    \item We further implement our methods on spectra from \citet{PriceFederrath2010}, \citet{2011ApJ...737...13K}, and \citet{2023MNRAS.518.6367B}, which span $13$~different simulations with very different numerical methods, including grid-based (with and without Riemann solvers, and different reconstruction schemes and orders) and smoothed particle hydrodynamics (SPH), and compare them to the main results of this work in \S\ref{ssec:num_scheme}. We find that the $\rek$ extracted from these spectral analyses agrees with $\rek$ from our best-fit models calculated at the corresponding $N$ and $\kturb$ to within a factor of $3$ in both subsonic and supersonic regimes.

    \item We show the target $\rek$, $\pem$, and $\rem$ from simulations with explicit viscosity and resistivity from various works in the literature together with the ILES data from this work and from \citet{2023ApJ...942L..34G}, and together with the respective model functions of $N$ in Fig.~\ref{fig:link}. Our relations together with the numerical method comparison presented in \S\ref{ssec:num_scheme} can be used to estimate the effective $\rek$, $\rem$, and $\pem$ for a given $N$, and to check whether a particular target Reynolds number can be achieved for a given $N$.

\end{itemize}

We conclude that for numerical simulations of turbulence to converge to the intended physical conditions, the explicit (target) Reynolds numbers ($\rek$, and for magnetised flows, also $\rem$) must be set such that they are below the corresponding numerical (ILES) Reynolds numbers obtained in this work for a chosen numerical grid resolution $N$. 

\section*{Acknowledgements}
We thank the referee for their thorough and insightful comments, which have significantly contributed to the clarity, rigour, and generality of this work. C.~F.~acknowledges funding provided by the Australian Research Council (Future Fellowship FT180100495 and Discovery Project grants DP230102280 and DP250101526), and the Australia-Germany Joint Research Cooperation Scheme (UA-DAAD). We further acknowledge high-performance computing resources provided by the Leibniz Rechenzentrum and the Gauss Centre for Supercomputing (grants~pr32lo, pr48pi and GCS Large-scale project~10391), the Australian National Computational Infrastructure (grant~ek9) and the Pawsey Supercomputing Centre (project~pawsey0810) in the framework of the National Computational Merit Allocation Scheme and the ANU Merit Allocation Scheme. The simulation software, \texttt{FLASH}, was in part developed by the Flash Centre for Computational Science at the University of Chicago and the Department of Physics and Astronomy of the University of Rochester.

%%%%%%%%%%%%%%%%%%%%%%%%%%%%%%%%%%%%%%%%%%%%%%%%%%
\section*{Data Availability}
The simulation data underlying this paper, which is a total of $\sim200\,\mathrm{TB}$, or specific parts thereof, will be shared on reasonable request to the authors.

%%%%%%%%%%%%%%%%%%%% REFERENCES %%%%%%%%%%%%%%%%%%

%%%%%%%%%%%%%%%%%%%%%%%%%%%%%%%%%%%%%%%%%%%%%%%%%%

%%%%%%%%%%%%%%%%% APPENDICES %%%%%%%%%%%%%%%%%%%%%

\appendix

\section{Compensated kinetic spectra}
\label{ap:comp}

\begin{figure*}
    \centering
    \includegraphics[width=1.0\linewidth]{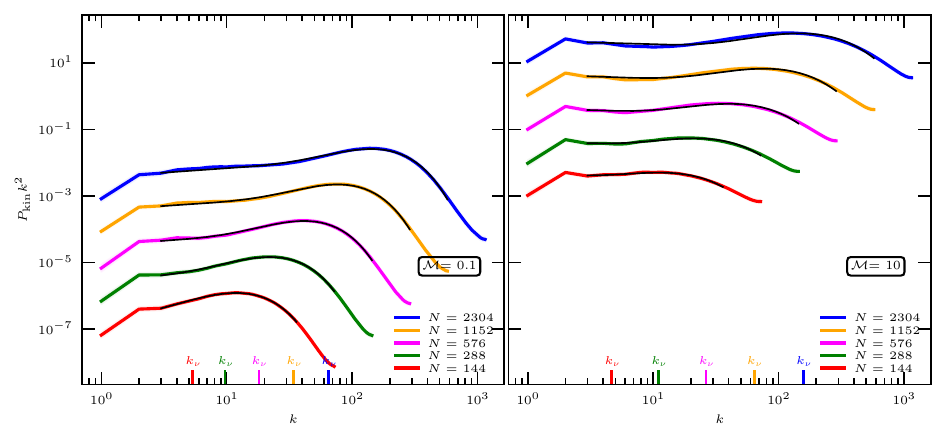}
    \caption{Same as the top panels of Fig.~\ref{fig:spectra}, but here the kinetic power spectra ($\Pkin$) are compensated by $k^{-2}$ to demonstrate the quality of the fit, including power-law section, Bottleneck, and dissipation range.}
    \label{fig:comp_spec}
\end{figure*}

Fig.~\ref{fig:comp_spec} shows the time-averaged kinetic power spectra, compensated by $k^{-2}$, to visualise the quality of the fit using Eq.~(\ref{eq:kinspectra}) to $\Pkin$. We see that Eq.~(\ref{eq:kinspectra}) is an excellent model for $\Pkin$, despite fixing the values of $\pkin$ and $\pnu$ (see Tab.~\ref{tab:fit_params}).

\section{Decomposed kinetic spectra in the supersonic regime}
\label{ap:decomp_spec}

\begin{figure*}
    \includegraphics[width=1.0\linewidth]{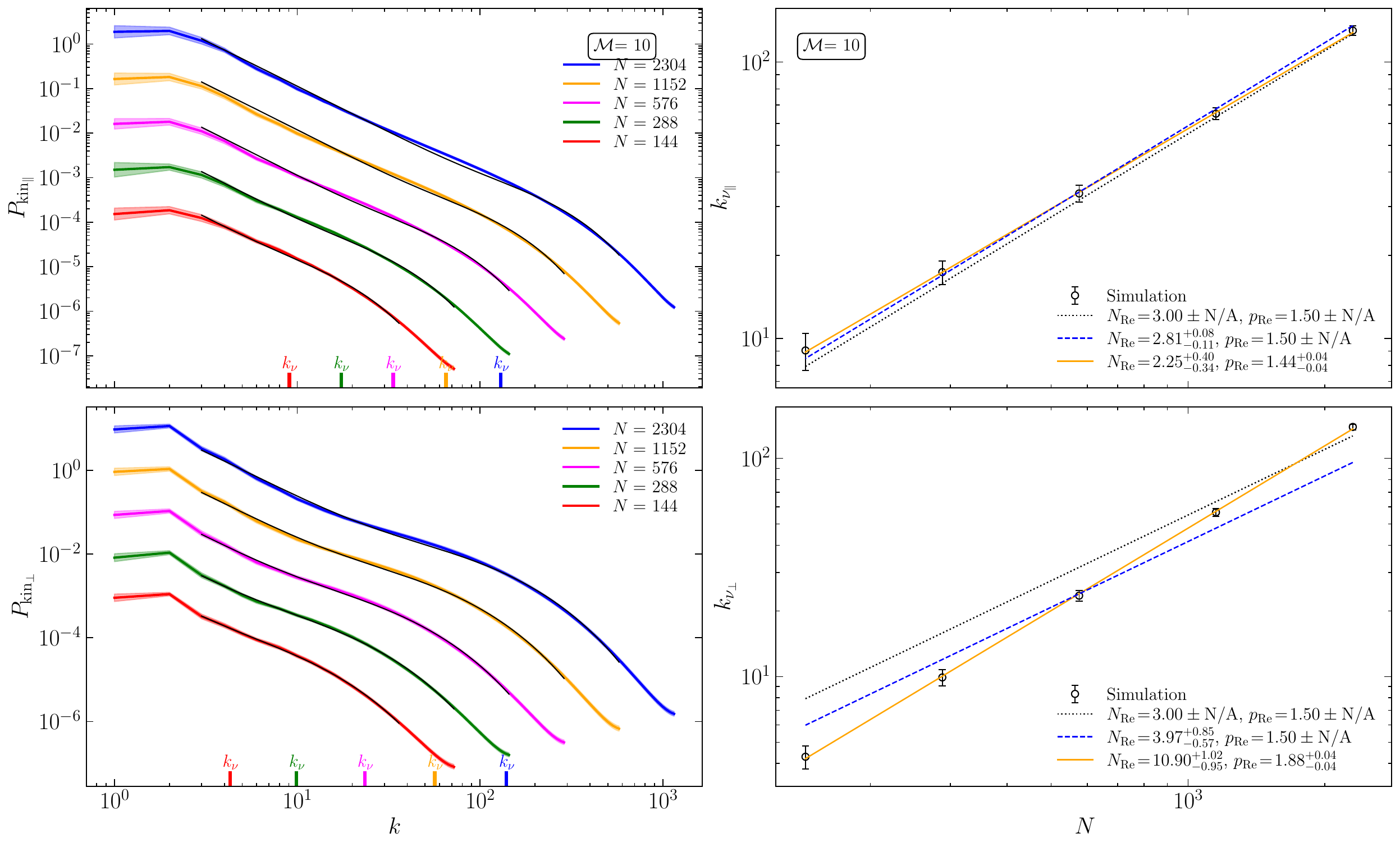}
    \caption{Same as the top right panels of Fig.~\ref{fig:spectra} and Fig.~\ref{fig:kvsN}, but here for the longitudinal (top panels; $P_{\mathrm{kin}_{\parallel}}$) and transverse (bottom panels; $P_{\mathrm{kin}_{\perp}}$) components of the kinetic spectra (left panels) in the supersonic regime, along with their respective derived viscous dissipation wave numbers (right panels; see column~7 of Tab.~\ref{tab:sep_fit_params} for $k_{\nu_{\parallel}}$ and $k_{\nu_{\perp}}$). The methodology of the fitting is the same as discussed in \S\ref{ssec:spec_fit} and \S\ref{ssec:kvsN}.}
    \label{fig:decomp_spec}
\end{figure*}

\begingroup
\setlength{\tabcolsep}{7pt} % 
\renewcommand{\arraystretch}{1.2}
\begin{table*}
	\caption{Kinetic parameters as in Tab.~\ref{tab:fit_params}, but here for the decomposition of the kinetic power spectrum of the kinematic phase into its longitudinal ($P_{\mathrm{kin}_{\parallel}}$) and transverse ($P_{\mathrm{kin}_{\perp}}$) components.}
	\label{tab:sep_fit_params}
\begin{center}
\begin{tabular}{r|ccccc|cc}
\hline
\multicolumn{8}{c}{$\mach = 10$}  \\
\hline
\multicolumn{1}{c|}{$N$} & \multicolumn{1}{c}{$\pkin$} & \multicolumn{1}{c}{$\pbn$} & \multicolumn{1}{c}{$\pnu$} & \multicolumn{1}{c}{$\kbn$} & \multicolumn{1}{c|}{$\knut$} &  \multicolumn{1}{c}{$\knu$} & \multicolumn{1}{c}{$\rek$} \\
\multicolumn{1}{c|}{(1)} & \multicolumn{1}{c}{(2)} & \multicolumn{1}{c}{(3)} & \multicolumn{1}{c}{(4)} & \multicolumn{1}{c}{(5)} & \multicolumn{1}{c|}{(6)} & \multicolumn{1}{c}{(7)} & \multicolumn{1}{c}{(8)} \\
\hline
\multicolumn{1}{c}{} & \multicolumn{5}{c}{\textbf{From $P_{\mathrm{kin}_{\parallel}}$}} & \multicolumn{2}{c}{\textbf{Derived}}\\
\hline
2304 & -2.0 & $0.00\pm0.00$  & 1.0 & $90.0\pm1.5$ & $130.0\pm2.6$ & $130.0\pm2.6$ & $2.2_{-0.5}^{+0.7}\times 10^{4}$ \\
1152 & -2.0 & $0.00\pm0.01$  & 1.0 & $48.2\pm0.9$ & $65.1\pm1.6$  & $65.1\pm1.6$  & $7.8_{-2.0}^{+3.0}\times 10^{3}$ \\
576  & -2.0 & $0.00\pm0.05$  & 1.0 & $25.1\pm0.6$ & $33.5\pm1.2$  & $33.5\pm1.2$ & $2.9_{-0.6}^{+1.0}\times 10^{3}$ \\
288  & -2.0 & $0.00\pm0.05$  & 1.0 & $13.5\pm0.4$ & $17.4\pm0.9$ & $17.4\pm0.9$ & $1.1_{-0.2}^{+0.4}\times 10^{3}$ \\
144  & -2.0 & $0.00\pm6.50$  & 1.0 & $7.54\pm0.41$ & $9.05\pm0.70$  & $9.05\pm0.70$  & $4.1_{-1.0}^{+1.0}\times 10^{2}$ \\
\hline
\multicolumn{1}{c}{} & \multicolumn{5}{c}{\textbf{From $P_{\mathrm{kin}_{\perp}}$}} & \multicolumn{2}{c}{\textbf{Derived}}\\
\hline
2304 & -2.0 & $0.00\pm0.00$  & 0.7 & $24.8\pm0.5$ & $31.7\pm0.3$ & $139.4\pm2.1$ & $2.5_{-0.5}^{+0.8}\times 10^{4}$ \\
1152 & -2.0 & $0.00\pm1.70$  & 0.7 & $14.8\pm0.3$ & $16.9\pm0.2$ & $56.6\pm1.1$  & $6.3_{-1.0}^{+2.0}\times 10^{3}$ \\
576  & -2.0 & $0.00\pm0.04$  & 0.7 & $9.17\pm0.27$ & $9.11\pm0.18$ & $23.5\pm0.7$ & 
$1.7_{-0.4}^{+0.6}\times 10^{3}$ \\
288  & -2.0 & $0.00\pm0.05$  & 0.7 & $5.74\pm0.24$ & $4.98\pm0.15$ & $9.91\pm0.42$  & $4.6_{-1.0}^{+2.0}\times 10^{2}$ \\
144  & -2.0 & $0.00\pm0.06$  & 0.7 & $3.74\pm0.22$ & $2.77\pm0.12$ & $4.29\pm0.26$  & $1.3_{-0.3}^{+0.5}\times 10^{2}$ \\
\hline
\end{tabular}
\end{center}
\end{table*}
\endgroup

In the main part of the study of the kinematic phase, we found that $\knu$ in the supersonic regime has a stronger scaling on $N$ than predicted by Burgers turbulence, i.e., we found $\prek=1.90$ instead of $\prek=3/2$. To elucidate this finding, we perform a Helmholtz decomposition and study the scaling of $\knu$ obtained from the longitudinal and transverse components of the kinetic power spectrum with $N$, separately. The methodology applied is otherwise the same as discussed in \S\ref{ssec:spec_fit} and \S\ref{ssec:kvsN}. The fit parameters from fitting the decomposed power spectra are listed in Tab.~\ref{tab:sep_fit_params}.

Fig.~\ref{fig:decomp_spec} shows the longitudinal (transverse) kinetic power spectrum $P_{\mathrm{kin}_{\parallel}}$ ($P_{\mathrm{kin}_{\perp}}$) and its viscous dissipation wave number $k_{\nu_{\parallel}}$ ($k_{\nu_{\perp}}$) along with their fits for our series of simulations in the top (bottom) panels.

We start by noting some of the theoretical expectations for the longitudinal ($P_{\mathrm{kin}_{\parallel}}$) and transverse ($P_{\mathrm{kin}_{\perp}}$) components of $\Pkin$. We emphasise that the transverse component is related to vorticity ($\nabla\times\textbf{u}$), which is dominant on small scales, while the longitudinal component relates to compression in shocks, which are correlated on relatively larger scales.

We now discuss the resulting $\knu$ vs.~$N$ plots for the decomposed spectra (right panels of Fig.~\ref{fig:decomp_spec}). Interestingly, for the longitudinal component ($k_{\nu_{\parallel}}$), we find that $\prek\sim3/2$, as theoretically expected for Burgers turbulence (c.f., Eq.~\ref{eq:pre_kin} and \S\ref{sssec:effecK_dic}). For the transverse component, we find practically the same results as in the main text, where we do not do the Helmholtz decomposition, i.e., where we only consider the total kinetic power. This easy to understand, because most of the kinetic power is in the transverse component even for supersonic turbulence \citep[see e.g., fig.~14, bottom left panel, in][]{2010A&A...512A..81F}.

A possible explanation for the unexpected value of $\prek$ obtained from $k_{\nu_{\perp}}$ is that, in the supersonic regime, the rotating fluid components (that exist on small spatial scales) are encompassed within the shocks caused due to compression (occurring at larger spatial scales). Therefore, the geometry and isotropic nature of the vorticity elements get distorted due to the stretching and compression. Therefore, the behaviour of $\prek$ may differ even within a spatial scale.

We note that even though the transverse component of the kinetic spectra is primarily guided by the vorticity of fluid elements, it is not entirely incompressible, as there still exists a continuous exchange of energy between the transverse and longitudinal components, which influences the nature of $P_{\mathrm{kin}_{\perp}}$. Therefore, it is to be wondered what the ratio of contribution of `purely' transverse and longitudinal components are to the nature of both $\Pkin$ and $\Pmag$.

Therefore, we conclude by saying that our study, while mainly focusing on the dissipation range of the spectra, has introduced questions about the behaviour of the longitudinal and transverse components of energies and the resulting parameters, which requires further research.

\section{Time evolution from the kinetic into the saturation phase} 
\label{ap:sat_reg_time_evol}

\begin{figure}
    \centering
    \includegraphics[width=1.0\linewidth]{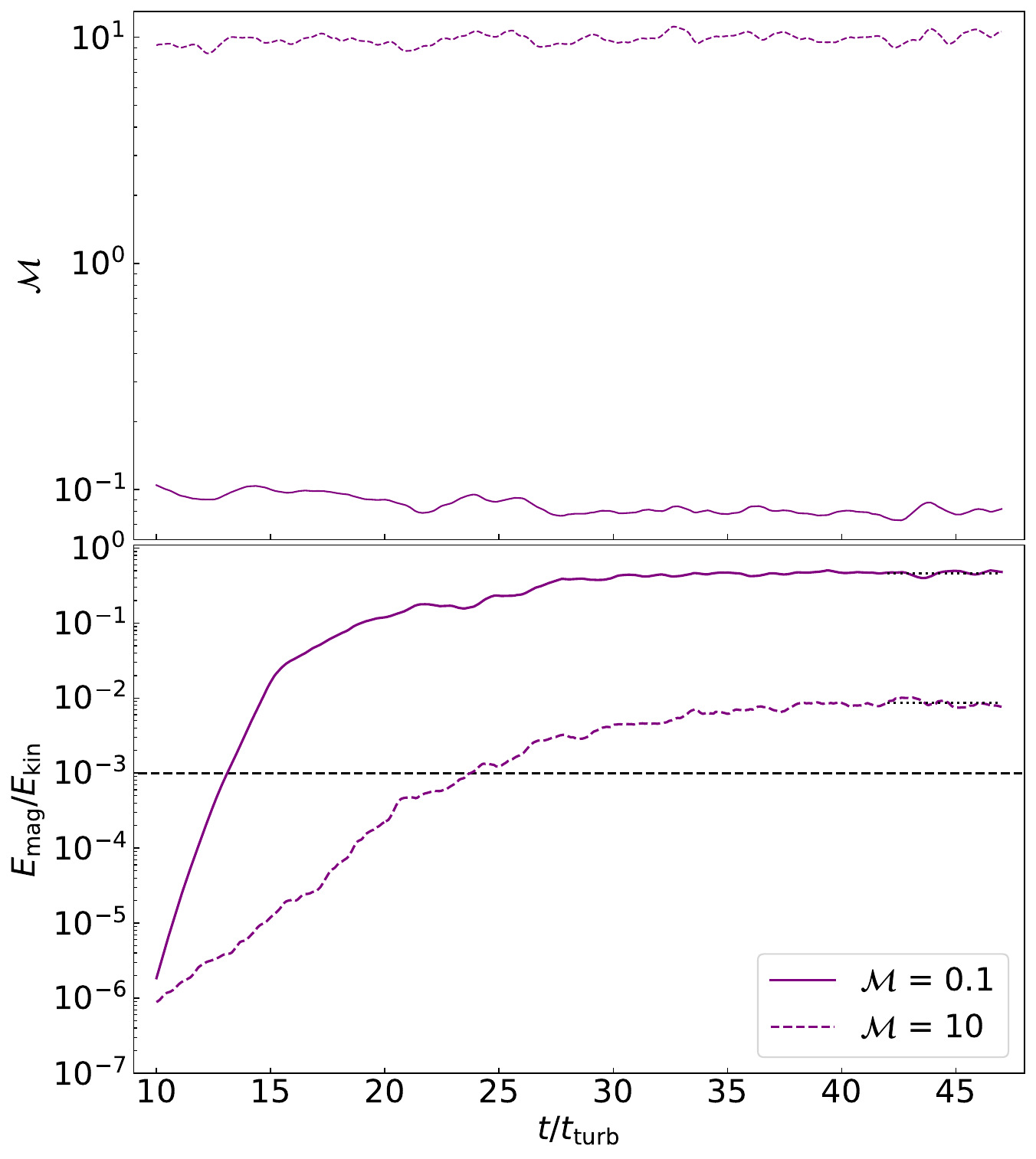}
    \caption{Same as Fig.~\ref{fig:Time_evol}, extended into the saturation phase of the dynamo, but only for $N=576$, in the $\mach = 0.1$ (solid) and $\mach = 10$ (dashed) regimes of turbulence. The saturation level is fitted in the saturation time window, $37\leq t/\tturb\leq 47$, and shown as the dotted lines (see col.~2 in Tab.~\ref{tab:sat_regime}). 
    }
    \label{fig:time_evol_sat}
\end{figure}

Fig.~\ref{fig:time_evol_sat} shows an extension of the time evolution shown in Fig.~\ref{fig:Time_evol} of the main text, but here followed into the saturation phase for the $N=576$ simulations at $\mach=0.1$ (solid) and $\mach=10$ (dashed). We clearly see how the kinematic phase transitions into the saturation phase. We further indicate the fits for the saturation level as dotted lines, in the saturation range, which we define as $37\leq t/\tturb\leq47$ based on visual inspection of the time evolution. We find $(\Emag/\Ekin)_{\mathrm{sat}}\sim0.46$ and $0.0087$ for the subsonic and supersonic regime, respectively. These are also listed in column~2 of Tab.~\ref{tab:sat_regime}. The values found here for $\mach=0.1$ and $10$, are consistent with earlier studies \citep{FederrathEtAl2011,2021PhRvF...6j3701S}, showing that the supersonic regime has substantially lower saturation levels compared to the subsonic regime of turbulence.

\section{$\cnu$ and $\ceta$ in the supersonic regime}
\label{ap:cons_calc}

\newcommand{\cnuktf}{c_{\nu\mathrm{,K25}}} % c_nu,K25
\newcommand{\cnusub}{\cnu^\mathrm{sub}}
\newcommand{\cnusup}{\cnu^\mathrm{sup}}
\newcommand{\cnuktfsub}{\cnuktf^\mathrm{sub}}
\newcommand{\cnuktfsup}{\cnuktf^\mathrm{sup}}

\newcommand{\cetaktf}{c_{\eta\mathrm{,K25}}} % c_eta,K25
\newcommand{\cetasub}{\ceta^\mathrm{sub}}
\newcommand{\cetasup}{\ceta^\mathrm{sup}}
\newcommand{\cetaktfsub}{\cetaktf^\mathrm{sub}}
\newcommand{\cetaktfsup}{\cetaktf^\mathrm{sup}}

We note that \citet[][eqs.~15 and~16, respectively]{2022MNRAS.513.2457K} provide values of $\cnu$ and $\ceta$ in the subsonic regime, but not in the supersonic regime. The more recent study by \citet[][fig.~6]{KrielEtAl2025} provides equivalent values for both the subsonic and supersonic regimes. However, \citet{2022MNRAS.513.2457K} and \citet{KrielEtAl2025} each use different definitions of $\knu$ and $\keta$, in that \citet{2022MNRAS.513.2457K} define $\knu$ and $\keta$ as we use it here based on spectral fits, while \citet[][]{KrielEtAl2025} define $\knu$ and $\keta$ based on the spectrum of $\rek$ and electric current, respectively, by locating the wave numbers where $\rek=1$ and where the current peaks, respectively. However, we need $\knu$ and $\keta$ in the previous definition of \citet{2022MNRAS.513.2457K}, i.e., those based on spectral fitting. Hence, we translate the coefficients from \citep[][fig.~6]{KrielEtAl2025} to the definition in \citet{2022MNRAS.513.2457K}. For this, we refer to $\cnu$ and $\ceta$, and $\cnuktf$ and $\cetaktf$ from \citet{2022MNRAS.513.2457K} and \citet[][fig.~6]{KrielEtAl2025}, respectively. Thus, for the subsonic regime we can directly use $\cnu=\cnusub=0.025$ and $\ceta=\cetasub=0.88$ from \citet[][eqs.~15 and~16, respectively]{2022MNRAS.513.2457K}. In the supersonic regime, we get
\begin{equation}
\cnu=\cnusup=\cnuktfsup\times\frac{\cnusub}{\cnuktfsub}=0.33\times\frac{0.025}{0.1}=0.083,
\end{equation}
and
\begin{equation}
\ceta=\cetasup=\cetaktfsup\times\frac{\cetasub}{\cetaktfsub}=0.33\times\frac{0.88}{0.53}=0.55.
\end{equation}
The corresponding errors of the coefficients were calculated accordingly.

\section{Relation between $\prek$ and $\pkin$} \label{ap:p_re_kin}

Here we derive the relation between $\prek$ and $\pkin$. We note that
\begin{equation}
    \Pkin(k) \propto \frac{du_k^2}{dk},
\end{equation}
where $u_{k}$ is the velocity of the gas at wave number $k$. As $\Pkin(k) \propto k^{\pkin}$ (see Eq.~\ref{eq:kinspectra}), we can write
\begin{equation}
    u_k^2 \propto k^{\pkin+1}.
\end{equation}
In terms of the length scale ($\ell$), this is equivalent to
\begin{equation}
    u_\ell \propto \ell^{-(\pkin+1)/2}.
\end{equation}
From the definition of the Reynolds number, we have
\begin{equation}
    \rek = \frac{\ell u_\ell}{\nu} \propto \ell^{(1-\pkin)/2}, \label{eq:Revsl}
\end{equation}
because $\nu=\mathrm{const}$. We further have $N\propto\ell/\ell_\nu$, where $\ell_\nu=2\pi/\knu$ is the length scale at which viscous dissipation takes over in the kinetic spectrum $\Pkin$. Therefore, from Eq.~(\ref{eq:fitRevsN}), we find
\begin{equation}
    \rek \propto \left(\frac{\ell}{\ell_\nu}\right)^{\prek} \propto \ell^{\prek}. \label{eq:Revsl_N}
\end{equation}
Thus, from Eqs.~(\ref{eq:Revsl}) and~(\ref{eq:Revsl_N}), we find
\begin{equation}
    \prek = \frac{1-\pkin}{2}.
\end{equation}

\section{Relation between $\prek$ and $\prem$}
\label{ap:p_re_rm}

Here we derive the relation between $\prek$ and $\prem$. Plugging Eq.~(\ref{eq:knugeneral}) with $\prek = \prekth$ and Eq.~(\ref{eq:pm_rel}) into Eq.~(\ref{eq:ketageneral}), and using Eqs.~(\ref{eq:fitRevsN}) and~(\ref{eq:fitRmvsN}), we find
\begin{align}
\keta &\propto \knu\pem^{1/2} \propto \rek^{1/{\prekth}}\left({\frac{\rem}{\rek}}\right)^{1/2} \propto \rek^{\frac{2-\prekth}{2\prekth}}\rem^{1/2},\nonumber\\
    &\propto \left({\frac{N}{\nrek}}\right)^{\prek\,\frac{2-\prekth}{2\prekth}}\left({\frac{N}{\nrem}}\right)^{\frac{\prem}{2}} \propto N^{\left(\frac{2-\prekth}{2\prekth}\right)\,\prek+\frac{\prem}{2}}.
    \label{eq:ketavsN_der}
\end{align}

As $\keta$ varies linearly with $N$ (see \S\ref{sssec:effecK_dic}), the exponent of $N$ in Eq.~(\ref{eq:ketavsN_der}) is unity. Therefore,
\begin{equation}
\left(\frac{2-\prekth}{2\prekth}\right)\prek+\frac{\prem}{2} = 1.
\end{equation}
For $\mach = 0.1$, given $\prekth = \frac{4}{3}$, we find
\begin{align}
\left(\frac{2-\frac{4}{3}}{2\times\frac{4}{3}}\right)\prek+\frac{\prem}{2} &= 1
\implies \frac{\prek}{4} + \frac{\prem}{2}  = 1,\\
\implies \prem = \frac{4-\prek}{2}. \label{eq:p_rel_sub}
\end{align}
Similarly, for $\mach = 10$, with $\prekth = \frac{3}{2}$, we have
\begin{align}
\left(\frac{2-\frac{3}{2}}{2\times\frac{3}{2}}\right)\prek+\frac{\prem}{2} &= 1 \implies \frac{\prek}{6} + \frac{\prem}{2} = 1,\\
\implies \prem = \frac{6-\prek}{3}. \label{eq:p_rel_sup}
\end{align}
Therefore, we find Eqs.~(\ref{eq:p_rel_sub}) and~(\ref{eq:p_rel_sup}) as the relation between $\prek$ and $\prem$ in the subsonic and supersonic regimes, respectively.

\section{Sonic Mach number dependence} 
\label{ap:ronmach}

\begingroup
\setlength{\tabcolsep}{1.5pt} % 
\renewcommand{\arraystretch}{1.4}
\begin{table*}
	\caption{Simulation parameters and main results of the Mach number study for $N = 576$.}
	\label{tab:mach_dep_params}
\begin{threeparttable}
\begin{tabular}{c|ccccc|ccc|ccccc}
\hline
\multicolumn{1}{c}{} & \multicolumn{5}{c}{\textbf{From $P_{\mathrm{kin}}$}} & \multicolumn{3}{c}{\textbf{From $P_{\mathrm{mag}}$}} & \multicolumn{5}{c}{\textbf{Derived}}\\
\hline
\multicolumn{1}{c|}{$\mach$} & \multicolumn{1}{c}{$\pkin$} & \multicolumn{1}{c}{$\pbn$} & \multicolumn{1}{c}{$\pnu$} & \multicolumn{1}{c}{$\kbn$} & \multicolumn{1}{c|}{$\knut$} & \multicolumn{1}{c}{$\pmag$} & \multicolumn{1}{c}{$\peta$} & \multicolumn{1}{c|}{$\ketat$} & \multicolumn{1}{c}{$\knu$} & \multicolumn{1}{c}{$\keta$} & \multicolumn{1}{c}{$\rek$} & \multicolumn{1}{c}{$\pem$} & \multicolumn{1}{c}{$\rem$}\\
\multicolumn{1}{c|}{(1)} & \multicolumn{1}{c}{(2)} & \multicolumn{1}{c}{(3)} & \multicolumn{1}{c}{(4)} & \multicolumn{1}{c}{(5)} & \multicolumn{1}{c|}{(6)} & \multicolumn{1}{c}{(7)} & \multicolumn{1}{c}{(8)} & \multicolumn{1}{c|}{(9)} & \multicolumn{1}{c}{(10)} & \multicolumn{1}{c}{(11)} & \multicolumn{1}{c}{(12)} & \multicolumn{1}{c}{(13)} & \multicolumn{1}{c}{(14)}\\
\hline
160     & -2.0   & $0.00\pm0.00$   & 0.7 & $7.89\pm0.35$ & $9.82\pm0.26$  & $2.00\pm0.10$    & $0.91\pm0.02$   & $21.2\pm1.4$    & $26.1\pm1.0$  & $28.5\pm4.2$   & $2.0_{-0.4}^{+0.4}\times 10^{3}$ & $3.9_{-2.0}^{+10.0}$	 & $7.8_{-4.0}^{+20.0}\!\times\! 10^{3}$ \\
80      & -2.0   & $0.00\pm0.03$   & 0.7 & $8.02\pm0.34$ & $9.81\pm0.26$  & $1.97\pm0.03$  & $0.91\pm0.02$   & $21.5\pm1.4$    & $26.1\pm1.0$  & $28.8\pm4.1$   & $2.0_{-0.4}^{+0.4}\times 10^{3}$ & $4.0_{-2.0}^{+10.0}$	 & $8.0_{-4.0}^{+20.0}\!\times\! 10^{3}$ \\
40      & -2.0   & $0.00\pm0.01$   & 0.7 & $8.20\pm0.33$  & $9.76\pm0.25$  & $1.80\pm0.03$   & $0.96\pm0.02$  & $26.5\pm1.6$    & $25.9\pm1.0$  & $30.5\pm4.3$   & $2.0_{-0.4}^{+0.4}\times 10^{3}$ & $4.6_{-3.0}^{+10.0}$	 & $9.0_{-5.0}^{+20.0}\!\times\! 10^{3}$ \\
20      & -2.0   & $0.00\pm0.01$   & 0.7 & $8.96\pm0.30$  & $9.72\pm0.22$  & $1.58\pm0.03$  & $0.95\pm0.02$  & $28.2\pm1.8$    & $25.7\pm0.9$  & $33.3\pm4.9$   & $1.9_{-0.4}^{+0.4}\times 10^{3}$ & $5.6_{-3.0}^{+10.0}$	 & $1.1_{-0.6}^{+3.0}\!\times\! 10^{4}$ \\
10      &-2.0  & $0.00\pm0.12$   & 0.7 & $10.5\pm0.3$ & $9.89\pm0.25$  & $1.43\pm0.03$  & $0.86\pm0.02$  & $22.9\pm1.7$    & $26.4\pm1.0$  & $37.8\pm6.6$   & $2.0_{-0.4}^{+0.4}\times 10^{3}$ & $6.7_{-4.0}^{+20.0}$ & 	$1.4_{-0.7}^{+4.0}\!\times\! 10^{4}$  \\
5       & -2.0   & $0.00\pm0.03$   & 0.7 & $12.1\pm0.4$ & $10.1\pm0.32$  & $1.46\pm0.02$  & $0.84\pm0.01$ & $18.4\pm0.7$   & $27.0\pm1.2$   & $31.4\pm2.5$   & $2.1_{-0.5}^{+0.5}\times 10^{3}$ & $4.5_{-2.0}^{+10.0}$ & 	$9.4_{-5.0}^{+30.0}\!\times\! 10^{3}$ \\
2.5     & -2.0   & $0.00\pm0.01$   & 0.7 & $10.6\pm0.2$ & $9.52\pm0.16$  & $1.73\pm0.01$ & $0.97\pm0.01$ & $22.1\pm0.3$   & $25.0\pm0.6$   & $24.5\pm0.7$  & $1.9_{-0.4}^{+0.4}\times 10^{3}$ & $3.1_{-2.0}^{+8.0}$ & 	$6.0_{-3.0}^{+20.0}\!\times\! 10^{3}$ \\
1.25    & -2.0   & $0.00\pm0.01$   & 1.0   & $11.1\pm0.2$ & $24.3\pm0.16$  & $2.11\pm0.01$ & $1.1\pm0.1$  & $25.4\pm0.3$   & $24.3\pm0.2$  & $21.6\pm0.4$  & $1.8_{-0.4}^{+0.4}\times 10^{3}$ & $2.6_{-1.0}^{+7.0}$ & 	$4.8_{-3.0}^{+10.0}\!\times\! 10^{3}$ \\
0.8     & -1.7 & $0.26\pm0.02$   & 1.0   & $13.1\pm0.2$ & $21.9\pm0.13$  & $2.22\pm0.01$ & $1.0\pm0.1$  & $23.8\pm0.3$   & $21.9\pm0.1$  & $20.9\pm0.6$   & $3.3_{-0.4}^{+0.4}\times 10^{3}$ & $1.2_{-0.4}^{+0.9}$         & $3.9_{-1.0}^{+3.0}\times 10^{3}$  \\
0.4     & -1.7 & $0.39\pm0.02$   & 1.0   & $12.9\pm0.1$ & $20.3\pm0.1$ & $2.36\pm0.01$ & $1.0\pm0.1$  & $21.0\pm0.3$ & $20.3\pm0.1$ & $20.3\pm0.7$  & $3.0_{-0.4}^{+0.4}\times 10^{3}$ & $1.3_{-0.5}^{+1.0}$         & $3.9_{-1.0}^{+3.0}\times 10^{3}$  \\
0.2     & -1.7 & $0.44\pm0.02$   & 1.0   & $12.3\pm0.1$ & $19.0\pm0.1$ & $2.42\pm0.01$  & $0.98\pm0.01$  & $18.4\pm0.3$   & $19.0\pm0.1$ & $19.4\pm0.7$  & $2.7_{-0.4}^{+0.4}\times 10^{3}$ & $1.3_{-0.5}^{+1.0}$         & $3.7_{-1.0}^{+3.0}\times 10^{3}$  \\
0.1     & -1.7 & $0.42\pm0.02$   & 1.0   & $11.5\pm0.2$ & $18.1\pm0.1$ & $2.56\pm0.02$   & $0.89\pm0.09$ & $12.8\pm0.5$   & $18.1\pm0.1$ & $17.5\pm1.2$   & $2.6_{-0.3}^{+0.3}\times 10^{3}$ & $1.2_{-0.4}^{+1.0}$         & $3.1_{-1.0}^{+2.0}\times 10^{3}$  \\
0.05    & -1.7 & $0.36\pm0.02$   & 1.0   & $10.4\pm0.2$ & $17.3\pm0.1$ & $2.71\pm0.03$  & $0.81\pm0.01$ & $8.47\pm0.4$   & $17.3\pm0.1$ & $13.8\pm1.2$   & $2.4_{-0.3}^{+0.3}\times 10^{3}$ & $0.82_{-0.3}^{+0.7}$        & $2.0_{-0.7}^{+2.0}\times 10^{3}$  \\
0.025   & -1.7 & $0.23\pm0.03$   & 1.0   & $8.90\pm0.28$  & $16.8\pm0.1$  & $2.89\pm0.04$  & $0.72\pm0.01$  & $4.63\pm0.4$   & $16.8\pm0.1$  & $8.44\pm1.20$   & $2.3_{-0.3}^{+0.3}\times 10^{3}$ & $0.32_{-0.1}^{+0.3}$        & $7.5_{-3.0}^{+6.0}\times 10^{2}$  \\
0.0125  & -1.7 & $0.00\pm0.04$ &  1.0   & $6.67\pm0.46$ & $16.6\pm0.2$  & $3.16\pm0.06$  & $0.62\pm0.01$ & $2.10\pm0.20$    & $16.6\pm0.2$  & $3.10\pm0.60$  & $2.3_{-0.3}^{+0.3}\times 10^{3}$ & $0.05_{-0.02}^{+0.04}$     & $1.0_{-0.4}^{+0.9}\times 10^{2}$  \\
0.00625 & -1.7 & $0.00\pm0.10$   & 1.0   & $7.22\pm0.72$ & $15.5\pm0.2$  & $3.47\pm0.06$   & $0.54\pm0.01$  & $0.75\pm0.09$ & $15.5\pm0.2$  & $0.59\pm0.12$ & $2.1_{-0.3}^{+0.3}\times 10^{3}$ & $0.00_{-0.00}^{+0.01}$ & $3.8_{-2.0}^{+3.0}$ \\
\hline
\end{tabular}
\begin{tablenotes}[flushleft]
\note{Same as Tab.~\ref{tab:fit_params}, but for varying $\mach$ at fixed grid resolution, $N=576$.}
\end{tablenotes}
\end{threeparttable}
\end{table*}
\endgroup

%%%%%%%%%%%%%%%%%%%%%%
\begin{figure*}
    \centering    
    \includegraphics[width=1.0\linewidth]{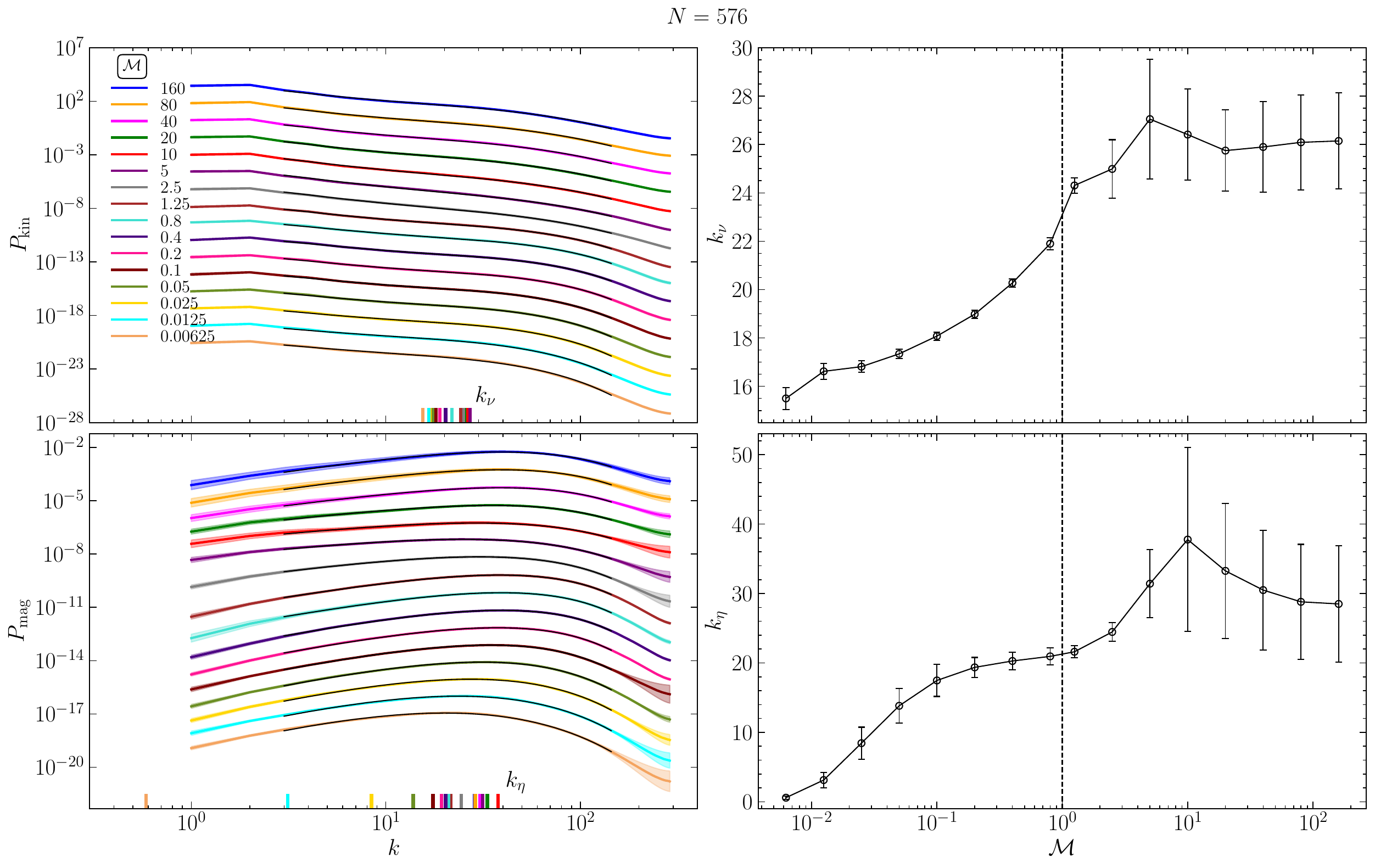}
    \caption{The left panels show the time-averaged kinetic power spectra ($\Pkin$; top) and magnetic power spectra ($\Pmag$; bottom) for a series of Mach numbers, $\mach$ (see legend) for $N = 576$. The right panels show the viscous dissipation wave number ($\knu$), obtained from $\Pkin$ (top; column~10 in Tab.~\ref{tab:mach_dep_params}), and the resistive dissipation wave number ($\keta$) obtained from $\Pmag$ (bottom; column~11 in Tab.~\ref{tab:mach_dep_params}), as a function of $\mach$, along with their 2-sigma variations as error bars.}
    \label{fig:spectra_mach}
\end{figure*}
%%%%%%%%%%%%%%%%%%%%%%%%%%%%%%%%%%%%%%%%%%%%%%%%%%

Here we repeat the analyses from the main part of the study for a series of $\mach$ values to evaluate the dependence of the numerical viscosity and resistivity on the sonic Mach number. We limit this study to a single grid resolution, $N=576$. For this purpose, we consider a wide array of $\mach$ values (column~1 of Tab.~\ref{tab:mach_dep_params}) from $\mach=0.00625$ to $160$, in multiples of $2$ (except for the step from $\mach=0.8$ to 1.25). We fit the same models, Eqs.~(\ref{eq:kinspectra}) and~(\ref{eq:magspectra}), to the averaged kinetic and magnetic spectra, respectively, displayed in the left panels of Fig.~\ref{fig:spectra_mach}. Tab~\ref{tab:mach_dep_params} lists the corresponding fit parameters (columns~2--9) and the dissipation wave numbers, $\knu$ and $\keta$ (columns~10 and~11, respectively), along with the derived $\rek$, $\pem$ and $\rem$ (columns~12,~13 and~14, respectively). Fig.~\ref{fig:spectra_mach} shows a summary of the spectra and the $\mach$ dependence of $\knu$ and $\keta$, with the black-dashed line demarcating the subsonic and supersonic regimes in the right-hand panels. 

We find that the maximum difference in $\knu$ across the 4 orders-of-magnitude in $\mach$ studied here, is just a factor of $\sim2$ (column~10 of Tab.~\ref{tab:mach_dep_params}; top-right panel of Fig.~\ref{fig:spectra_mach}). Additionally, in the supersonic regime, we see that $\knu$ is nearly a constant within the 1-sigma variations. The subsonic regime and up to $\mach\sim5$ shows a monotonic decline in $\knu$ as $\mach$ decreases, with a potential indication of a plateau towards very low $\mach$. In contrast to $\knu$, the maximum difference in $\keta$ is nearly 2 orders-of-magnitude across $\mach$. However, $\keta$ is nearly a constant within the 2-sigma variations in the supersonic regime, similar to $\knu$. In fact, a drastic change in $\keta$ is seen only in the lower $\mach$ regime. In the subsonic regime, till $\mach\sim0.1$, the maximum change in $\keta$ is only a factor of $\sim1.5$. Therefore, considering only data down to $\mach\sim0.1$, the maximum difference in $\keta$ is only a factor of $\sim2$ across 3 orders-of-magnitude in $\mach$, implying only a relatively weak dependence of $\keta$ on $\mach$.

It is only for $\mach<0.1$ that we find significant differences arising, with both $\knu$ and $\keta$ decreasing strongly, which implies a significantly elevated level of dissipation in the regime of very low Mach numbers. This is not surprising, as low-Mach-number flows exhibit substantial numerical dissipation \citep{BirkeEtAl2021,LeidiEtAl2022}. Our results suggest that turbulence simulations with $\mach<0.1$ are significantly affected by numerical dissipation, and special-purpose solvers designed to reduce numerical dissipation in the low-Mach regime are required to achieve high Reynolds numbers.

\bsp	% typesetting comment
\label{lastpage}
\end{document}